\documentclass[letterpaper,twocolumn,10pt]{article}
\usepackage{usenix-2020-09}

\usepackage{tikz}
\usepackage{amsmath}
\usepackage[]{cite}

\usepackage{filecontents}
\usepackage{datetime}
\begin{document}

\title{\Large \bf Key-Value Stores on Flash Storage Devices: A Survey}

\author{
{\rm Krijn Doekemeijer}\\
Vrije Universiteit, Amsterdam and Universiteit van Amsterdam
\and
{\rm Animesh Trivedi}\\
Vrije Universiteit, Amsterdam
} 

\maketitle

\begin{abstract}
    Key-value stores (KV) have become one of the main components of the modern storage and data processing system stack. With
    the increasing need for timely data analysis, performance becomes more and more critical. In the past, these stores were
    frequently optimised to run on
    HDD and DRAM devices. However, the last decade saw an increased interest in the use of flash devices because of their
    attractive properties. Flash is cheaper than DRAM and yet has a lower latency and higher throughput than HDDs. This
    literature survey aims to highlight the changes proposed in the last decade to optimise key-value stores for flash devices
    and predict what role these devices might play for key-value stores in the future.
\end{abstract}


\section{Introduction}
\label{sec:introduction}
It is estimated that we will generate over 175 zettabytes of data globally by the year 2025\cite{patrizio2018idc}. This is
mainly because of the ever-increasing interest in big data, the cloud and the internet of things\cite{patrizio2018idc, syed2013future, khan2014big}. 
As the size of the datasets keeps increasing, so do the demands of the systems that are used to store and process this data.
This in turn has caused for an increased interest in optimising the data processing stack. A big part of this stack is used
by key-value stores. It is therefore beneficial to look into how key-value stores can be optimised.

Key-value stores are a means of storing data and are radically different from the more traditional RDBs, also known as
relational databases\cite{harrington2016relational}. Key-value stores store data as a single collection, where each key is
unique and leads to one value. Data can be accessed using these keys with basic operations such as: get, put, delete and
scan. Key-value stores can be used for all sorts of applications and are not limited to a particular size or hardware. Some
common applications include caching systems\cite{eisenman2019flashield}, messaging applications\cite{cao2020characterizing},
games\cite{debnath2011skimpystash}, web shops\cite{risch2015introduction}, SQL backends \cite{dong2017optimizing} and time
series management\cite{kondylakis2020coconut}.

Traditionally the main storage medium used to store key-value stores was the \textit{Hard Disk Drive}
(HDD)\cite{daim2008forecasting}. Most data structures and algorithms were thus optimised around the physical properties of
these devices. These were among others high latencies, symmetric read and write speeds, slow random access and an infinite
number of reads and writes for each block on the HDD. This caused certain HDD specific optimisations such as trying to
always write and read 
sequentially\cite{o1996log,ghemawat2011leveldb,escriva2012hyperdex,chang2008bigtable,menon2014cassandra}. 

However, as flash devices became cheaper, many data centres and consumers alike transitioned to flash
devices\cite{lee2008case, andersen2010rethinking}. This made it important to ensure that applications can still be properly
used with flash devices and are in addition also optimised for these devices. (Un)fortunately, most of the properties and
assumptions that hold for HDDs, do not hold for flash devices. Flash devices have lower latencies, have asymmetric read and
write speeds, do not allow for in-place updates, have an all-new erase operation, are indifferent to random reads on the
cell level and individual cells have a finite life cycle. The finite life cycle, commonly known as \textit{wear levelling}
(WL), can in particular be problematic if unattended. If applications carelessly keep writing to the same cells, the cells
will eventually stop working correctly. Lower latencies are also important as lower latencies are becoming more critical for
applications\cite{barroso2017attack}. Yet, at the same time lower latencies on flash result in the latency overhead moving
to other parts of the key-value store, such as the software that is executed on the host, and therefore require different
design considerations\cite{barroso2017attack}.  Because of such idiosyncrasies, properly and efficiently using these devices
requires a transition\cite{he2017unwritten}. 

This survey tries to highlight the changes proposed in the last decade for using key-value stores on flash. We will look
into various optimisation strategies that can be used to use key-value stores more efficiently on flash. However, first we
will take a look at flash and key-value stores themselves. We will then combine the two topics and take a look at the main
design concerns that occur when combining them. After having defined the problem space, we will show how these problems can
be solved. We will start by covering the commonly used data structures for key-value stores. This will be followed by how
key-value stores and flash communicate with each other and how this communication can be optimised. Then we will look at
software optimisations, data related optimisations and the current role of flash. Lastly, we will identify forward-looking
trends on the role of flash for key-value stores.

\section{Related work}
\label{sec:relatedwork}
Key-value stores and flash are by no means new technologies, both already appearing in the 20th
century\cite{dipert1993flash, sharma2012sql}. In this section we will look at other surveys on these topics and cover a few
adjacent works. There have been a few surveys that cover flash
storage\cite{Fevgas2019IndexingIF,chung2009survey,yang2014garbage} and there have been a few surveys on key-value
stores\cite{idreos2020key,Cattell2011ScalableSA,davoudian2018survey}. Most of the related works on key-value stores focus on
NoSQL\cite{idreos2020key,Cattell2011ScalableSA,davoudian2018survey}. It is important to note that key-value stores are
considered to be a type of NoSQL, but that there also exist various other NoSQL stores that are by some considered key-value
stores and by others not. Further on, some data stores satisfy the requirements of multiple types of NoSQL. This makes the
exact definition and nomenclature of key-value stores blurry. To give an example, Cassandra can be considered as a
wide-column store\cite{gupta2017nosql} but is at the same time also frequently referred to as a key-value
store\cite{menon2014cassandra}. In this survey we will refrain from using the term NoSQL.

To give a few examples of NoSQL surveys: Idreos et al. cover various key-value storage engines\cite{idreos2020key}, Cartell
et al. cover scalable SQL and NoSQL data stores\cite{Cattell2011ScalableSA}, and Chen et al. did a survey on
NoSQL\cite{davoudian2018survey}. There have also been literature studies on various storage techniques for commonly used
data structures in key-value stores, such as on LSM-trees\cite{luo2020lsm}.

To the best of our knowledge, there has not been a publically available academic survey that covers key-value store
techniques or general optimisations for key-value stores driven by flash. There do exist a few works that aim to help with
understanding key-value stores, such as a practical overview by Seegers et al\cite{seeger2009key} on a few key-value stores.
Other works that we found only used key-value stores. Such as a survey on multimedia systems\cite{pouyanfar2018multimedia},
data storage in the cloud\cite{mansouri2017data} and a survey on big data systems\cite{davoudian2020big}.

For this survey, we are also interested in flash storage. There have been various surveys on parts of flash storage. Such as
surveys on indexing in flash storage\cite{Fevgas2019IndexingIF}, the flash translation layer\cite{chung2009survey} or
garbage collection and wear levelling\cite{yang2014garbage}. We will provide a more comprehensive introduction to these
topics in \autoref{sec:background}.

This survey aims to build the bridge between key-value stores and flash storage. It will thus combine topics from key-value
store (NoSQL) surveys and flash storage surveys. In other words, the focus will not be on the technologies separately, but
on the combination of the two.

Lastly, we shortly address some adjacent works on new storage technologies. Recently there have been hardware trends that
introduced various new storage devices that could also be used as memory; such as Optane memory. A lot of effort has already
been put into moving key-value stores to these devices\cite{benson2021viper, kaiyrakhmet2019slm, zhang2021chameleondb}. A
further study can look into these devices with regards to key-value stores. Similarly there have been new types of SSDs,
made specifically for key-value stores, which are aptly named as KVSSDs. This could also be a valuable target for further
investigations and has already garnered a healthy amount of 
interest\cite{wu2018kvssd, jung2021gpukv, qin2021kvraid, jin2017kaml, lee2019ilsm, im2020pink}.

\section{Study design}
\label{sec:design}
In this section we will describe the research goal and how this goal has been reached. In particular, we will go over the
research questions, the scope of the research project and the methodology used to achieve the research goal.

\subsection{Research Goal}
This study aims to get an overview of how various key-value stores are optimised for flash storage. The main research
question is:
``What is the impact of flash storage on the design choices for key-value stores?".
The following sub-questions were asked to aid in answering our research question:
\begin{itemize}
    \item RQ1: What is the current role of flash for key-value stores?
    \item RQ2: How has flash influenced the design of key-value stores over the decade?
    \item RQ3: What are the main challenges involved in using key-value stores on flash and how can they be mitigated?
    \item RQ4: How will flash contribute to key-value stores in the future?
\end{itemize}

\subsection{Scope}
\label{sec:Scope}
 This study is about persistent key-value stores for flash devices that specifically use the block interface. Therefore,
 it does not cover caching or other non-persistent workloads. It also does not cover other promising novel non-flash
 solutions such as persistent memories. Key-value SSDs will also not be covered since they are object-addressable and are
 thus not a block-based technology. Further on, we will mainly focus on how the key-value stores are optimised for flash
 devices. We will for example not cover how key-value stores are optimised for the network, even if flash is involved.
 Lastly, we only cover relatively recent contributions. In this case from approximately 2010 up to early 2022, but
 exceptions can be made to create context.
 From these requirements, we set up the following inclusion \& exclusion criteria:
\begin{itemize}
    \item I.1: The work is novel. 
    \item I.2: The work should be either about how flash impacts the design of key-value stores or how key-value stores
    can be optimised for flash.
    \item I.3: The key-value store must match the criteria of a persistent key-value store. 
    \item E.1: The work is about ephemeral key-value stores such as caching. 
    \item E.2: The work targets only non-flash solutions such as Optane memory or hard disk drives. 
    \item E.3: The work merely uses flash or key-value stores but is not about them. 
    \item E.4: The work is not relatively recent, main points should be from 2010 and up; papers from before 2010 can only
    be used for historical context. 
\end{itemize}

Not every paper that satisfies these requirements is included. Papers are more likely to be included, in order of
importance, if flash and key-value stores are the main topics, if they are cited frequently, if they are novel or if they
were published recently.

\subsection{Methodology}
To accurately answer the research questions, it is important to select and evaluate a large selection of papers.
Nevertheless, it is a Herculean effort to read every paper. Thus, only a selection of all papers related to the topic will
be examined. A common approach is the \textit{Snowball methodology}. With this methodology, you start with a few seed
papers and recursively find related papers. This can be done by either looking at the references used in the paper or by
verifying which papers have cited these papers themselves. It is also a good idea to start with at least a few recent
papers, considering that they might bring up some novel ideas or terminology that did not exist before. 
The seed papers used in this research are:
\begin{table}[h!]
\centering
 \begin{tabular}{||l l r||} 
 \hline
 Paper & Source & Year \\ [0.5ex] 
 \hline\hline
 NVMKV\cite{marmol2015nvmkv} & ATC & 2015  \\ 
 LOCS\cite{wang2014efficient} & EuroSys & 2014  \\
 SILT\cite{lim2011silt} & SOSP & 2011  \\
 SILK\cite{balmau2019silk} & ATC & 2019  \\
 WiscKey\cite{lu2017wisckey} & ACM TOS & 2017  \\
 PebblesDB\cite{raju2017pebblesdb} & SOSP & 2017  \\
 RocksDB space amplification \cite{dong2017optimizing} & CIDR & 2017  \\ [1ex] 
 \hline
 \end{tabular}
  \caption{Seed papers. Paper titles are shortened to make space.}
\end{table}

There are also a few conferences that typically reserve a few spots on storage topics. These were scraped from 2021 to
2012. This was done by iteratively checking the papers of the conferences and verifying if they might be relevant.
Relevance is determined according to the rules defined in \autoref{sec:Scope}. The names of the conferences considered
were: \textit{USENIX ATC}, \textit{FAST}, \textit{NSDI}, \textit{EuroSys}, \textit{OSDI}, \textit{SOSP}, \textit{Systor},
\textit{HotStorage}, \textit{ASPLOS}, \textit{SIGMOD}, \textit{VLDB}, \textit{SoCC}, \textit{ICDCS} and
\textit{Middleware}. 
Lastly, a few papers were found by simply querying Google Scholar (sorted on relevance and limited up to 2022), Connected
Papers(using Semantic Scholar), USENIX, ACM, DBLP and Semantic Scholar with keywords. The keywords were picked based on
words that are considered important in the seed papers and conferences. For some keywords, we also tried various synonyms
and closely related words. The used keywords are:
\begin{itemize}
    \item Flash Key-Value 
    \item Flash KV
    \item Flash NoSQL
    \item NVMe Key-value
    \item NVMe KV
    \item KVSSD
    \item SSD key-value
    \item SSD KV
    \item persistent key-value store flash
    \item LSM SSD
    \item LSM NVMe
    \item Btree SSD
    \item Btree NVMe
    \item Btree Key-value store SSD
    \item Key-value garbage collection
    \item Key-value wear levelling
    \item Key-value write amplification 
    \item Key-value read amplification
    \item Key-value space amplification
\end{itemize}

At times the search results were irrelevant to the research because they failed to meet the requirements. Frequently they
were about caching, NVM as memory and other devices such as SMR disks or Optane. These have to be filtered out by hand.
Many papers reoccurred across different search tools, which should make it more reproducible to get a part of the papers.
Nevertheless, for best results, it is advised to set the range in years to a maximum of 2022. The literature study was
conducted in January 2022 and can thus not contain any more recent papers. Also note, that the order of results can
fluctuate between the years and different people might assess papers differently on the provided criteria. We aim to
provide sufficient details about our methodology to aid the reproduction of work.

\section{Background}
\label{sec:background}
Before delving into various optimisation techniques, we take a look at the problem space.
We will first take a look at key-value stores and flash technologies separately. Then we look at the combination of the
two and formulate the main performance concerns that arise when combining the two.

\subsection{Persistent key-value stores}
\label{sec:kv}
Key-value stores are ubiquitous\cite{idreos2020key,dong2021evolution,debnath2011skimpystash,risch2015introduction,cao2020characterizing}.
They are a type of NoSQL store, but we will in general refrain from using that term in this survey since its exact
definition can be blurry and there exist multiple NoSQL stores that have a KV-like design. Such as can be seen in a
survey by Davoudian et al.\cite{davoudian2018survey} on NoSQL. Therefore, we will provide a different definition.
Key-value stores revolve around a flat structure with key-value pairs. Each key is unique and points to data. This data
can in general be anything and restrictions depend entirely on the specifications of the store at hand. Key-value stores
are considered to be lightweight and to implement only a few standard actions. These actions are specified in
\autoref{tab:kvops}.

\begin{table}[h!]
    \centering
    \begin{tabular}{||l l l||}
        \hline
        Operation & Explanation & Required \\
        \hline \hline
            put & Stores key-value pair & Yes \\
            \hline
            get & Retrieves key-value pair & Yes \\
            \hline
            delete & Deletes key-value pair & No \\
            \hline
            scan & Retrieves range of key-value pairs & No\\
         \hline
    \end{tabular}
    \caption{Key-value operations.}
    \label{tab:kvops}
\end{table}

Key-value stores should at least implement a \textit{put} and a \textit{get} operation. Put associates a key with a
value and set retrieves said value if it exists. The \textit{delete} action is also regularly included, but can also be
achieved with a put with an empty value. Lastly, most key-value stores define a \textit{scan} operation to get a range
of key-value pairs, but this is not required. An advantage of a scan operation is that if you already know that you need
a range, you do not need to send multiple get requests. The key-value store can then also already optimise for such an
operation. Lastly, not all key-value stores have to be persistent. They can for instance also be used for caching. We
will only focus on persistent solutions. These in general follow BASE\cite{pritchett2008base}, but some also follow
ACID\cite{rusinkiewicz1995specification}. This means that key-value stores are in general less strict on consistency.
Some have even fewer restrictions than BASE. Some examples of popular key-value stores are
RocksDB\cite{dong2021evolution}, LevelDB\cite{ghemawat2011leveldb} and DynamoDB\cite{sivasubramanian2012amazon}.  With
such a small set of requirements, it can and is used in all sorts of different applications. It can for instance be used
as a backend for relational databases\cite{dong2021evolution}, for gaming services\cite{debnath2011skimpystash}, for web
shops\cite{risch2015introduction}, for machine learning\cite{cao2020characterizing} and the social
web\cite{cao2020characterizing}.

That different workloads are used has performance implications. Not every application needs low latency or high
throughput. For some it might be more beneficial to invest as little as possible in storage devices. This means that
there is space to optimise key-value stores for different goals, which is also reflected in the literature that we will
cover in this survey. Certain stores also optimise for a certain ordering of data. We will see a few of these
optimisations later on.

Key-value stores have few constraints on the actual implementation and therefore contain a wide variety of
implementations. They all make use of \textit{data structures}. Data structures are structures that can be used to store,
organise and access data. An example data structure would be an array. In key-value stores, the main data structure is
used for \textit{indexing}.  Indexing means a method to get a reference to a value, think about getting a value based on
a key. Indexing structures reserve a part of the memory. For example, it is very typical to keep the references in memory
and store the actual values on storage. This can become problematic when the amount of memory required for each entry
increases, as it hinders the maximum amount of key-value pairs that can be stored. It is therefore typical to also move a
part of the index to storage. Moving a part of the index to storage does have a disadvantage. It requires more read and
write operations to storage for each key-value store operation. That is because the index now also needs to be read and
written. The amount of extra I/O operations needed for each logical operation is known as I/O amplification. Index
structures have to make a trade-off between memory usage and I/O amplification. This trade-off is also highly dependent
on the storage medium. For example, if the storage is slow, it becomes less advantageous to move more of the index to
storage. Therefore, key-value stores that are optimised for flash have to consider the flash specifics for optimal
performance, specifics which we will see in the next section. Commonly, additional data structures are used on top of
the index to minimise the amount of I/O, such as buffers. Some common data structures that are used for indexing in
key-value stores are LSM-trees, B-trees and hash tables. All three of these data structures will be explained in further
detail in \autoref{sec:Datastructures}. For now, it is enough to know that each of these has its advantages and
disadvantages.

\subsection{Flash storage}
\label{sec:flash}
There are quite a few characteristics that make flash memory stand out, and flash memory itself is not just limited to
one design. In order to properly optimise key-value designs for flash, it is imperative to first define what types of
flash are used in key-value stores and what characteristics these types of flash poses. This should then allow us, to
properly define the main performance concerns. For more in-depth explanations on flash, there exist various more
in-depth explanations on flash, including explanations with more technical 
details\cite{bez2003introduction, micheloni2010inside, aritome2015nand, chung2009survey, yang2014garbage, lee2008case}.

Flash is a non-volatile storage medium where data is stored as an electronic charge on a floating gate between a control
gate and the channel of a CMOS transistor. We will not go to deep on what such technicalities entail. In fact, for this
survey we only need to understand a couple of details. The number of bits the earlier mentioned gate can store depends
on the internal cell technology. For example there are \textit{single-level cells} (SLC) and \textit{multi-level cells}
(MLC) technologies. When the number of levels in a cell increases, so does the number of bits it can store. These
technologies are as of now in order of their density: SLC, MLC, TLC, QLC and PLC\cite{ma2020low}. Multi-level cell
technologies allow us to store more data in a smaller area, but also have  downsides. Multi-level cell technologies
namely result in lower throughput and an increased wear levelling\cite{alsalibi2018survey}. Wear levelling means that
after each write, the cell slowly degrades. This results in a decreased performance and eventually leads to a dead cell.
It is therefore considered harmful to do excessive writes on flash. Cell degradation is also not just limited to writes,
reads can also degrade the performance of cells, albeit to a lesser degree. This is known as read
disturbance\cite{liu2015read}. 

NAND flash is a technique that can be used to pack the earlier mentioned flash cells together. NAND packs the cells
together densely, which is a good fit for commercial mass storage but only allows addressing cells on the block
level\cite{cornwell2012anatomy}, which can be any size such as 4 KiB. NAND flash does not allow writing to an already
written block. It first has to be erased, before it can be written. Therefore updating a single bit in a block, will
result in writing an entirely new block and at some point in time erasing the old block. A NAND package is organised
hierarchically in pages, that are stored in blocks, that are stored in \textit{planes}, that are stored in
\textit{dies}. NAND packages can be processed in parallel. Planes can also process requests in parallel. Further on,
NAND cells themselves can also be packed vertically, resulting in 3D NAND. This is a property that also should be
addressed since the heat of one area also propagates vertically and can therefore increase the wear levelling of
neighbouring cells\cite{wang2020temperature}.

NAND flash can then be used within a storage device, such as \textit{Solid State Drives}
(SSD)\cite{cornwell2012anatomy}. Note that SSDs can also use different types of storage other than flash. SSDs that use
flash internally allow for parallelism through the number of independent NAND flash chips, which allows it to process
multiple write, read and erase requests in parallel\cite{wang2014efficient}. This parallelism can be exposed to the host
in the form of flash \textit{channels}. SSDs do not just include non-volatile storage, they come with a controller that
allows managing the device and usually come with a small bit of DRAM. SSDs are known as solid-state disks because they
contain no moving parts. This is in contrast to more traditional storage such as HDDs, which require moving a platter
and an arm. This allows them to deliver both adequate sequential and random I/O, which would not be possible if parts
had to move. There also exist various ways to connect SSDs; such as AHCI through SATA and NVMe through
PCIe\cite{landsman2013ahci}. Depending on the connection, different protocols can be used and different levels of
throughput, latency and concurrency can be achieved.

Lastly, SSDs frequently contain firmware. These are typically a \textit{File Translation Layer} (FTL) and a
\textit{Garbage Collector} (GC) among others. An FTL maps virtual addresses to physical addresses\cite{chung2009survey}.
This allows the SSD to determine what it considers an optimal location to store a block. It also removes the need for
the host to issue erasure commands. For the host, overwrites are still allowed. The FTL determines how to manage such
I/O internally. The Garbage collector (GC) is then used to erase dead blocks and move blocks around to more beneficial
locations\cite{yang2014garbage}. An optimised program should account for the internal logic of this firmware. There also
exist a few alternatives that allow the host to drop the need of an internal FTL and GC (these must then be defined on
the host) such as open-channel SSDs\cite{bjorling2017lightnvm} and ZNS\cite{bjorling2021zns}. Open-channel SSDs are
already used in a few key-value designs as we will see in \autoref{sec:ocssd}. Note that we have intentionally left out
some additional challenges for flash devices. \textit{The Unwritten Contract of Solid State Drives} by He et al. is an
interesting read on some additional challenges\cite{he2017unwritten}.

\subsection{Performance challenges}
\label{sec:perf}
In \autoref{sec:kv} and in \autoref{sec:flash} we respectively defined key-value stores and flash storage. We have
already seen some challenges that are involved for flash and key-value stores separately, now we will describe the
challenges involved in combining the two. We identify the following challenges that are commonly referred to in the
literature: write amplification, read amplification, space amplification, garbage collection overhead, memory footprint
and software overhead.

\textit{Write amplification}: Write amplification (WA) means that the physical amount of data that is written on the
device is more than the logical data that is written. For example writing 64 KiB of data to the device, when the
application issued only 2 KiB. This can be because of the block size, garbage collection and internals of various data
structures. Reducing write amplification seems to be an especially prevalent problem in studies. That is because write
amplification can significantly reduce both throughput and increase latency. Another problem is wear levelling, which
can degrade the flash devices and eventually make them unusable. Write amplification causes wear levelling and therefore
inevitably also increases the monetary cost of using these devices. The effect of write amplification is also bigger in
multi-cell technologies, such as MLC and TLC. 

\textit{Read amplification}: There are also a few works on reducing read amplification (RA). Read amplification means
that the amount of device-internal data that is read is more than the user-visible data that is read. For example,
issuing a read of 2 KiB and in the end reading a total of 64 KiB. This can for example be because the application needs
to find the data, requiring more reads, or because of the block size of flash. This can be problematic in read-heavy
stores that require low latency for reads. Nevertheless, write amplification is generally more problematic than read
amplification because writes cause wear levelling, which is more harmful than read disturbance, and because of the
asymmetric nature of I/O, which makes write operations more expensive than read operations. 

\textit{Space amplification}: Space amplification and space efficiency can also be a problem. Space amplification is an
application problem and means that the key-value store stores more data on storage than there are key-value pairs. This
is inevitable, as some metadata is always needed for persistency. Further on, some key-value stores also reserve space
for background operations, further limiting the amount that can be used for key-value data. The main concern is to keep
the amount of extra data needed to a minimum. It is also a possibility to resort to compression, to reduce the amount of
data that needs to be stored, sometimes creating negative space amplification. Space amplification can be problematic
because it results in a need to invest in more flash because there is not enough space. 

\textit{Garbage collection}: Garbage collection is also frequently stated as a problem. In this case, we refer to
garbage collection in the device, as described in \autoref{sec:flash}. Garbage collection causes erasure operations and
moves data around in the background. Since the data is moved around, this can cause further write amplification. The
background operations also introduce significant latency fluctuations. Works that focus on this problem aim to either
reduce the latency fluctuations, use hints to help the garbage collection or properly separate hot and cold data, which
helps the garbage collection by only moving hot data.

\textit{Memory footprint}: Reducing the memory footprint is an issue that is mostly stated in situations where the usage
of DRAM is more expansive or not a lot of DRAM is available in the first place. This is more prevalent in studies of the
early 2010s. Generally, most of the data such as values are already stored on flash, but a part of the index and some
caching is typically stored in memory, as described in \autoref{sec:kv}. When memory is scarce, it is not feasible to
keep full index data structures or big caches in memory, so these works try to reduce memory footprint with more
lightweight index data structures and move more data to flash.

\textit{Software overhead}: Lastly, some works aim to reduce software overhead on the host. These works mainly consider
flash devices with lower latency and higher throughput, such as NVMe SSDs. In this case, the CPU on the host can become
the performance bottleneck. Therefore, it can be beneficial to streamline the software stack and focus on performance
techniques on the host side.

From these challenges we formulate the following main concerns:
\begin{itemize}
    \item Q1: How to reduce write amplification? 
    \item Q2: How to reduce read amplification?
    \item Q3: How to reduce space amplification?
    \item Q4: How to deal with garbage collection?
    \item Q5: How to reduce the memory footprint?
    \item Q6: How to reduce software overhead?
    \item Q7: How to integrate flash into an architecture and use each component efficiently?
\end{itemize}
The challenges and concerns introduced in this section answer research question RQ3: ``What are the main challenges
involved in using key-value stores on flash and how can they be mitigated?". In the next sections, we will cover various
solutions that aim to find solutions to these concerns.  We will find that there is no such thing as a free lunch.
Indeed no solution is the solution to all our problems. On the contrary, most solutions we find will come with a
trade-off or might be specific to only a few use cases. 

 We will give a short overview of the papers covered in the section and what concerns they address at the beginning of
 each section. Concerns will be marked with a corresponding Q number as given in the concern list we have just given.
 For example, concerns about write amplification will be referred to with \textit{Q1}.


\section{Data structures for flash}
\label{sec:Datastructures}
\label{sec:datastruct}
\begin{table}[h!]
  \centering
  \begin{tabular}{||l l l||}
    \hline
    Problem   & Solution        & Examples                                            \\
    \hline \hline
    Q2        & AMQs            & BloomStore\cite{lu2012bloomstore}                   \\
    \hline
    Q1,Q2,Q3, & Traditional     & BufferHash\cite{anand2010cheap},                    \\
    Q5,Q6&hash&
    ChunkStash\cite{debnath2010chunkstash},                                           \\
    &solutions&FlashStore\cite{debnath2010flashstore},\\&&SkimpyStash\cite{debnath2011skimpystash},
    \\&&FAWN\cite{andersen2009fawn},\\&&uDepot\cite{kourtis2019reaping}\\
    \hline
    Q1,Q2,Q4, & Multi-stage     & SILT\cite{lim2011silt}                              \\
    Q5        & hash solution   &                                                     \\
    \hline
    Q1,Q2,Q4, & B-tree          & FlashDB\cite{nath2007flashdb},                      \\
    Q5,Q6&solutions&TokuDB\cite{tokutek2013tokudb},\\&&WiredTiger\cite{nguyen2018optimizing}
    ,\\&&Tucana\cite{papagiannis2016tucana},\\&&ForestDB\cite{ahn2015forestdb,lee2021boosting},
    \\&&SplinterDB\cite{conway2020splinterdb}\\
    \hline
    Q1,Q2     & LSM-tree        & LSM-tree\cite{o1996log}                             \\
    \hline
    Q1,Q2,Q4  & LSM-tree:       & WiscKey\cite{lu2017wisckey},                        \\
    & key-value&
    SardineDB\cite{dong2021sardinedb},                                                \\
    &separation&RocksDB\cite{dong2021evolution},\\&&HashKV\cite{chan2018hashkv},\\&&
    DiffKV\cite{li2021differentiated}\\
    \hline
    Q1,Q2,Q3, & LSM-tree:       & PebblesDB\cite{raju2017pebblesdb},                  \\
    Q4,Q5,Q6&reduce& PTierDB\cite{liu2021ptierdb},\\&compaction&RocksDB\cite{dong2021evolution}
    ,\\&overhead&SifrDB\cite{mei2018sifrdb},\\&&Accordion\cite{bortnikov2018accordion},\\&&
    SILK\cite{balmau2019silk},                                                        \\
    &&LDC\cite{chai2019ldc},\\&&TRIAD\cite{balmau2017triad}\\
    \hline
    Q1,Q2,Q4, & LSM-tree:       & Kreon\cite{papagiannis2018efficient}                \\
    Q6        & Different index &                                                     \\
    \hline
    Q3        & LSM-tree:       & RocksDB\cite{dong2021evolution,dong2017optimizing}, \\
              & reduce space    & FlashKV\cite{zhang2017flashkv},                     \\
              & amplification   & LOCS\cite{wu2003efficient}                          \\
    \hline
  \end{tabular}
  \caption{Overview of papers covered in \autoref{sec:datastruct}.}
  \label{tab:hybridworks}
\end{table}

Various data structures can be used to implement persistent key-value stores. Together they represent the internals of
the key-value store. Nevertheless, there have been three data structures in particular that have been most prevalent for
key-value stores in the last decade. These are all index data structures as explained in \autoref{sec:kv}. The data that
they index to are the key-value pairs. Keys can also be used as the pointer. The three data structures are \textit{Hash
  Tables}, \textit{B-trees}, and \textit{LSM-trees}. Many ideas have been proposed to either optimise these data structures
for flash or use them as a part of their key-value store. Therefore, we will take a closer look at these structures and
go into detail about how they can be optimised for flash. Of these three data structures, LSM-trees seem to have become
the most commonly used and investigated. LSM-trees will therefore also be discussed most prominently. However, we will
first take a short look at \textit{approximate membership queries} (AMQ), which can be used along with all three data
structures to increase performance.

\subsection{AMQ}
\label{sec:amq}
\textit{Approximate membership queries} (AMQ), also known as filters, are a family of data structures that are used in
close to all key-value solutions we will cover. They are data structures that are meant to prevent expensive I/O
operations by approximating if a key is present in the key-value store or not\cite{bender2011don}. An important
characteristic is that they can lead to false positives, but not to false negatives. This characteristic can be used to
prevent lookups. If the AMQ returns that it is not present, it is also definitely not present on flash. However, if the
AMQ does return that it is present, it might be present on flash, necessitating a lookup.

One might already wonder about the use case for such queries, as they only seem to increase the amount of work necessary
for each key-value operation. The reason is that AMQs are generally smaller than the data structure they approximate. The
amount of space they require can be tuned, in this case space is traded for the accuracy of the AMQ. A smaller size
allows it to be stored on a faster device that has less space and to be cached. If a device is fast enough such as DRAM,
it can therefore eliminate the cost of accessing more expensive I/O such as flash. AMQs thus aim to solve issues with Q2.
Some AMQs that we have seen in key-value stores optimised for flash include \textit{bloom
  filters}\cite{rottenstreich2014bloom}, \textit{cuckoo filters}\cite{fan2014cuckoo} and \textit{quotient
  filters}\cite{bender2011don}. Bender et al. also proposed a \textit{Cascading Filter} that combines multiple quotient
filters and should also be efficient when parts of the filter are stored on flash\cite{bender2011don}, but AMQs should
generally be kept completely in memory. Take note that many designs that we will see later on, can be improved by the use
of such AMQs to prevent I/O operations. Lu et al. even propose BloomStore\cite{lu2012bloomstore}, a key-value design that
revolves around using bloom filters as the main index structure.

\subsection{Hash tables}
\label{sec:hash}
\textit{Hash tables} contain a wide array of different solutions, but they are all based on a common principle; they
maintain a table of keys with a mapping to the address of their value, and can therefore also be referred to as
\textit{mapping tables}. This fits very closely to the definition of key-value stores, as key-value stores are also a
collection of keys mapping to values. Each mapping in the mapping table occupies a location, which we will refer to as a
\textit{slot}. In all solutions we cover all values are maintained on flash, so only the offset to values is maintained
in the mapping tables. Keys are stored along with the values but are regularly also a part of the mapping table itself.
If keys are not part of the mapping table, either a hash of the key is used or a part of the key is used in the mapping
table. The mapping table is cached at least partially in memory and can be constructed from the stored data. Hash table
designs become problematic when a lot of key-value pairs are stored since the table increases linearly with the number of
key-value pairs. Therefore not the entire hash table can fit in memory and a part needs to be repeatedly stored and
loaded from persistent storage. This swapping between storage and memory can lead to read amplification when there is not
a lot of memory. Further on, hash tables are also known to have problems with scan operations since keys and values are
(typically) not sorted on flash. In this section we will see that a part of these problems can be mitigated. We will
cover a few hash table designs, even some that consist out of multiple substructures.

\subsubsection{Buffering}
One of the first optimisations we will cover is \textit{buffering}, which is deployed in many hash
designs\cite{anand2010cheap, debnath2010chunkstash, debnath2010flashstore, debnath2011skimpystash, lim2011silt}.
Buffering is a technique that is not unique to hash designs, and a design we will also see in the other two index
structures later on; see \autoref{sec:btree} and \autoref{sec:LSM}. The general idea is to keep items in memory until
they collectively reach a certain size, which can be set as a threshold. To guarantee persistence, items should also be
flushed to storage after a short amount of time has passed and the size threshold has not yet been reached. Buffering is
done because most SSDs are block-based and force a minimum write size. Writing small strips of data would therefore
require adding padding, which contains bytes that will not be used by the actual key-value store. This padding thus leads
to write amplification. Buffering also removes the need for sending each operation to the device, essentially sending one
big request. Sending each operation would incur a lower throughput, latency and more garbage collection (Q4) in the end.
It would also incur a software overhead as each operation will need to be processed from beginning to end. Thus buffering
can help by avoiding small I/O and batching operations (aiding Q6).

\subsubsection{Reducing the memory footprint}
\begin{figure}[h]
  \centering
  \begin{minipage}{.25\textwidth}
    \centering
    \includegraphics[width=1\linewidth]{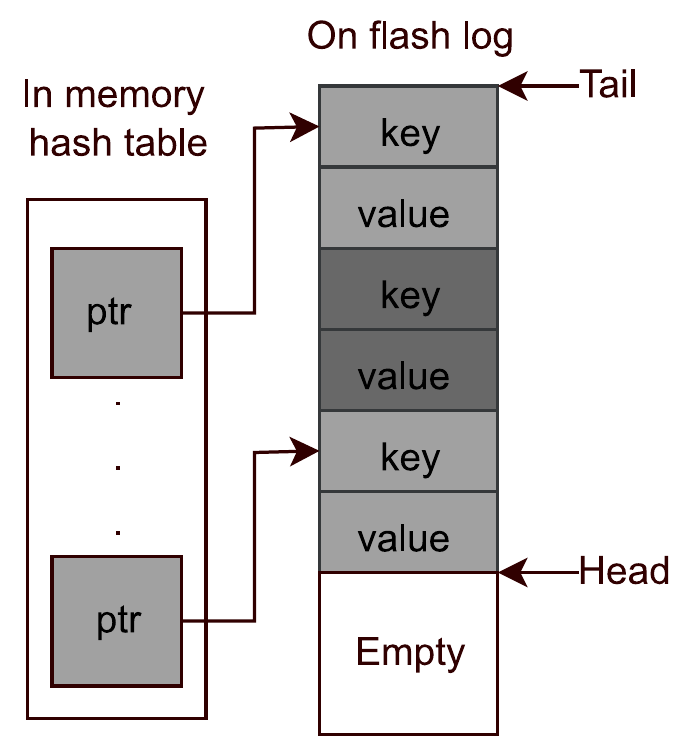}
  \end{minipage}%
  \caption{An example of a typical hash table. Pointers to key-value pairs are stored in RAM and values are stored in an
    append-only log. Darker key-value pairs can be reclaimed.  }
  \label{fig:hash}
\end{figure}

Another problem is keeping the memory footprint to a minimum and at the same time keeping read amplification low (solving
Q2 and Q5 issues). Key-value stores typically store all the values on flash in an append-only data structure, known as a
\textit{log}. Alternatives do exist but are not as common. SILT for example also stores values in a sorted log among
others\cite{lim2011silt}. We will mainly focus on the log design, because of its frequency in
usage\cite{debnath2010flashstore, debnath2010chunkstash, debnath2011skimpystash, andersen2009fawn}. After the buffer is
flushed, the data is appended to the log and the offset to said data is saved in the hash table. This forces sequential
I/O on the storage device, which is considered better than random I/O. When a value is deleted or updated, the old entry
in the log is marked as invalid. Eventually, such values \textit{do} need to be erased and removed from the log, probably
with a separate garbage collection process. Such a hash table design is visible in \autoref{fig:hash}.

A naive implementation would then simply conclude by making the hash table uniquely link the entire key to the offset of
a value in the log, but we are not there yet. This implementation directly indexes full keys to value offsets.
Unfortunately, full keys also take multiple bytes and this would not scale when billions of keys need to be saved.
Instead, we need to find a way to only keep part of the keys or a part of each key in memory. For example what if we have
10 bytes of RAM and 5 keys of each 5 bytes. We would not be able to fit the entire hash table in memory. We either need
to keep part of the key-value mappings on flash such as only 1 key, or we need to reduce the size of each mapping. We
will cover a few examples of both approaches.

Our first solution is proposed by Andersen et al. in the form of \textit{FAWN}\cite{andersen2009fawn}. FAWN reduces the
memory footprint by only storing a part/fragment of each key in the hash table, reducing the memory footprint for each
individual key-value pair. This allows more key-value mappings to remain in memory or to simply reduce the amount of DRAM
used on the host. Such a design does have two problems. As only a fragment of each key is stored, it can lead to
\textit{hash collisions}, which means that two mappings end up in the same slot. Hash collisions can occur because two
keys that are different can have exactly the same fragment. This would then force them to the same slot in the hash
table, requiring multiple value offsets to occupy the same slot. Similarly, since you only know the fragment, you can not
reliably tell if the key-value pair exists in the first place by only looking at the occupancy of the slot. FAWN solves
this by storing the keys along with the values on flash, which allows verifying if it was indeed the correct key-value
combination on lookup. Hash collisions are resolved in FAWN with the help of hash chaining. Slots in the hash table then
contain chains of offsets instead of individual value offsets. Colliding keys are then added to the same slot in a chain
in memory. Lookups can therefore issue multiple reads to find the correct pair because the entire chain might be read
(creating Q2 problems). This is dependent on how big of a part is chosen from the key as a fragment, but Andersen et al.
report numbers such as 1 in 1000 and 1 in 32,768 for having to do more than 1 lookup in their use
case\cite{andersen2009fawn}, indicating that extra lookups are rare. At the same time, this solution still requires some
bytes in memory for each entry, even if it alleviates some memory issues by using fragments.

Debnath et al. take it one step further with the design of \textit{SkimpyStash}\cite{debnath2011skimpystash}, achieving
on average less than 1 byte for each entry, while still using only one hash table. To achieve this they make use of a
hash chain, similar to FAWN. However, instead of storing the chain in memory, they store the entire chain on flash. The
entry in the hash table will point to the first entry of the chain. To get the next entry in the chain, some additional
metadata needs to be stored on flash. Therefore on flash we will also write a pointer to the next entry in the chain
along with each key-value pair. Such a design can however lead to even more read amplification. It can also lead to read
imbalance, especially when the chain is long. Long chains cause read imbalance because they can require significantly
more reads than short chains. Lim et al. report an average of 5 lookups for each key-value pair\cite{lim2011silt}
(increasing Q2). To mitigate such issues it does try to keep parts of the chain sequentially stored on flash to minimise
the number of needed reads. It also adds a bloom filter as an AMQ for each chain, needing fewer reads for each chain on
average.

\textit{BloomStore}\cite{lu2012bloomstore} takes it yet another step further, by completely moving the hash index
structure to flash along with the key-value pairs. To achieve this, it uses a bloom filter to great effect. BloomStore
keeps another bloomfilter based index in memory that can point to the hash indexes.

\textit{uDepot}\cite{kourtis2019reaping} tries a different route. It drops the idea of using just one hash table and uses
a two-level hopscotch hash table instead. Hopscotch is a hashing method\cite{herlihy2008hopscotch}, that tries to move
colliding keys into a \textit{neighbourhood} close to the key, which has caching benefits among others. The key of a
key-value pair is used for both hash tables. A part of the key is then used to access the location of a
\textit{directory}, the first level hash table. The other part of the key is used to access the index in this directory
to retrieve an offset, the second level hash table. We thus have directories of offsets. Such a multi-structure design
does incur more reads and more writes for each entry, but it does keep both memory overhead and read amplification
relatively low. It is essentially a trade-off.

\textit{Small Index Large Table} (SILT)\cite{lim2011silt} also uses multiple data structures. In fact it uses a total of
three and they are not all hash tables, which we will explain soon. SILT also keeps a low memory footprint but does not
result in large unpredictable read amplification like some of the previous designs. Lim et al. recognised that other data
structures might be beneficial as well, instead of only using a hash design\cite{lim2011silt}. Different data structures
have different advantages and disadvantages, combining multiple data structures can therefore combine the best features
of such structures. At the same time, it comes with its own set of challenges. Such as efficient translations between
data structures.

This idea lead to SILT, which used a \textit{multi-stage design} (MS) in the end, which is essentially a series of data
structures that are chained in sequence like a funnel. An LSM-tree, which we will cover in \autoref{sec:LSM} is also an
example of a multi-stage design. Each successive data structure in SILT will keep fewer data in memory, but more data on
flash. Such a design makes the average size of each key-vale pair in the indexing structure still small but also keeps
performance up.

\begin{figure}[h]
  \centering
  \begin{minipage}{.45\textwidth}
    \centering
    \includegraphics[width=0.5\linewidth]{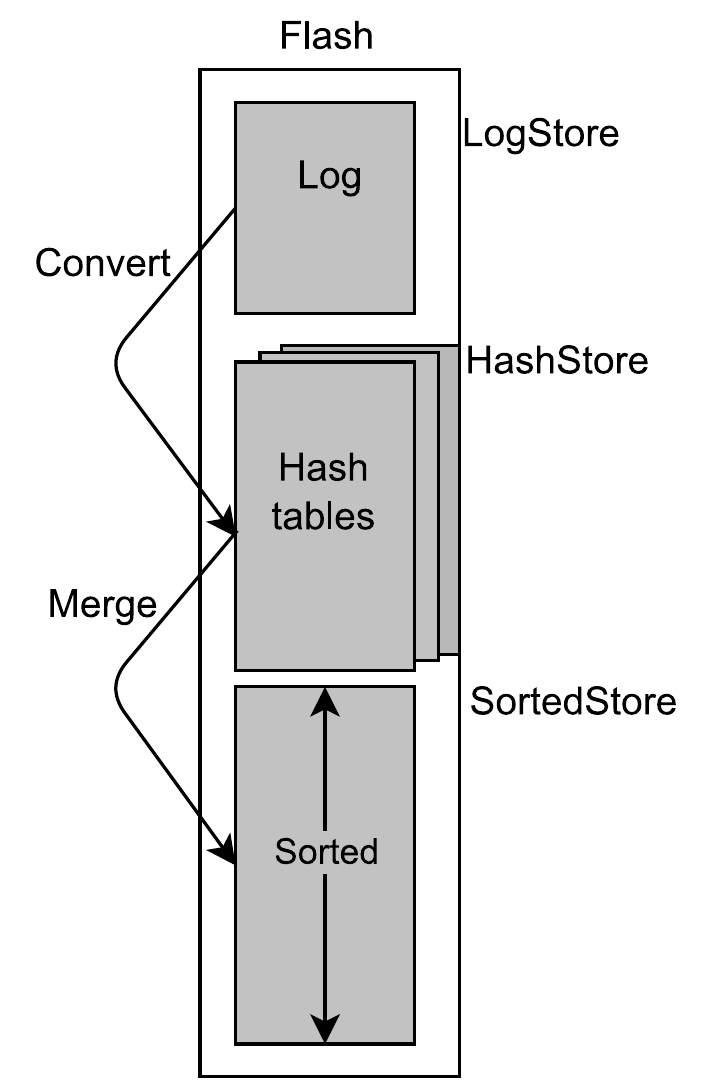}
  \end{minipage}%
  \caption{The multitier design of SILT. Reworked figure of the one made by Lim et al.\cite{lim2011silt}}
  \label{fig:silt}
\end{figure}

SILT uses in order a \textit{LogStore}, a \textit{HashStore} and a \textit{SortedStore}, visible in \autoref{fig:silt}.
The LogStore uses a hash table design that is close to the design we already discussed earlier. HashStores are simply
LogStores stored to disk, indexes included. This requires minimal translation. SILT allows multiple HashStores on flash
to enable some buffer space. The last store, the SortedStore stores all of its keys in order. There can only be one
SortedStore present at the same time. SortedStore also adds compression, which allows SILT to significantly reduce space
amplification (Q3). Entries slowly transition from structure to structure and each structure takes a little longer to
read and write. This does incur write amplification, which is a major drawback of multi-stage designs. Further on, the
transitions from HashStore to SortedStore can be quite expensive, which can cause some latency issues.

In short, we have seen various hash-based designs and each has its own advantages and disadvantages. There is a trade-off
to be made between the amount of the indexing structure that is maintained in memory and on flash. The more stored on
flash, the bigger the read amplification and read latency (trading Q2 for Q5). At the same time multi-staged designs
could be employed at the cost of more write amplification and wear levelling, but a possibly lower memory occupation and
acceptable read amplification.

\subsection{B-trees}
\label{sec:btree}
\begin{figure}[h]
  \centering
  \begin{minipage}{.40\textwidth}
    \centering
    \includegraphics[width=1\linewidth]{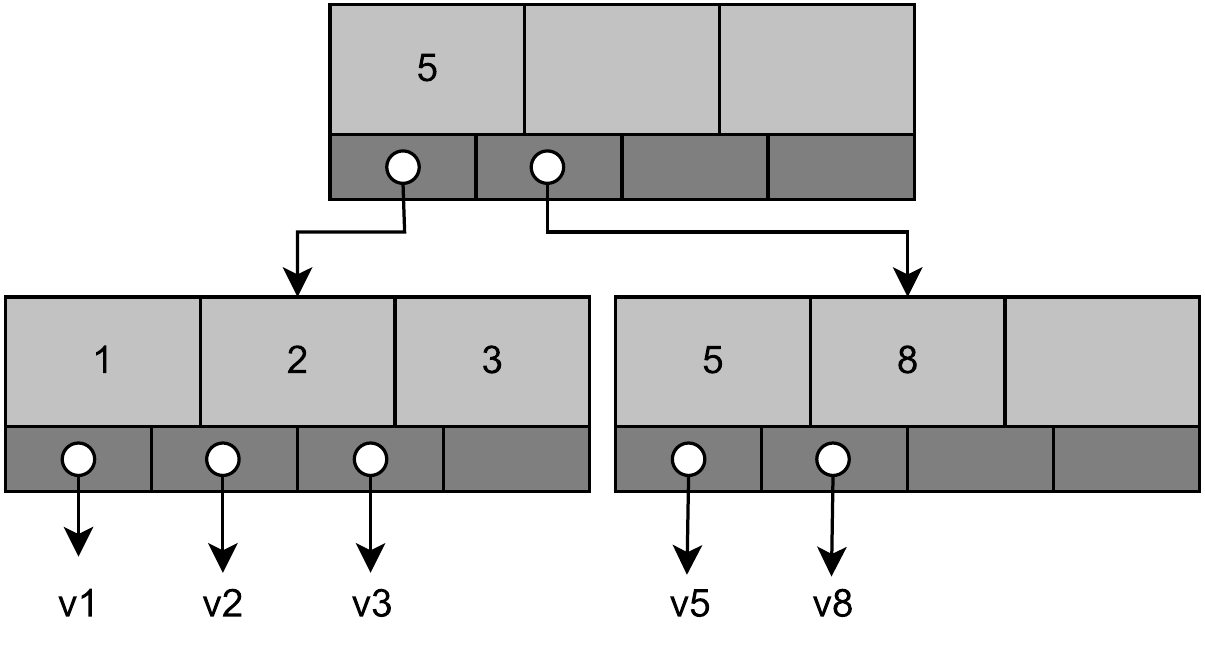}
  \end{minipage}%
  \caption{B+-tree with three nodes. Values are only stored in the leaves.}
  \label{fig:bplustree}
\end{figure}
The \textit{B-tree} is a tree data structure that is used widely in various databases. Especially in the field of
traditional relational databases\cite{graefe2011modern, comer1979ubiquitous, kudale11575258b+} and read-heavy key-value
stores\cite{nguyen2018optimizing}. They are optimised for read performance and were used frequently in the early days of
flash-based key-value stores, leading to among others \textit{FlashDB} and
\textit{TokuDB}\cite{nath2007flashdb,tokutek2013tokudb}. However, they are still used in some modern designs such as
\textit{WiredTiger}, \textit{ForestDB} or
\textit{Tucana}\cite{nguyen2018optimizing,ahn2015forestdb,papagiannis2016tucana}. They are also used at times as
substructures in for example LSM-trees\cite{papagiannis2018efficient}. They are known to have a good read performance
because read operations have an upper bound in regards to the height of the tree, requiring only one read for each level
in the tree. B-trees can thus be beneficial if a key-value store is read-heavy.

B-trees have a long history and have at least already been deployed to flash storage systems as least as early as
2003\cite{wu2003efficient} and were already considered ubiquitous in 1979\cite{comer1979ubiquitous}. When a key-value
store uses a B-tree as its main key-value pair indexing data structure, it typically actually refers to the
\textit{B+-tree}, which is also called a \textit{$B^*$-tree}\cite{comer1979ubiquitous}. B+-trees continue on the
original design of B-trees and such a design can be seen in \autoref{fig:bplustree}.

Both the B-tree and the B+-tree are M-ary trees and both are designed to be self-balancing. This is accomplished by
automatically merging and splitting nodes on insertions and deletions. All of the leaves are on the same height. This
ensures upper bounds on the required number of reads to be equal to the height of the tree.

What differs between the two is what is stored in nodes and in leaves. In the B-tree design nodes and leaves both contain
keys and values and keys are not duplicated. In B+-trees on the other hand, nodes can only contain keys in the form of
pivots. Keys can be duplicated all the way from root to the leaf, also visible in \autoref{fig:bplustree} for key
\textit{5}. Leaves store the actual value pointers. In addition, the leaves can be linked. This linking has a major
advantage because sequential access can be improved. Scan operations for example can directly go on to the next linked
leave and do not require traversing the entire tree again for each subsequent read. Yet another advantage of the B+-tree
is that values are only stored in leaves and not on a random level, adding search efficiency. The key-value pairs are
also sorted in the data structure, allowing the client to only read one node on each level of the tree. The earlier
stated read performance is thanks to this sorted and sequential characteristic.

B+-trees store each node and leave separately on different blocks on the flash device. There is no guarantee that nodes
and leaves will fill entire blocks, which can waste space and cause write amplification (Q1) as the entire block still
needs to be written\cite{kuszmaul2014comparison}. This can be reduced with compression as we will see in
\autoref{sec:comp}. Data is also not stored sequentially because inserts and updates change parts of the B-tree rather
than appending their changes to the end. Reading, inserting and updates thus incur random I/O as data needs to be written
to random locations. Random I/O causes a performance overhead, but not just because of the storage medium. We will show
why this causes write amplification by looking at inserts. Parts of index data structures are typically cached, for
example for B-trees a few blocks and leaves are cached. On an insert, data needs to be written to leaf blocks. If this
block is in the cache, the write can simply be done in the cache. If the block is not in the cache, the block must first
be loaded from I/O before being altered. The old block that was present in the cache is then generally written back to
I/O as there is only a limited amount of memory. It is stated by Kuszmaul et al. that in the worst case, each insertion
requires rewriting an entire block\cite{kuszmaul2014comparison}. If the data is stored sequentially, multiple inserts can
be written in one go. Lastly, B+-trees do not come with a lot of buffering by default, making it harder to amortise such
operations.

There also exist various other B+-tree designs, which are tested on both flash and key-value stores, but we will only
look at two B-tree designs: modified \textit{HB+-tries} and \textit{$B^\epsilon$-trees}. However,  first we will look at
problems that occur when B-trees are directly stored on flash storage.

\subsubsection{Implementing a B-tree directly on flash}
It is possible to directly build B-trees on flash devices, without an extra layer between the B-tree and the storage.
This could for example be an FTL (see \autoref{sec:background}) that uses a B-tree, but some devices, which we will
discuss in more detail in \autoref{sec:ocssd}, can also directly allow the key-value store to run on flash. Directly
building a B-tree on flash comes with a few challenges, challenges which we do not have when there is some logic between
the key-value store and the storage. We will explain these problems shortly.

B-trees store each node and leave separately on different blocks on the flash device. On an update this entire block
needs to be rewritten as flash itself does not allow in-place updates (this is traditionally resolved with the help of an
FTL). This inevitably brings us to a major crux of the B-tree when it is used directly on flash: if any node changes, its
location changes. Therefore the parent node also needs to be updated because its pointers are no longer accurate, which
recurses all the way back to the root. This is commonly referred to as the \textit{wandering tree}\cite{kang2007mu}.
Operations on B-trees can therefore cause a non-negligibly high write amplification (Q1).

A way to reduce the wandering tree problem is by packing multiple nodes and leaves into one page, allowing the cost of
traversing writes to be reduced. Such as is done in the \textit{$\mu$-tree}\cite{kang2007mu}, which packs nodes and
leaves together in a manner optimised for flash. This has not been tested on key-value stores as far as we know and we
have not seen a key-value store that itself directly uses a B-tree on flash storage. Nevertheless, it is still
interesting as the underlying FTL might use this functionality, which has implications for the key-value store on top.
Similarly, it leaves options open for future key-value stores that aim to build a B-tree directly on flash storage.

\subsubsection{HB+-trie}
\textit{HB+-tries} were originally implemented in ForestDB\cite{ahn2015forestdb}. It uses a \textit{trie}, which is a
kind of data structure, where each node of the trie is a B+-tree. Leaves point either to another B+-tree node or to
values. For more information on tries, we recommend looking at the ForestDB paper\cite{ahn2015forestdb}.

The HB+-trie comes with two optimisations that help with increasing the performance for flash. Updates to data instead of
the data itself are stored in an append-only log, which avoids in-place updates. This helps flash because it reduces the
need for random I/O. It also uses a small write buffer index. This index buffers references the log and is only flushed
to the HB+-trie after a certain number of writes. This helps with amortising expensive I/O operations (addressing Q1).

Unfortunately, the append-only log requires regular cleaning, which necessitates a separate garbage collection process
known as \textit{compaction}. This requires copying old data over to new locations. This introduces more write
amplification and temporarily reduces the latency and throughput for the client operations during the compaction. Lee et
al. propose to make these compactions more optimised for SSDs by properly using the available
parallelism\cite{lee2021boosting}. It uses a parallel-fetch, which they call \textit{p-fetch}. It submits multiple
parallel reads, which can be linked to different flash channels. To make sure that the host does not have to wait for all
operations, it makes use of asynchronous I/O (addressing Q6), which we will cover further in \autoref{sec:async}. This
reduces the software overhead and the time compactions would take. We conclude by stating that such an approach could
also be reused in other log designs.


\subsubsection{$B^\epsilon$-tree}
\label{sec:betree}
\begin{figure}[h]
  \centering
  \begin{minipage}{.45\textwidth}
    \centering
    \includegraphics[width=1\linewidth]{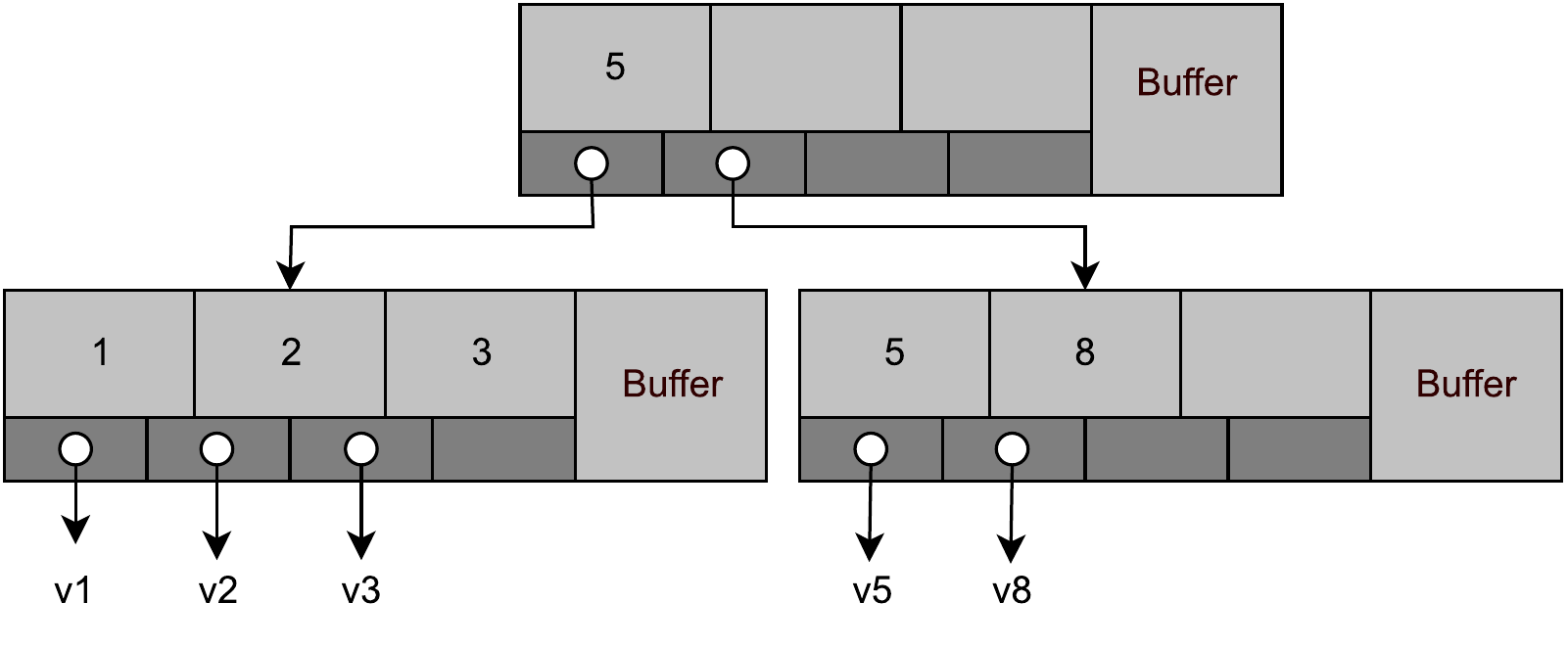}
  \end{minipage}%
  \caption{Example of the standard $B^\epsilon$-tree. Each node also has some buffer space. }
  \label{fig:betree}
\end{figure}
The \textit{$B^\epsilon$-tree} adds buffers to all intermediate nodes of the B+-tree. If we were to adapt our original
B+-tree design, visible in \autoref{fig:bplustree}, with buffers it would look like \autoref{fig:betree}. These buffers
can be used to store operations such as writes for new data or updates for older data and allow amortising expensive I/O
operations (addressing Q1). When a buffer is filled, it \textit{spills} its data and moves the data downwards until it
reaches the leaves, which have no buffer. $B^\epsilon$-trees allow for a write-heavy workload but do not suffer from
compaction operations such as LSM-trees. At the same time, it does require more random I/O. Such a design is therefore
only effective on devices that provide fast random I/O, such as NVMe SSDs (note that not every SSD is equal). It is also
most effective when the entire index structure fits in memory (conflicting with Q5), which can be big. Such a design,
albeit modified and further optimised, was implemented in \textit{Tucana} by Papagiannis et
al.\cite{papagiannis2016tucana} and is shown to be able to reduce the software overhead and increase throughput. Similar
to WiscKey, which we will cover later, it is possible to separate the index structure from the values. This allows the
index to be smaller and reduces the write amplification.

A remaining problem with $B^\epsilon$-tree is that it can cause significant write amplification (Q1) in spite of the
buffering. That is because every time a buffer needs to be altered the entire buffer needs to be rewritten. Conway et al.
therefore proposed yet another data structure, The \textit{STBe-tree}\cite{conway2020splinterdb}. This covers ideas from
both the LSM-tree and the $B^\epsilon$-tree. Instead of using direct buffers in nodes, it uses multiple sub-indexes in
the form of B-trees, replacing the log buffers of the $B^\epsilon$-tree. Each of these B-trees can be updated
independently, reducing write amplification. Similarly to the main B-tree, each sub B-tree should also come with an AMQ,
such as a bloom filter, to reduce the number of reads necessary for these trees (addressing Q2).

Another novel idea proposed for the STBe-tree is \textit{flush-then-compact}\cite{conway2020splinterdb}.
Flush-then-compact enables more I/O and CPU parallelism. This reduces Q6 and the immediate effects of Q1. Traditionally
data is moved from parent to child with a process known as \textit{flushing}. Think about data moving from node to node.
Data in the target node then needs to be reorganised, which is known as a \textit{compaction}. This value copying
requires locks to be held while all values move and temporarily introduces expensive I/O operations that need to be
completed in one go. With flush-then-compact, a \textit{pointer-swing} is performed instead.  References to the child's
active branches are then copied from parent to child. This allows locks to be held very shortly, enabling better
concurrency. Operations to copy the values can then be scheduled later on asynchronously without requiring locks. This
also allows scheduling them in parallel trivially. Such a design, could in general also be used in other designs that
make use of such a compaction operation.

\subsection{LSM-trees}
\label{sec:LSM}
On flash storage writes are more expensive than reads, which makes it attractive to look for a data structure that is
optimised for writes instead of reads. These types of data structures are commonly referred to as \textit{Write Optimised
  Indexes} (WOI). An example of such a data structure is the \textit{log-structured merge-tree} (LSM)\cite{o1996log}. A big
part of the modern flash-optimised key-value stores make use of
LSM-trees\cite{dong2017optimizing,raju2017pebblesdb,lakshman2010cassandra,lu2017wisckey,Dgraph2022}. Nevertheless, just
putting a plain LSM-tree on flash storage does not immediately make it optimised. There are a few optimisations that can
be done to increase its performance, most with trade-offs. A few of the more novel solutions will be discussed in more
detail. We will first discuss the general LSM-tree structure, followed by various optimisations and trade-offs.

\subsubsection{What is an LSM-tree}
\begin{figure}[h]
  \centering
  \begin{minipage}{.25\textwidth}
    \centering
    \includegraphics[width=1\linewidth]{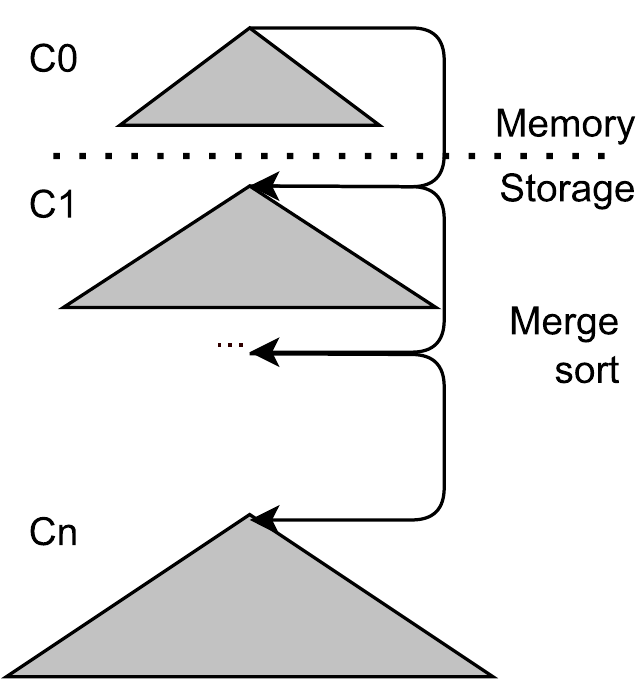}
  \end{minipage}%
  \caption{Original LSM-tree design in 1996, based on a figure by Lim et al.\cite{o1996log}}
  \label{fig:LSM}
\end{figure}
\begin{figure}[h]
  \centering
  \begin{minipage}{.35\textwidth}
    \centering
    \includegraphics[width=1\linewidth]{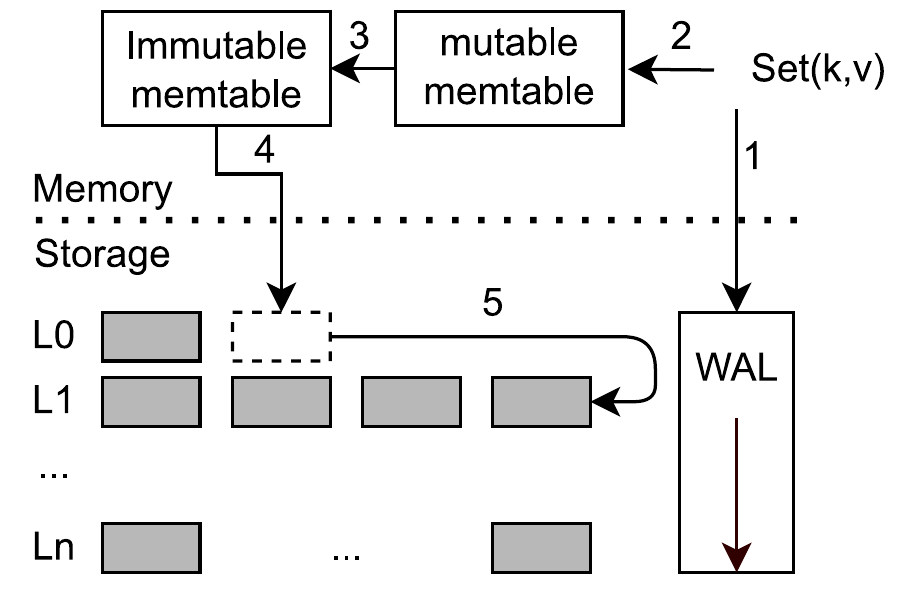}
  \end{minipage}%
  \caption{Common LSM-tree design, originally popularised by LevelDB. The numbers represent the steps in order of a
    operation that updates/inserts/deletes a pair. It is based on a figure made by Lu et al.\cite{lu2017wisckey}}
  \label{fig:LSMLevelDB}
\end{figure}

The LSM-tree was originally intended to be used for HDDs, all the way back in 1996\cite{o1996log}. Back then, HDDs were
commonplace and DRAM was becoming big enough to allow for buffering. Many of the properties of the LSM-tree were thus
optimised for the characteristics of this system. Such a design is seen in \autoref{fig:LSM}. This design starts with a
small part of data in memory and then uses multiple layers on disk. Each incremental layer is bigger and merge sorts into
the next when it becomes too big. This relatively simple model has since been heavily optimised/modified, and it it is
this modified design that we actually see back in modern literature for flash key-value stores. This design is visible in
\autoref{fig:LSMLevelDB} and was popularised by Ghemawat et al. in LevelDB\cite{ghemawat2011leveldb}. The data structure
consists of four substructures, chained together in a \textit{multi-stage} (MS) structure. A \textit{sorted in-memory
  table}, a \textit{sorted immutable in-memory table}, a \textit{write-ahead log} (WAL) on storage and multiple
\textit{sorted string tables} (SSTables) that are stored on storage. We will take a look at how data propagates through
the structure by addressing all steps in \autoref{fig:LSMLevelDB} by numerical order.

(1): On an insert or update, every key-value pair is first inserted into the WAL, which is an append-only log. This
functions similar to journaling in traditional databases and guarantees persistence and
consistency\cite{rothermel1989aries}. On recovery, the WAL can then be replayed as it contains all changes in order.

(2): After the write to the WAL is complete, the same data is written again, but then to the in-memory table, which is
typically small such as a few MBs\cite{balmau2017triad}. The in-memory table functions as a buffer that can be used to
avoid small writes to disk (reducing Q1). This is similar to buffering in hash tables and $B^\epsilon$-trees, see
\autoref{sec:hash} and \autoref{sec:betree}.

(3): When the in-memory table reaches a certain size, it is converted to an immutable in-memory table. These immutable
tables should eventually be written to storage. At the same time, a new mutable memtable is created or an old memtable is
reused. The immutable in-memory tables are used to allow writes to continue to memory without a prolonged wait.
Otherwise, the key-value store has to wait for the entire I/O operation to complete. New data is written to the mutable
memtable, while the old immutable table is written to storage.

(4): The immutable in-memory table is written to storage as is. It is written as a sorted file, known as a Sorted String
Table (SSTable). Such an operation is known as a \textit{flush}. An SSTable is sorted on keys and contains a subset of
the total range of keys. The SSTable itself is thus essentially an immutable ordered collection of key-value pairs,
sorted on keys. The SSTables are stored on storage in a series of levels. Reaching from level 0, commonly referred to as
\textit{L0}, up to level N, commonly referred to as \textit{LN}. The number of levels depends on the design and can be
tweaked. Each level can only hold a certain amount of SSTables. This amount increases with each level, so higher levels
can hold more data. LevelDB for example maintains a growth factor between levels of 10x by
default\cite{ghemawat2011leveldb}. For correctness, only 2 levels would be enough, but adding more levels helps with
amortising step (5).

(5): Data can be moved to a higher level, with a process known as \textit{compaction}. Compaction moves a number of
SStables from one level to the next level. This is done with an \textit{n-way merge} on the SSTables in the next level.
In this context n-way merge approximately refers to merge sorting on multiple SSTables in the next level. Since each
SSTable is limited to a certain size, this can also introduce new SSTables and in the worst case create yet another
compaction for the next level. This process thus leads to cascaded write amplification as writes have to be done for each
level\cite{sun2020cascaded}, instead of writing them just once in total. This write amplification can reach numbers as
least as high as 50x in LevelDB\cite{balmau2017triad}. More than 2 levels help with amortising compactions because
updates move down multiple levels, not forcing an n-way merge with all data on every compaction.

Because of compaction, we can be sure that each level contains non-overlapping SSTables, except for L0, the first level.
On L0, tables are simply appended. So whenever a read operation is performed, this can lead to one read in the mutable
memtable, one in the immutable memtable, one for each SSTable in L0 and one for each additional level. Read operations
thus have significant read amplification (Q2). This is often mitigated with the usage of AMQs such as bloomfilters.

LSM-trees are considered favourable for flash because of their append-reliant implementation and their buffering
capabilities. Erase and write operations are expensive on flash and can be reduced with the usage of LSM-trees. It also
reduces problems that occur with the usage of random I/O on flash as it favours sequential I/O. Some problems that can
occur when random I/O are used, are expensive erasures and garbage collection. In addition, LSM-trees limit small writes
with the help of buffering in DRAM, increasing throughput and lowering latency and write amplification.

At the same time, they also come with various disadvantages. LSM-trees have for example significant write and read
amplification because of their levelled design (causing Q1 and Q2 issues). They also suffer from the expensive compaction
procedure, which is expensive because it reduces latency of other operations, reduces available throughput and has a high
software overhead.



\subsubsection{Separating keys and values}
\label{sec:sep}
\begin{figure}[h]
  \centering
  \begin{minipage}{.35\textwidth}
    \centering
    \includegraphics[width=1\linewidth]{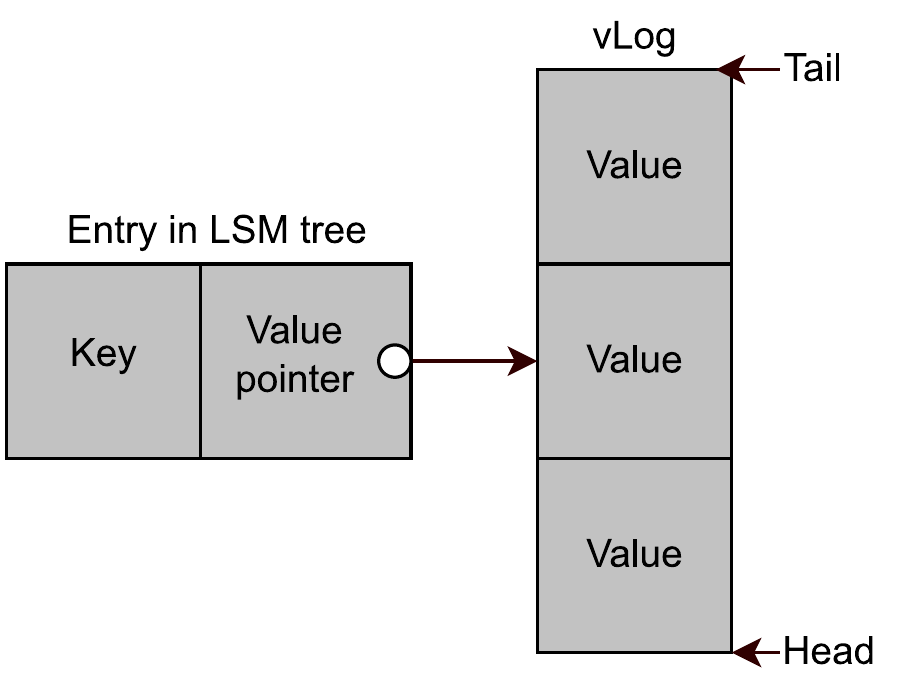}
  \end{minipage}%
  \caption{Separation of keys and values as used in WiscKey.}
  \label{fig:WiscKey}
\end{figure}
Traditional LSM-trees batch their keys and values together sequentially. Whenever a key-value pair needs to be read or
written, the entire pair is read or written. This occurs not only in client operations such as update and delete but also
in background operations such as compactions. This is beneficial for devices that suffer from random I/O as data stays
close together, but flash is known to suffer less from this problem and suffers more from the alternative; the
alternative being a higher write amplification (Q1).

Lu et al. therefore proposed \textit{WiscKey}\cite{lu2017wisckey}. WiscKey separates keys and values, which prevents
rewriting the values unnecessarily. Values are instead stored in a separate data structure. The LSM-tree itself is then
limited to keys and data pointers only. This also has the added side effect that the tree itself becomes smaller and is
thus both easier to cache and requires fewer operations to read. This design is shown to both reduce write and read
amplification and increase throughput. The idea is also already implemented in various other key-value
stores\cite{dong2021sardinedb, dong2021evolution,Dgraph2022,xanthakis2021parallax,chan2018hashkv,li2021differentiated}.
Nevertheless, it also results in new problems and is not a silver bullet. Most problems have to do with the separate data
structure that is used to store the values.

The separate data structure that values are stored in, can in theory be anything. However, a common design is to use a
circular log, called a \textit{vLog}, which is also used in WiscKey\cite{lu2017wisckey}. The vLog fits naturally on flash
storage because of its append-only structure. An additional benefit is that in theory when a log is used, this log can
also function as part of the WAL\cite{dong2021sardinedb}.

Nevertheless, this structure creates problems for scanning operations, because the data is not sorted (increasing effects
of Q2). This is in various designs resolved with the usage of prefetching further scan iterations
beforehand\cite{lu2017wisckey}, but this does not remove the actual problem. Therefore, Li et al. propose to store a part
of the values in another LSM-like tree\cite{li2021differentiated}, known as a \textit{vTree}, that is partially sorted to
get better scan performance. However, the vTree would need additional management and is still not completely sorted.

Using a vLog also causes other problems. Various works namely reported that performance can degrade when the store is
update and/or delete heavy and a vLog is used at the same time. That is because updates and delete operations result in
garbage collection in the vLog. The overhead of this garbage collection increases, when the values stored are small. We
expect similar problems when a different data structure is used. A solution that is data structure agnostic,
differentiates between small, medium and large values\cite{xanthakis2021parallax,chan2018hashkv,li2021differentiated}.
Large values are stored separately and small values are stored in place as in the traditional approach.

The garbage collection procedure remains a major performance bottleneck, even when differentiating between the value
sizes. Therefore Lu. et al, proposed a more garbage collection friendly data structure\cite{chan2018hashkv}. They
proposed mapping values into partitions by hashing the keys. So instead of using one big log, they make use of multiple
partitions. This adds isolation of hot and cold data and determinism of where key-value pairs are stored. The same key is
hashed to the same partition next time. Separating hot and cold data and determinism make it easier for a garbage
collector (Q4) to move fewer data and thus reduces write amplification (Q1).

To conclude, key-value separation is a common optimisation that allows reducing write amplification (Q1) and better
caching of the LSM-tree, but does come with additional garbage collection and can cause lower scan performance.

\subsubsection{Reducing compaction overhead}
\label{sec:compact}
Compactions are known to be one of the main bottlenecks of LSM stores. They incur many writes and reads in a short matter
of time and can cause space amplification during the merge (Q1, Q3 issues). Chan et al. for example mention write
amplifications at least as high as 50x during compactions\cite{chan2018hashkv}. In addition compactions consume CPU
resources (Q6 issues) \cite{bortnikov2018accordion}. Since compactions generally happen at intervals, this causes spikes
in latency and makes stable performance problematic. Therefore, various designs opt to postpone compactions or reduce the
impact of compactions.

\textbf{Tiering and levelling}: There exist two major LSM models. These models are referred to as \textit{levelled} and
\textit{tiered} designs\cite{liu2021ptierdb} or alternatively as LSM-tree and \textit{LSM-forest}
designs\cite{mei2018sifrdb}. Levelled designs follow the standard model we defined earlier, where each level above L0
contains non-overlapping sorted SSTables. However, there also exist various LSM-trees that are less strict on this rule.
These go by multiple names. We will simply refer to them as LSM forests. They are known as forests because they can
contain multiple logical LSM-trees on each level instead of just one. This can occur when during compaction, the trees
are not fully merged, but only partially or not at all.
Raju et al. noticed that in a write-dominated setting it is more important to reduce the write amplification, than it is
to reduce read amplification\cite{raju2017pebblesdb}. Forest designs trade read amplification for lower write
amplification, addressing Raju's concern. They thus trade Q2 for Q1.

The common idea is to postpone compaction all the way up to the last level. This increases read amplification and space
amplification and reduces the scan performance; since this would require reading all SSTables on each level. Further on,
it can also cause a more expensive compaction on the long run because there is one big compaction at the end. On the
other hand, it does reduce the write amplification significantly in many cases. PebblesDB for example reduced write
amplification by 2.3-3x compared to RocksDB\cite{raju2017pebblesdb}. Mei et al. show similar results for PebblesDB for
sequential workloads and favourable results for two other LSM-forest designs, SifrDB and size-tiered
Cassandra\cite{mei2018sifrdb}. Mei et al. do note that \textit{partitioned forest} designs such as PebblesDB and
size-tiered Cassandra have higher write amplification for sequential workloads and will thus not achieve the goal of
reducing write amplification for such workloads. We will get to the definition of partitioned forest, but we will first
look at some additional forest designs.

\begin{figure}[h]
  \centering
  \begin{minipage}{0.45\textwidth}
    \centering
    \includegraphics[width=1\linewidth]{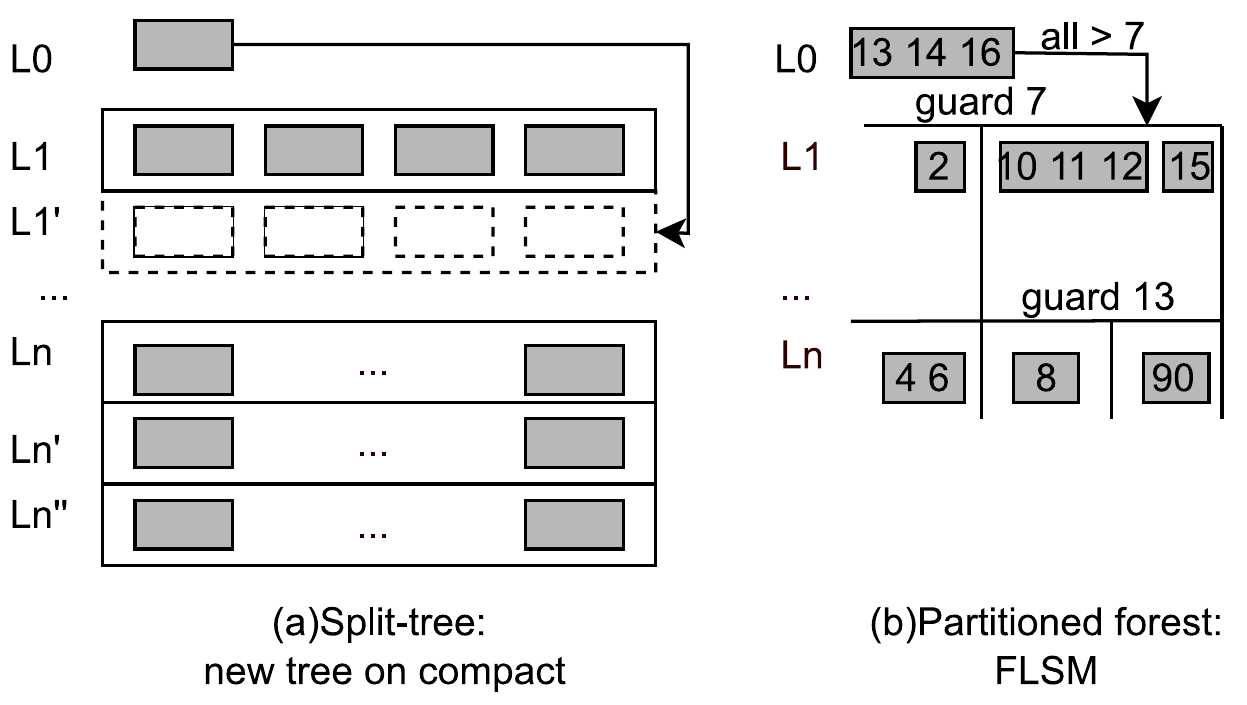}
  \end{minipage}%
  \caption{Multi-stage taxonomy and figure as given by Mei et al.\cite{mei2018sifrdb}}
  \label{fig:forest}
\end{figure}

To mitigate the issues that are related to reads with forest designs, PebblesDB\cite{raju2017pebblesdb} introduced the
\textit{fragmented LSM-tree} (FLSM-tree), visible in \autoref{fig:forest}(b). This design adds a new data structure to
LSM-trees known as \textit{guards}. Each level contains guards in addition to the trees. The guards themselves have a key
range similar to SSTables, but are strictly non-overlapping, decreasing the number of reads necessary on each level and
essentially creating a layered index design. To explain why it is a layered design, we explain what happens on a read. On
a read, first the guards are investigated rather than the SSTables. The guard that matches the key-value pair of the
guard will be picked and all SSTables of that guard will need to be read. Thus two different types of ranges are in
sequence.  On a compaction, the SSTables are moved to their corresponding guards on the next level. In
\autoref{fig:forest}(b) we see such an operation with the SSTable moving to L1 and directly into the guard having the
range $7\sim\infty$. The FLSM design is still suitable for read-heavy situations unlike the standard forest design
because of the usage of guards.

FLSM is known as a \textit{partitioned LSM-forest} design. This means that trees on each LSM-tree level can overlap. For
example, SSTables can move to already existing guards at a higher level on compaction. On a compaction, SSTables can also
be split among guards. This can incur overhead when only a small part of an SSTable falls into a guard and the rest in
another, which increases I/O amplification (in the form of Q1). It also increases write amplification on completely
sequential workloads. That is because on completely sequential workloads, no merges are necessary on compactions. After
all, each flushed SSTable has a new range. In that case, compaction is cheaper than creating new trees. Alternatively, a
\textit{split LSM-forest} design can be used. This has non-overlapping trees on each level. Such an idea is used in
\textit{SifrDB}, proposed by Mei et al.\cite{mei2018sifrdb} and visible in \autoref{fig:forest}(a). This design aims to
solve many of the earlier mentioned problems. The trees are stored separately and each tree itself is properly sorted.
This allows the design to read trees in parallel, which can increase read performance and properly use the internal flash
parallelism of SSDs. It also allows the store to already remove parts of the data during a merge, reducing space
amplification. This removal is not possible with a standard tiered design or a levelled design. Lastly, split forests do
not have problems with increased write amplification on sequential workloads. The decision to use either a tree or forest
design is hard to make, also when only considering flash. In general, if the goal of a key-value store is to either
process a lot of writes or reduce wear levelling caused by compaction, it might be beneficial to look at forest designs.

\textbf{In memory compactions}: It is also possible to leverage memory capabilities to do compactions, reducing the need
for expensive I/O. Such an idea is proposed by Bortnikov et al\cite{bortnikov2018accordion}. Their design,
\textit{Accordion}, uses a pipeline of segments in memory instead of a single in-memory table. There is always one
segment mutable and it is converted to an immutable segment once a certain size is reached. An immutable segment is
implemented as a different data structure that has a lower memory footprint, but sacrifices modification which it no
longer needs. It thus uses only what is needed for immutable segments, good read performance. A lower footprint allows
multiple immutable memtables to exist. Once a number of immutable segments exist, they are merged in memory and flushed
to L0. This allows an earlier merge, already moving part of the problem that would occur in L0. It also reduces wear
levelling as unnecessary writes are removed.

\textbf{Delay compactions}: Another idea is to delay compactions. Simply delaying compactions reduces the effect on the
short run, but can have aggregating effects and can eventually cause a significant spike in long-tail
latency\cite{balmau2019silk, chai2019ldc}. Instead delaying compactions should be based on evidence. For example,
\textit{TRIAD}, proposes to delay compactions until the overlap of SSTables in L0 is big enough\cite{balmau2017triad}.
This avoids writing a large number of duplicates on compaction and therefore reduces write amplification. It can,
however, cause the read performance in L0 to decrease or can cause more expensive compaction later on. TRIAD also aims to
reduce the impact of flushing by reusing the WAL. Instead of rewriting all of the data from the WAL, which is, in
essence, a replay, it reuses the WAL on a flush. The data in the LSM-tree levels will then hold references to the data in
the WAL, instead of copying the data itself. This works especially well for uniform workloads but can introduce some
fragmentations. It can cause fragmentations because the WAL also contains data that is no longer valid, think for example
about updates to data, which invalidates old parts of the WAL.

\textbf{Scheduling compactions}: The final idea we will cover is properly scheduling compactions. This idea does not aim
to remove compactions, but to do compactions at the right time. It should be scheduled in such a manner that read and
write operations receive as little hinder as possible from this background operation. For example, if compactions are
performed too late, flushes have to wait before L0 is properly compacted to L1. On the other hand, if they are performed
too early, compactions might not fully use the buffering capabilities available and compact duplicate data (hot entries
that keep reoccurring).

Such a scheduling idea is proposed by Balmau et al. in the form of SILK\cite{balmau2019silk}. It properly links the
compaction operations to the internal flash parallelism. It always prioritises compactions on L0 because these
compactions are most relevant for client operations and it allows preempting compactions. This means that for example a
compaction on L2 can be stopped, to perform a compaction on L0 instead. Compactions also only get a part of all available
flash bandwidth, allowing clients to always make some progress, even when multiple compactions happen at the same time.
This idea is shown to stabilise the latency of the key-value store and shows no latency peaks like would happen without
such a scheduling procedure.

\subsubsection{Adding indexes to LSM-tree levels}
Typically LSM-trees use sorted buffers on each level of the tree, organised in SSTables. It is also possible to use a
more involved data structure, such as using index structures, of which we have already seen three examples. This
radically changes the characteristics of the individual levels. It for example reworks the entire compaction logic needed
between each level. If chosen adequately it can avoid many of the compaction problems that we discussed in
\autoref{sec:compact}.

Papagiannis et al. for example propose \textit{Kreon}\cite{papagiannis2018efficient}, which uses a B-tree on each level,
accompanied with a completely different compaction operation (addressing Q1, Q2, Q4 and Q6).  The design still uses an
LSM-tree, but uses B-trees as indexes for each level, combining advantages of both data structures. Compactions for the
B-tree indexes work as follows: instead of completely rewriting the next level on compaction, it only \textit{spills} a
part of a B-tree to the next LSM level. To explain spilling, we take a look at how compaction is done for LSM-trees.
Compactions for LSM-trees read the entire next level (except for tiering) and merge sort the entire level into this next
level. Spills read the current level in lexigraphic (sorted) order and inserts it into the B-tree of the next level. This
prevents the reading and sorting of all of the data in the next level, only requiring parts of the tree to be updated. As
a further optimisation, spilling can be done in batches. This is enabled by the lexicographic order and applies multiple
changes to the B-tree before writing it back to storage, effectively amortising multiple spill operations. Such a design
reduces CPU overhead but does introduce many random I/O requests. It performs many random I/O requests because of the
nature of B-trees, see \autoref{sec:btree}. So it only makes sense when CPU and I/O amplification are the main overhead.
Similar to ordinary LSM-trees values can be stored separately, this is also done in Kreon\cite{papagiannis2018efficient}.
This proves that despite switching to an index on each level, many other optimisations meant for standard LSM-trees can
still be performed. Using indexes on each level leverage the idea that it is also a valid idea to add index logic to the individual levels of LSM-trees. Future work, could look at other indexing structures for the LSM-tree levels.

\subsubsection{Reducing space amplification}
\label{sec:space}
Space amplification (Q3) is an important goal of many key-value stores. For some even more important than throughput and
latency. Storage can be expensive and efficient usage of the space can keep the monetary costs down. For example, at
RocksDB they shifted the focus from reducing writing amplification to reducing space
amplification\cite{dong2017optimizing,dong2021evolution}. Dong et al. proposed a few techniques to achieve better space
utilisation and reduce space amplification for LSM-trees.

In the multi-stage design of LSM-trees, the next level might be only marginally bigger than the previous level, which can
waste space. Therefore an idea is to make use of \textit{dynamic levelled
  compaction}\cite{dong2017optimizing,dong2021evolution}. This ensures that each level is at most \(\frac{1}{x}\) of the
next level. The ``x" is configurable; the bigger the ``x", the lower the space and read amplification, but the bigger the
write amplification (trading Q1 for Q2 and Q3).

Another idea that is commonly used to keep space usage down is lossless compression. This can reduce the size of each
file to only a fraction of its normal size. This does come with the disadvantage that data needs to be compressed on
writes and decompressed on reads, which generally is done on the CPU, adding software overhead (Q6). Later on in
\autoref{sec:comp} we will also discuss SSDs that have built-in compression support and therefore do not need to use the
CPU of the host anymore for compression. Dong et al. proposed a \textit{tiered compression} design meant for
LSM-trees\cite{dong2017optimizing}. Tiered compression uses different compression on different levels. Top levels are
smaller and accessed more frequently. Compression on these levels would have less effect on the space but might hamper
throughput. Therefore top levels have no compression, mid levels have light compression and bottom levels have the
highest compression. Lastly, cached pages should always remain uncompressed as they are frequently accessed.

The last idea we will cover is simply dropping some structures that might not be necessary in all cases. Dong et al. for
example argue that bloom filters can be dropped on the last LSM-tree level\cite{dong2017optimizing}. Bloom filters can
use quite a bit of space for each entry and the last level contains quite a few entries. At the same time, these are
accessed infrequently. So in certain use cases, it can be viable to drop these filters. This idea can be generalised to
all AMQs on the last level, so also for cuckoo filters or quotient filters.

A similar approach can be taken for the WAL, as it can also take a bit of space. Therefore, some designs assume a small
bit of NVM, PCM or battery-packed DRAM is present\cite{wang2014efficient,zhang2017flashkv}, which can store the metadata
instead of the SSD. This also causes the SSD to use no space for WAL at all. If no such hardware is available, it might
also be possible to drop the WAL entirely under certain conditions\cite{dong2017optimizing}. The WAL can be dropped if
there already exist other measures to keep the structure persistent and recoverable. Such as replication of the key-value
store (duplicate data as a fail-safe) or databases on top of the key-value store that already provide persistency. Lastly
in TRIAD data in the WAL can be reused in the SSTables\cite{balmau2017triad}, which prevents data duplication, but does
not guarantee lower space amplification.

So in short, some common techniques we have seen for reducing space amplification on LSM-trees include: space-friendly
compaction strategies, compression strategies and dropping unnecessary data structures.

\section{Integrating with the flash interface}
\label{sec:interface}
\label{sec:interface}
\begin{table}[h!]
    \centering
    \begin{tabular}{||l l l||}
        \hline
        Problem & Solution & Examples \\
        \hline \hline
        Q1,Q4,Q7 & Multistream  & Cassandra\cite{huen2017performance},\\&SSDs&RocksDB\cite{yang2015optimizing},\\&&
        MongoDB\cite{nguyen2018optimizing}\\
        \hline 
        Q1,Q2,Q4, & Rely on & NVMKV\cite{marmol2015nvmkv}\\
        Q5,Q6,Q7&FTL capabilities&\\
        \hline
        Q1,Q2,Q4, & Open-channel & LOCS\cite{wang2014efficient},\\Q7&SSDs& NoFTL-KV\cite{vinccon2018noftl},\\&&FlashKV\cite{zhang2017flashkv}\\
        \hline
        Q1,Q2,Q4, & ZNS SSDs & LSM GC\cite{choi2020new},\\Q7&&ZenFS\cite{bjorling2021zns}\\
        \hline
        Q1,Q2,Q4, & Near data & Co-KV\cite{sun2018co}\\
        Q5,Q6,Q7&processing&\\
        \hline
    \end{tabular}
    \caption{Overview of papers covered in \autoref{sec:interface}.}
    \label{tab:interfaceworks}
\end{table}

\begin{figure}[h]
\centering
\begin{minipage}{0.25\textwidth}
  \centering
  \includegraphics[width=1\linewidth]{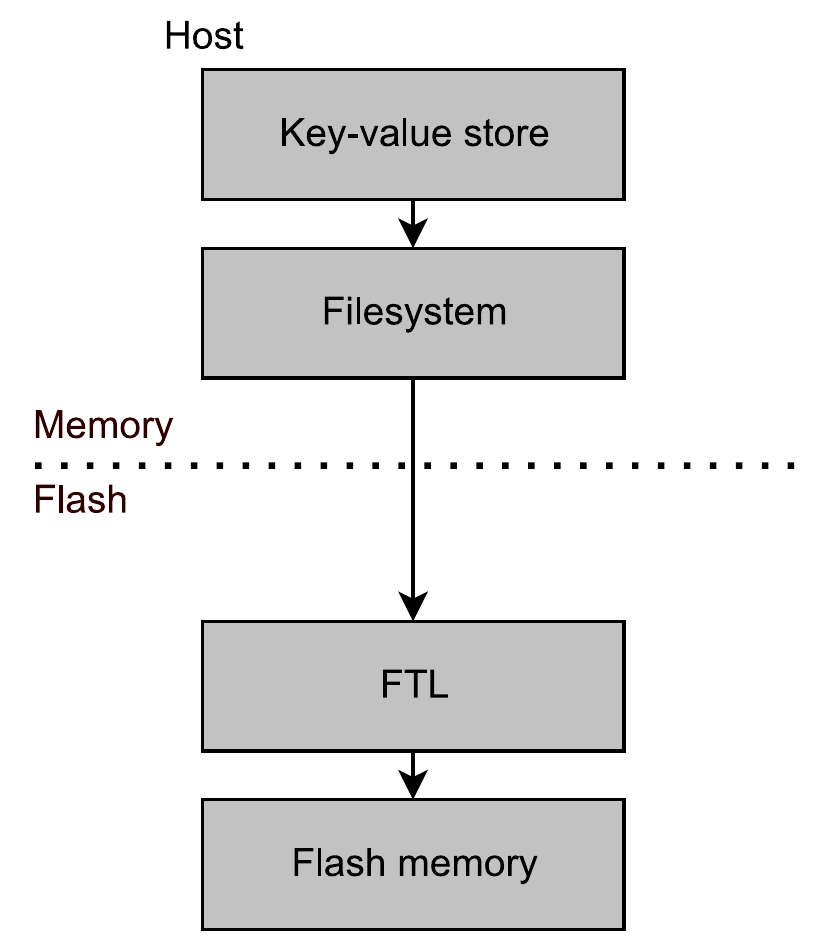}
\end{minipage}%
\caption{The traditional layered approach. Some layers have been abstracted away such as the block layer commonly used in
operation systems such as GNU/Linux.}
\label{fig:layers}
\end{figure}

Key-value stores are commonly stored on a file system. This file system itself then traditionally added on top of another
layer, known as the \textit{Flash Translation Layer} (FTL), which is managed on the flash device itself. For more
information about the FTL, see \autoref{sec:background}. This leads to a layered design, which is visualised in
\autoref{fig:layers}. This design is essentially a double-edged sword. It hides implementation details and allows both
the key-value store and the file system to function with a well defined interface. However, at the same time there is a
\textit{semantic gap} between all of the layers. For example it is not always possible to communicate with the underlying
layer what the difference is between hot and cold data. This in turn leads to duplicate work and leads to
\textit{auxiliary write amplification} (AWA) in the device, making optimisations non-trivial.

Therefore there has been both an academic and an industrial push to reduce the amount of layers needed and to better
integrate the already existing layers. These ideas will be discussed in this section. It is important to note that all of
these ideas are tied to certain devices and can therefore not be used in all cases, which is one of their main
weaknesses.

\subsection{Multi-stream SSD}
\textit{Multi-stream SSDs} are SSDs that allow the host to mark write commands with a stream hint. Normal SSDs have a
single append point, also known as a \textit{stream}, where all new data is stored. Multi-stream SSDs allow writing to
multiple different streams\cite{kang2014multi, bhimani2017enhancing}. These streams can then be handled differently by
the underlying FTL. This allows differentiating data and can thus help with separating hot and cold data, which in turn
helps in bridging the aforementioned semantic gap (solving issues with Q1, Q4). It focuses on improving the communication
between layers but does not remove any layers of itself (this communication helps with Q7). The main advantages of this
improved communication will be a more efficient garbage collection and a reduced AWA.

The technology is promising, but not a lot of key-value stores are optimised specifically for such devices. Nevertheless,
it has been tested on a few already existing key-value stores. For example on the wide-column store \textit{Cassandra}.
For more information on Cassandra, we recommend the paper written by Lakshman et al.\cite{lakshman2010cassandra} on the
data store. Cassandra uses an LSM-tree and can use multi-stream SSDs to give different \textit{stream ids}, used to
differentiate between hot and cold streams, to different parts of the store\cite{huen2017performance, kang2014multi}. For
example, the WAL can be given to one stream and bloom filters, cache and metadata to another. Lastly, as some readers
might already expect they also used different streams for different LSM-tree levels. This design was shown to
significantly reduce write amplification and make garbage collection more efficient. Such ideas were also at some point
in time tested on RocksDB; using similar approaches as Cassandra\cite{dong2021evolution, yang2015optimizing}.

Multi-stream SSDs are also in use for B-tree designs such as WiredTiger in \textit{MongoDB}\cite{nguyen2018optimizing}.
MongoDB makes use of files to store data. They note that mapping different files to different streams is inadequate for
their use case, as it does not address internal fragmentation of data. Hence they proposed an alternative,
\textit{boundary-based stream mapping}, where parts of individual files can be sent to different streams, allowing parts
of files to be differentiated. Parts of a file can for instance be hot or cold.

We conclude by stating that this technology is not bound to any data structure and could also have benefits for other
data structures such as hash tables. It is, for now, unsure if this technology will become bigger for key-value stores.

\subsection{Native FTL capabilities}
\label{sec:native}
There exist various SSDs with FTLs that already provide various transactional operations and persistency guarantees.
Instead of adding layers on top of these FTLs with similar functionality, it can be advantageous to make direct use of
the FTL. This does not only avoid duplicate work, but it also leaves optimisations to the SSD itself (Q7). Not every SSD
is made the same, so it does have advantages to leave optimisations up to the SSD vendor (this can help with Q1, Q2, Q4,
Q5, Q6).

\begin{figure}[h]
\centering
\begin{minipage}{0.25\textwidth}
  \centering
  \includegraphics[width=1\linewidth]{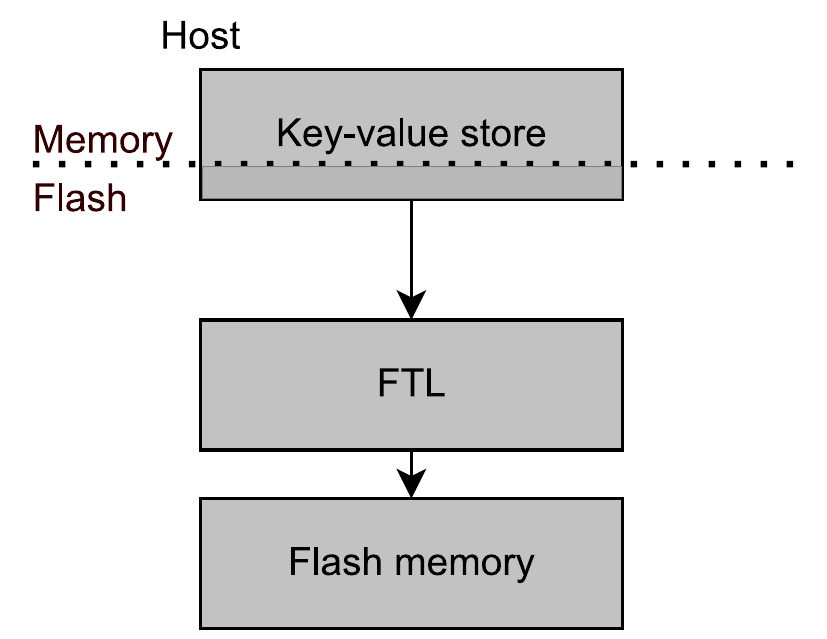}
\end{minipage}%
\caption{The layered approach when using native FTL capabilities. Part of the key-value store is managed by the
underlying FTL.}
\label{fig:nvmkv}
\end{figure}

This idea was pursued in \textit{NVMKV}\cite{marmol2015nvmkv}. This design only works for devices that have implemented
certain capabilities such as \textit{atomic multi-block writes}, \textit{persistent trimming} operations, \textit{exist}
operations and \textit{iterate} operations. 
NVMKV then maps all key-value operations such as get, put and delete to these FTL operations.  Guarantees such as
persistence and recoverability can already be encapsulated by existing FTL operations. Keys are in such designs mapped to
hashes which are interpreted as logical addresses by the device, which the FTL then maps to the actual physical location.
The device thus has complete control over where a hash is stored. Hash collisions can be resolved with operations such as
linear probing, but are in general let up to the design of the store and are not relevant for this survey. Atomic
multi-block writes can be used to write one file atomically, persistent trim operations can be used to make deletes
persistent, iterate operations can be used for scan operations, et cetera. This makes the design of the key-value store
more simplistic. The key-value store essentially becomes a thin layer on top of the FTL, or in other words, the FTL is
also a kind of key-value store. Such a design can be seen in \autoref{fig:nvmkv}.  

Advantages of a thin layer on top of the FTL are a lower CPU usage and a lower DRAM footprint on the host (resolves Q5
and Q5). It also reduces AWA (Q1) since the device is now directly responsible for the application. Lastly, it reduces
the need for multiple levels of key-value management, which reduces the complexity.

NVMKV also addresses another issue, that is absent in most other key-value stores that we cover: \textit{Multi-tenancy},
the idea of running multiple applications on the same device. Multi-tenancy is favourable as it removes the need to use
different storage devices for each application. It is also possible for multiple key-value stores to run on the same
machines, which already happens in designs such as FAWN\cite{andersen2009fawn}. These other key-value stores also make
use of the available flash channels, the DRAM and the cache of the device and the storage of the device. These key-value
stores can therefore hamper each other's performance and interfere with the scheduling of operations. NVMKV aims to solve
such issues with the use of pools. FTLs can provide a pool abstraction that allows binding operations to a group, a
different group can be used for each application. KV pairs of different pools will then be spread evenly across the
address space of the device.

\subsection{Open-channel SSD}
\label{sec:ocssd}
\begin{figure}[h]
\centering
\begin{minipage}{0.25\textwidth}
  \centering
  \includegraphics[width=1\linewidth]{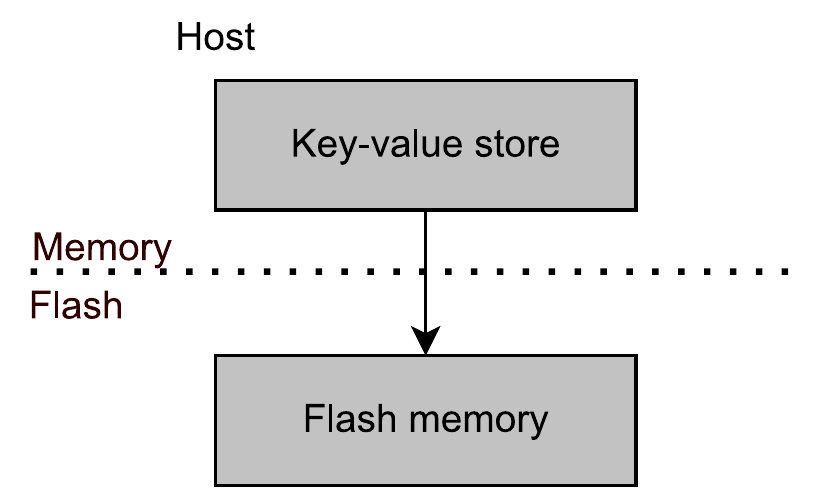}
\end{minipage}%
\caption{An example of a stack on an open-channel SSD. It directly interfaces with the memory.}
\label{fig:locs}
\end{figure}

\textit{Open-channel SSDs} (OCSSD) are SSDs that allow developers to directly control the SSD\cite{bjorling2017lightnvm}.
They do not come with FTL interfaces and instead entirely rely on the host device managing the SSD; minus a few device
specific functionalities such as bad block management. This means that applications, such as key-value stores, themselves
can assert direct control over page mappings and garbage collection (solving Q7 and reducing Q1, Q2, Q4). This leads to
the lean stack visible in \autoref{fig:locs}, provided that the file system is dropped as well. The Linux kernel allows
interfacing with such an SSD through \textit{LightNVM}\cite{bjorling2017lightnvm}. We will cover a few designs that make
use of such devices. 

A key-value store that popularised the idea of running key-value stores on OCSSDs is a design known as \textit{LSM-tree
based Key-Value Store on Open-Channel SSD} (LOCS)\cite{wang2014efficient}. This store has been implemented on
\textit{Software Defined Flash} (SDF)\cite{ouyang2014sdf}, which is a type of SSD with open-channel like capabilities
made by Baidu. SDF, unlike ordinary SSDs, exposes each channel as a different device to the host. This should make it
possible to directly map operations to internal flash channels (see \autoref{sec:flash}) and to schedule the individual
operations.

LOCS internally uses an LSM-tree, see \autoref{sec:LSM}, and proposes various scheduling techniques that more adequately
map LSM-tree operations to the individual device channels. Wang et al. discovered that it is beneficial to take the type
of action into account when doing such scheduling\cite{wang2014efficient}. They propose assigning a different cost to
reads, writes and erasures and schedule each of these operations accordingly. Scheduling techniques could then include
simple scheduling techniques such as Round Robin Scheduling\cite{rasmussen2008round}, but also more involved ones that
try to keep the queue size of each channel to a minimum, with queue size based on the sum of all assigned operations and
their costs.

\textit{FlashKV}\cite{zhang2017flashkv} tries some completely different scheduling techniques but still uses an LSM-tree.
It does not directly assign a cost to actions, but instead prioritises actions over other actions. It for example
prioritises read operations overwrite operations as these take less time. It also chooses to give erasures the lowest
priority, until there is not enough space left. This is done because of the non-critical nature of the erasure operation
when there is enough space available and the high cost of said operation. It also includes separate scheduling techniques
for compaction, based on the workload. When the workload is write-heavy it favours compactions. Similar to other LSM-tree
scheduling techniques such as proposed in SILK\cite{balmau2019silk} (covered in \autoref{sec:compact}), L0 compactions
are considered to be the most important compactions and are prioritised.

FlashKV\cite{zhang2017flashkv}, LOCS\cite{wang2014efficient} and SILK\cite{balmau2019silk} all come with different
scheduling ideas and yet all achieve significant performance gains. This raises the concern that scheduling operations
and compactions are important for getting adequate throughput and latency. Key lessons that could be learned from these
designs revolve around the principles of scheduling each operation differently, differentiating between client and
background operations and how operations can be scheduled across channels to achieve better throughput.

Another idea that is interesting to try out is how files and data are spread across the different channels of the device.
For example, writing a file to just one channel or to stripe the file's blocks across multiple channels at the same time.
Such an idea was proposed by Zhang et al.\cite{zhang2017flashkv} and can improve the read and write performance of
individual files significantly. That is because each channel can read a part of a file in parallel on a read request.
However, such a design does come with its own metadata and management. LOCS also showcased that it is profitable to store
key ranges that are present on multiple LSM-tree levels on different channels. For example by storing a particular range
on channel 0 for L0 and channel 1 for L1. Storing on different channels has benefits because this allows parallel reads
on multiple LSM-tree levels\cite{wang2014efficient}. This avoids having to do all reads on one channel for a read
operation.

Vin{\c{c}}on et al. focus on an entirely different optimisation with OCSSDs. They state that with such devices it is
possible to separate hot and cold data (addresses Q4). For example by sending different levels of an LSM-tree to
different channels\cite{vinccon2018noftl}. This is possible because all data can be managed from the side of the host,
which allows the host to make decisions ranging from where data is stored to what data is stored. This could allow more
efficient garbage collection or access patterns.

In short, we have seen various optimisation techniques for OCSSDs. Key-value stores can be made more efficient on such
devices by considering how both operations and data are spread among the individual flash channels and scheduling
operations accordingly.

\subsection{ZNS SSD}
Another device interface is the \textit{Zoned Namespace} (ZNS) SSD\cite{bjorling2019open}. ZNS SSD builds further on the
\textit{NVMe} specification with the \textit{NVMe ZNS} specification and is a successor of open-channel SSDs. Such
specifications allow standardising approaches and make optimisations transferable between devices. The main idea of the
ZNS specification is to divide the capacity of the devices into distinct zones. Each zone can be read in any order, but
can only be written sequentially. If a zone needs to be changed, it needs to be erased first. This specification closely
follows the characteristics of flash and can thus lead to better internal placement (aiding Q4), higher write throughput,
lower QoS and higher capacity. Unlike standard block interfaces and similar to open-channel SSDs, most logic can also be
moved to the host (addressing Q7). Such as a \textit{host-managed FTL} (HFTL). Optimisations discussed in
\autoref{sec:ocssd} can thus also be applied to ZNS devices, with some translations to make use of zones instead.

Since the host can control most logic and it has an append-friendly structure, it can be beneficial to also store
key-value stores on ZNS SSDs. Especially append-heavy structures such as LSM-trees are promising (potentially aiding with
Q1, Q2 and Q4). As a matter of fact an LSM-tree has already been ported to ZNS, but then for a garbage collection scheme
and not for a key-value store\cite{choi2020new}. ZNS is also considered promising because ZNS is not tied to a certain
device and should in principle be able to run on all types of SSDs and other hardware such as HM-SMR HDDs, provided that
they implement the interface.

A key-value store that runs on a ZNS SSD was proposed by Bj{\o}rling et al.\cite{bjorling2021zns} by letting RocksDB use
\textit{ZenFS}, a ZNS-friendly file system. This increased the throughput and lowered the write amplification in
comparison to other block-interface file systems such as \textit{F2FS} and \textit{XFS}; mainly because the effect of
garbage collection was significantly less.

\begin{figure}[h]
\centering
\begin{minipage}{0.25\textwidth}
  \centering
  \includegraphics[width=1\linewidth]{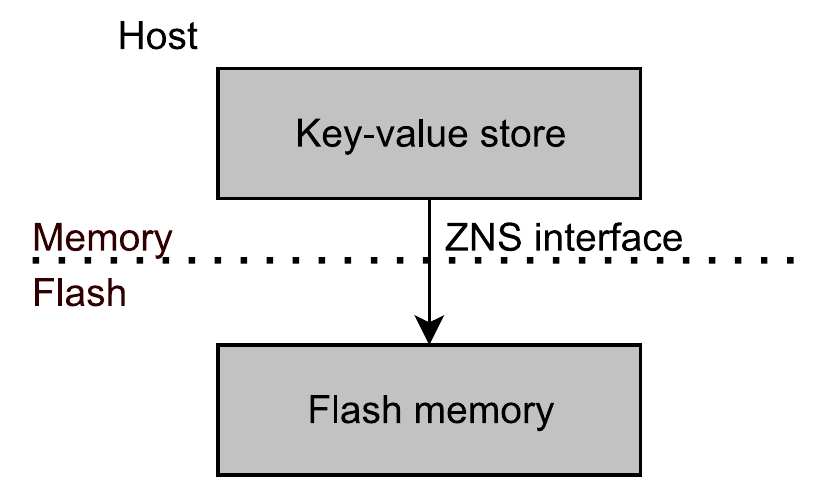}
\end{minipage}%
\caption{A potential design that could be used for ZNS devices.}
\label{fig:zns}
\end{figure}

This proofs that ZNS is a viable candidate; not just for file systems, but also for key-value stores. Further research
could look into the merit of optimising a key-value store for such a device without the help of any file system, similar
to the designs we saw in \autoref{sec:ocssd}. This could lead to a stack as seen in \autoref{fig:zns}.

\subsection{Near data processing}
\begin{figure}[h]
\centering
\begin{minipage}{0.25\textwidth}
  \centering
  \includegraphics[width=1\linewidth]{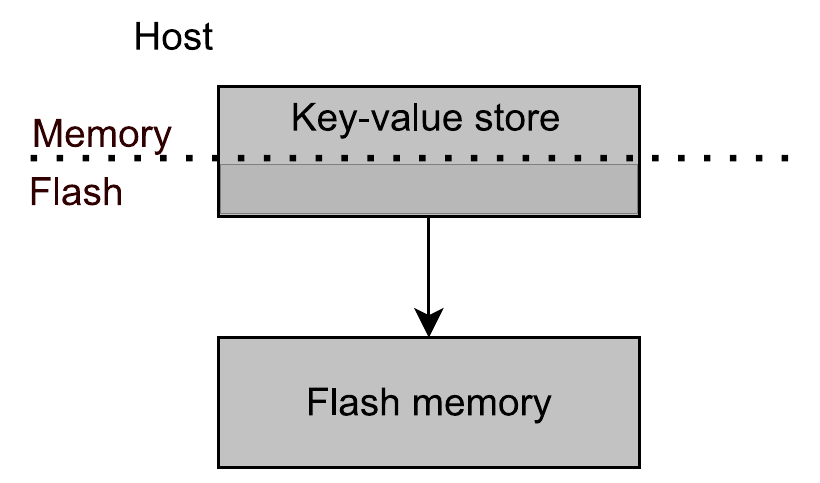}
\end{minipage}%
\caption{A key-value store design that uses near data processing.}
\label{fig:neardata}
\end{figure}

\textit{Near data processing} is the practice of moving the computation close to the data. SSDs could for example come
with their own computation units such as ARM chips or FPGAs. Such devices are also referred to as computational storage
and various other names\cite{lukken2021past}. It allows the logic to be almost completely moved to the device with the
benefits of having the computation close to the data and a lower lookup time (answers Q7 and can reduce Q1, Q2 and Q4). A
beneficial side effect is not having to rely too much on the computing power of the host (indirectly solving Q6, Q5).
Therefore, it takes the complete opposite direction of LOCS and ZNS. In \autoref{fig:neardata} a potential stack for near
data processing is illustrated. For more information about exact definitions and current works on such devices, we
recommend a survey by Lukken et al\cite{lukken2021past} on such devices. We will not cover such attempts in-depth as they
mostly fall outside of the scope of this survey; we will merely mention them and explain the general idea with an
example. 

\textit{Collaborative-KV} (Co-KV) is a proposal to use near-data processing for key-value stores proposed by Sun et
al\cite{sun2018co}. Co-KV uses LSM-trees and offloads the compaction to the device. This should remove the need to move
data between host and device solely for the need of compaction. We have already seen compaction to be a performance
bottleneck for I/O, CPU and memory in \autoref{sec:compact}. This design therefore also reduces the CPU and memory
footprint for compaction on the host. However, such a design does require efficient coordination between the host and the
device itself. 

\section{Reducing software overhead}
\label{sec:CPU}
\begin{table}[h!]
    \centering
    \begin{tabular}{||l l l||}
        \hline
        Problem & Solution              & Examples                               \\
        \hline \hline
        Q6      & Shared nothing        & Kvell\cite{lepers2019kvell},           \\&&
        PrismDB\cite{raina2020prismdb},                                          \\&&SplinterDB\cite{conway2020splinterdb}\\
        \hline
        Q6      & Polling or interrupts & uDepot\cite{kourtis2019reaping},       \\&&
        SpanDB\cite{chen2021spandb}                                              \\
        \hline
        Q6      & I/O paths             & uDepot\cite{kourtis2019reaping},       \\&&
        SpanDB\cite{chen2021spandb},                                             \\&&Kvell\cite{lepers2019kvell},\\&&Kreon\cite{papagiannis2018efficient},\\&&Tucana\cite{papagiannis2016tucana}\\
        \hline
        Q1,Q6   & Efficient commit log  & SplinterDB\cite{conway2020splinterdb}, \\&&
        Tucana\cite{papagiannis2016tucana},                                      \\&&Kreon\cite{papagiannis2018efficient},\\&&
        SpanDB\cite{chen2021spandb},                                             \\&&Kvell\cite{lepers2019kvell}\\
        \hline
    \end{tabular}
    \caption{Overview of papers covered in \autoref{sec:CPU}.}
    \label{tab:cpuworks}
\end{table}
SSDs are known to be significantly faster than earlier storage devices such as HDDs and tape. SSDs themselves however also
do not have a homogeneous performance. There can be a significant difference in the latency and concurrency capabilities
of SSDs. Some devices such as \textit{Z-NAND} and NVMe SSDs are known to be able to achieve \textit{ultra-low latency}
(ULL) and are therefore at times referred to as ULL devices\cite{koh2018exploring}. To give an example of how low low-latency is:
Z-NAND can achieve a memory-read latency of 3µs, which is reported to be 8x times faster than the fastest
page access latency of modern multi-level cell flash storage\cite{koh2018exploring}. Kourtis et al. come with similar
numbers, reporting that fetching a 4KiB block on NVMe SSDs took 80µs and only 12µs on Z-NAND\cite{kourtis2019reaping}.
They even state that storage starts to challenge the performance of network I/O, as a common round-trip latency of a TCP
packet over 10 Gigabit Ethernet is 25-50 µs.

These fast devices are generally directly connected with \textit{PCIe} and the \textit{NVMe} protocol instead of with a
more traditional approach such as \textit{SAS}/\textit{SATA} and \textit{AHCI} to achieve optimal performance. SAS and
SATA do not match the performance that can be reached with SSDs, limiting the bandwidth that can be achieved with SSDs.
PCIe is better able to match the performance of the device. Many different SSD vendors implement their SATA, SAS or PCIe
communication differently requiring an abstract host-to-controller interface such as AHCI\cite{xu2015performance}. AHCI
causes some performance bottlenecks and challenges, such as a single point of aggregation, some additional latency and a
complicated software stack. NVMe removes this point of aggregation, simplifies the software stack (aids Q6), shortens the
hardware data path and allows the software to make use the full device parallelism\cite{xu2015performance}.

For these types of devices, many assumptions about flash and storage in general no longer hold. The result is that if we
are to optimise key-value stores for such devices, we should focus on keeping software overhead down to relieve the CPU.
Instead of trading storage for CPU, we should trade CPU for storage. We will cover a few of such ideas, which will all
focus on solving Q6. Some ideas we have already covered before and will not repeat, such as the $B^\epsilon$-tree and
ideas to reduce compaction overhead in \autoref{sec:compact}. Also note that a large part of these ideas, seem to
contradict what we have determined earlier. That is because optimisations depend on the complete system architecture. So
before optimising, always check where the bottleneck of the current underlying architecture is.

\subsection{Shared nothing}
A design that is frequently used to increase concurrency capabilities, is the \textit{shared nothing} approach. Each
thread has its own data structures and sharing of data between threads is kept to a minimum. Such an approach, ensures
that threads do not have to content for the same resource. This essentially removes locking issues and allows each thread
to use its full capabilities. Some of such ideas used for key-value stores include: dividing the key-value stores index
structures across threads\cite{lepers2019kvell,raina2020prismdb}, letting each thread handle a subset of the key
space\cite{lepers2019kvell} and per thread WALs\cite{conway2020splinterdb}.

Kvell uses a data structure that is optimised around the idea of shared nothing\cite{lepers2019kvell}. They propose giving
each thread a subset of the key space and its own B-tree index structure. The B-trees are supposed to be
completely stored in memory, negating ordinary B-tree drawbacks. Such a design, can require some extra DRAM (worsening
Q5), which can be reduced by only storing the prefixes of keys in the B-tree. This is similar to what we have seen for
hash designs in \autoref{sec:hash}. The design of Kvell also does not force storing data in sequence in any way, reducing
the need of communication among threads, but instead allowing more random I/O to occur.

Another interesting idea, which is adjacent but not the same as shared nothing, is to make use of multiple memtables in
the LSM-tree design. For example using two mutable MemTables for each individual channel\cite{wang2014efficient} instead
of only one mutable and one immutable table in total. This makes better use of both the internal flash parallelism and of
the CPUs parallelism and does not rely on a single point of operations. With the increasing performance of storage, we
can only expect to see more of such works in the future.

\subsection{Polling or interrupts}
A classic dilemma is the decision to use \textit{interrupts} or \textit{polling}. Polling does a request to the device
and then regularly checks if the request is ready; all the time keeping the thread busy. Interrupts on the other hand do
a request to the storage device and then temporarily context switch. Once the data is ready, the context is switched back
to the original context. Interrupts make sense when the I/O is slow because it is inefficient to keep a thread busy while
other work could be done at the same time. However, they do come with one major disadvantage: the need to context switch.
These context switches take time and can trash the cache. Typically I/O was slow, so this decision was given less thought
for storage. Yet, with the onset of faster SSDs, it raises serious questions about the use case of interrupts for flash
as interrupts might actually reduce the performance.

Kourtis et al. addressed this concern and discuss using polling for fast devices\cite{kourtis2019reaping}. Some other
works also make use of asynchronous I/O, which we will discuss further in \autoref{sec:async}. Unfortunately, the line is
blurred between the two concepts and we will therefore mainly focus on one of the two: asynchronicity. For now, we
conclude by stating that asynchronous I/O can make use of polling and that the decision to use polling or interrupt will
remain relevant.

\subsection{I/O path}
\label{sec:path}
There exist a multitude of ways to interface with the storage device. Most of these go through the kernel. This adds
software overhead as it adds additional complexity and depends on expensive operations such as \textit{system calls}
(syscalls). Lepers et al. even mention that with fast devices, focusing on reducing the number of syscalls instead of
focusing on reducing random I/O is more lucrative\cite{lepers2019kvell}. I/O that goes through the kernel is also
dependent on the Operation System and is unaware of the key-value stores' characteristics. This causes a semantic gap
similar to what we have seen in \autoref{sec:interface} between the key-value store and the FTL. Similar approaches for
solving the semantic gap might therefore be beneficial, such as rewriting layers for the key-value store or removing
layers.
In this section, We will cover a few common ways to access I/O and what problems and solutions have been identified for
these approaches.

\subsubsection{Asynchronous I/O}
\label{sec:async}
Traditionally applications made use of \textit{synchronous I/O}. Synchronous I/O means that whoever calls the action has
to wait for the action to complete. This is in the case of a local key-value a thread. In order to perform multiple
operations concurrently, the application typically adds more threads. In order to achieve full performance, all flash
queues need to be saturated, which are quite a few. Unfortunately, this adds expensive context switches and does not
scale. \textit{Asynchronous I/O} on the other hand means queuing a request, but not waiting for the request to finish;
allowing something else to be done concurrently without requiring a context switch.

Asynchronous I/O makes it possible to simply add a request, such as a get to a request, and at the same time move on to
different tasks in the same program.  This more efficiently fills the device queues and allows the application to make
progress. Linux allows both, a synchronous I/O backend\cite{kourtis2019reaping} and asynchronicity with
AI/O\cite{bhattacharya2003asynchronous, kourtis2019reaping} and io\_uring\cite{lee2021boosting}.

Key-value stores like Kvell\cite{lepers2019kvell}, SpanDB\cite{chen2021spandb}, uDepot\cite{kourtis2019reaping} and
others\cite{lee2021boosting} all report using asynchronous I/O to increase performance. Using multiple asynchronous
operations allows satisfying the individual queues of the flash device better.

\subsubsection{User-space I/O}
\textit{user-space I/O} is another way to address I/O. Kourtis et al.\cite{kourtis2019reaping} mention that such a design
moves the logic entirely to the user-space, removing the kernel from the data path. This in turn reduces context
switches, copying overheads and scheduling overheads. In short, it removes layers and thus potential overhead.

Kourtis et al.\cite{kourtis2019reaping} therefore opt to use user-space I/O where possible. Unfortunately, they also
mention that such a design requires access to the device, which is not always possible. This still necessitates
traditional approaches for many use cases and leads to a major reason to still use synchronous I/O, compatibility.
Therefore, they propose their own scheduler, known \textit{TRT}, that can use various I/O backends, allowing the
key-value store to use what is most efficient and available. A key-value store should therefore look into using the
most-efficient backend, but to be compatible needs to keep its options open for now. An example of a user-space I/O
framework is SPDK\cite{yang2017spdk}.

\subsubsection{Memory-mapped I/O}
A common way to access data on the disk is by using \textit{memory-mapped I/O} (mmap). Mmap directly translates virtual
addresses to logical addresses on flash and allows applications to treat storage like it is part of the memory. This can
in principle be done with any of the earlier mentioned I/Os such as synchronous or asynchronous I/O.

Such a design can be a handy interface but is unfortunately hard to control. That is because the mmap implementation
determines what is in memory and what is on disk at any given time, which leads to a semantic gap again. It also comes
with extra operations such as translation operations. Therefore Papagiannis et al. propose using a custom implementation
of mmap, known as \textit{Kmmap}\cite{papagiannis2018efficient}. Kmmap allows assigning priorities to pages, which
essentially separates hot and cold data. Separating hot and cold data helps with assigning what pages to evict. It also
comes with several optimisations such as asynchronicity for writes, which make it a performant choice for key-value
stores. We thus see that memory-mapped I/O is still viable for fast devices, but it might need some slight modifications
to be optimal for flash and key-value stores.

\subsubsection{Direct I/O}
Another type of I/O is \textit{Direct I/O}. Direct I/O directly accesses the device, which can again be both synchronous
and asynchronous. This allows the key-value store to directly determine all of the I/O operations. This can essentially
remove a big part of the semantic gap, as the programmer is given more control.

Unfortunately, such a design is known to not be able to fully saturate the disk queues when requests are done
synchronously. This is because as discussed in \autoref{sec:async}, synchronous I/O has to wait for completion.
Therefore, Lepers et al. state that it would in such a case be more beneficial to \textit{batch} operations
\cite{lepers2019kvell}. Batching issues multiple operations at the same time. We suspect that batching can also have
gains for asynchronous I/O. This not only allows more operations to be completed in one go, but it also reduces the
number of syscalls. This is actually very similar to buffering techniques we have seen in data structure optimisations in
\autoref{sec:datastruct}. To reduce expensive syscalls, we should group multiple operations into one batch operation.

\subsubsection{Caching pages in memory}
\label{sec:cache}
Caching is a general optimisation technique that is also used heavily in key-value stores. It is a common practice to
cache pages retrieved from storage to memory. Contrary to what one might believe, such caching can also reduce
performance. Caching can for example increase software overhead, especially when cache misses are frequent. There is no
clear consensus on how caching should be done. Therefore, we discuss a few topics and hope to make clear that there is no
silver bullet.

Caching is typically done by using the page cache of the underlying system, such as the \textit{virtual file system}
(VFS). Such a design is inefficient for a number of reasons\cite{papagiannis2018efficient, lepers2019kvell}. Firstly, it
introduces duplication since data needs to be copied between kernel and user space. Secondly, the page cache uses its own
eviction policy. Such a policy is not optimised for key-value stores and can keep files in memory that might not be
necessary or remove files that are in fact necessary. Thirdly, it can only issue one read at a time for a single thread
or use locking in the multi-threaded case. Various works also mention various other
disadvantages\cite{papagiannis2018efficient, lepers2019kvell, kourtis2019reaping}, but the gist is that such a cache is
not aware of the use case and is not optimised for such a case in the first place.

Therefore an idea is to create a cache optimised for key-value stores and cut out the middleman.
Kvell uses such an approach with its own internal page cache, which uses a \textit{Least Recently Used} (LRU)
order\cite{lepers2019kvell}. SplinterDB also uses a user-level cache, optimised for
concurrency\cite{conway2020splinterdb}, and SpanDB uses a user-level cache made for its own file system,
\textit{TopFS}\cite{conway2020splinterdb}.

It is also possible to use mmap for caching to avoid having both a copy in the kernel and user
space\cite{papagiannis2016tucana}. \textit{Kreon} uses a modified version of mmap, Kmmap (see \autoref{sec:path}), to
bypasses the Linux page cache\cite{papagiannis2018efficient}. It uses a priority-based FIFO replacement policy. Such a
priority-based policy allows keeping important data in the cache, such as in the case of LSM-trees: L0 and the WAL. It
also used independent banks, that are completely independent and allow for fine-grained locking, which in turn allow for
full parallelism.

Common among all these cache replacements is that they add semantic knowledge to the cache and opt for more concurrency
capabilities, which we assume helps with increasing the performance of key-value stores.
However, we can also ask ourselves if we even need a data cache in the first place (excluding CPU caches). Such a
consideration is made possible because of the speed of modern SSDs. Cached data is typically stored in DRAM because of
its lower latency compared to storage, but the gap is closing. Key-value stores use less memory and have a lower software
overhead when the cache is dropped (Q5 and Q6 in one). Designs without caching such as uDepot\cite{kourtis2019reaping}
have shown that such a design can have potential.
In short, many designs that consider reducing the software overhead, focus on moving the cache away from the kernel path
and towards the user-space or opt to remove the cache entirely.

\subsection{Efficient commit log}
\label{sec:commit}
We already discussed the usage of a \textit{write-ahead log} (WAL) as a commit log for data structures such as LSM-trees
in \autoref{sec:LSM}. Such a WAL does not only cause write amplification, it also incurs extra software overhead. There
exist alternatives for the traditional WAL that are less of a burden on the CPU.

The first idea we will cover is turning the WAL into a concurrency friendly data structure. The WAL is by default one
single shared log. This leads to resource contention when multiple threads are used. Therefore Conway et al. opted for a
per-thread WAL for the design of SplinterDB\cite{conway2020splinterdb}. In order to ensure order on recovery, they make
use of \textit{cross-referenced logs}. Each operation that is performed increments a generation number that is logged
along with the data in the individual WALs, enabling a global order. SpanDB takes a similar approach by using multiple
parallel WAL write streams\cite{chen2021spandb}.

The second idea we will cover is using \textit{copy-on-write} (CoW) for persistence instead of a
WAL\cite{papagiannis2016tucana,papagiannis2018efficient}. WAL causes each write to be done twice. Once to the WAL and
once in-place to the actual data structure that is used. However, such a design does force I/O to be sequential. The main
idea is therefore to trade more writes for sequential I/O (increasing Q1). CoW never does in place updates. Instead on a
write, an entire structure is immediately rewritten out-of-place. This guarantees versioning and persistence because both
the old and the new data structure remain. At some point in time, the old entry can then be removed. CoW therefore only
writes new data once, but it does require more random I/O. As the data is not stored sequentially as would be the case
with a WAL. It thus trades fewer writes for random I/O (solving Q1 issues). We already know that flash has fewer problems
with random I/O and performance can be gained, especially with faster SSDs.

The last idea we will cover is simply not using a commit log in its entirety. We have seen similar ideas in
\autoref{sec:space} for LSM-trees. We now cover commit logs in general. For example, Kvell completely drops the commit
log because it does not buffer updates at all. Buffering does not always make sense when devices are fast enough. Instead
in such a design updates are immediately flushed to disk, preventing unnecessary I/O (Q1)\cite{lepers2019kvell}.

\section{Data related optimisations}
\label{sec:data}
\begin{table}[h!]
  \centering
  \begin{tabular}{||l l l||}
      \hline
      Problem & Solution & Examples \\
      \hline \hline
      Q1,Q2,Q4 & Transform  & Key reshaping\cite{kim2021optimizing}\\
      & access patterns & \\
       \hline
      Q1,Q2,Q3 & Use spatial & EvenDB\cite{gilad2020evendb},\\&locality&
      SlimDB\cite{ren2017slimdb},\\&&RocksDB\cite{dong2017optimizing}\\
      \hline
      Q1,Q2,Q4 & Skewed data & EvenDB\cite{gilad2020evendb},\\&&TRIAD\cite{balmau2017triad},
      \\&&elasticbf\cite{li2019elasticbf}\\
      \hline
      Q1,Q2,Q3,Q4, & Compression & KallaxDB\cite{chen2021kallaxdb},\\Q5,Q6,Q7&SSDs&B-tree \\&&compression\cite{qiao2021closing}\\
      \hline
  \end{tabular}
  \caption{Overview of papers covered in \autoref{sec:data}.}
  \label{tab:dataworks}
\end{table}

Various novel optimisations only make sense under certain circumstances or are very specific and are not necessarily tied
to a data structure. For example optimisations for certain types of datasets or certain access patterns. In this section,
we will discuss these types of optimisations.  

\subsection{Access patterns}
It has already been argued that the differences between sequential and random accesses are less severe on flash storage
than on traditional storage media such as HDDs. Nevertheless, flash still performs better on sequential access patterns
than on random patterns. That is because most SSDs are still limited to a block device, which means that if multiple
values are stored on one block, fewer reads are necessary (reducing Q2). Further on, various data structures such as
B-trees and LSM-trees perform better with sequential operations. Random operations on B-trees might for example require
going back from the top of the tree to the bottom of the tree for each random operation, in the process also tampering
with the cache. The difference becomes more pronounced with scan operations. Stores such as LSM-trees (and especially
tiered LSM) can also result in many reads for each individual read through read amplification. 

Various works have therefore proposed to prefetch results beforehand\cite{lu2017wisckey,chan2018hashkv}, which can
mitigate the performance issues of random I/O (Q2). Another solution that has been proposed, is to make the key pattern
more sequential, removing the problem entirely. This can for example be done with key reshaping proposed by Kim et
al.\cite{kim2021optimizing} Key reshaping uses an additional B-tree to translate keys to sequential keys, which also
requires an additional key for each unique key. On an insert, a new sequential key is created, which is used for the
actual put operation in the key-value store. At the same time, there is a separate store, where the old key is inserted.
The old key then points to the earlier created sequential key. On a lookup, the sequential key is retrieved through this
extra B-tree, which is then used on the key-value store. So each operation requires one extra operation in a translation
tree. This additional tree also needs to be stored but is relatively small as keys are generally
small\cite{cao2020characterizing}. Key reshaping is shown to significantly increase read and write performance (effecting
Q1, Q2 and Q4)\cite{kim2021optimizing}. It should be noted that when a key-value store is only exposed to sequential
patterns, some optimisations that we discussed in earlier sections no longer make sense. For example for LSM-trees it
becomes possible to directly copy SSTables to the next level on compaction instead of merging into the next level, making
optimisations such as tiering (partitioned forests) inefficient.

We will also shortly cover a unique problem that can occur with \textit{3D NAND} devices. 3D NAND devices stack multiple
layers of flash on top of each other, allowing flash storage to become denser. Unfortunately, this can also cause excess
heat to move from one layer to another, causing vertical wear levelling. This necessitates moving data around to
different places when using such technologies. Such optimisations for LSM-trees are proposed by Wang et
al\cite{wang2020temperature}. The lesson to learn here is that wear levelling issues can differ between devices and still
needs to be accounted for.

\subsection{Spatial locality}
\label{sec:spat}
There exist a variety of patterns that are common in key-value stores. Properly using such patterns can improve
performance. For example, \textit{composite keys} are common, which are essentially keys existing out of multiple parts.
Think for example about multiple files in the same directory, where the prefix is the directory and the suffix is the
filename. For such types of data, spatial locality typically has a meaning. However, this is often not accounted for.
LSM-trees for example group data based on temporality and not on locality. Further on, such temporal grouping is
indiscriminate with regards to key popularity, causing wear levelling by excess writing of cold data (worsening Q1).

Gilad et al. propose a data structure, known as \textit{EvenDB}, that combines the batching of LSM-trees with spatial
locality\cite{gilad2020evendb}. To achieve this they make use of \textit{chunks} as the unit of I/O, which hold
contiguous
key ranges. This means that instead of caching or writing pages, always a range of key-value pairs is written. Each chunk
has its own key-value range, SSTable and WAL, allowing related data to stay together. On a read, the appropriate
key-value
range must be found, so the appropriate chunk. If the chunks index is in memory, the data is read directly from this
chunk. Otherwise, the index first needs to be loaded into memory. The data can then be retrieved both from the chunks
SSTable or the WAL. Since a range is used, scan performance can be significantly increased as a part of the range to scan
can be brought directly into memory (reducing Q2 effects). This can significantly reduce write amplification and thus
wear
levelling, but such a design would not work when data has no spatial locality.

Composite keys are also a type of \textit{semi-sorted data}; data where sorting does matter, but frequently only on a
part of the data. For example, if you want to query all data of one user that all use the same prefix in the keys.
Semi-sorted data typically exists out of a prefix and a suffix, such as a directory and a file path for file metadata.
Key-value stores are generally not optimised for such data. This can cause significant read amplification since each
key-value pair can be stored elsewhere. RocksDB allows using bloom filters made specifically for
prefixes\cite{dong2017optimizing}, removing part of the bottleneck for reading. Ren et al. propose a key-value store
entirely optimised for such a situation, \textit{SlimDB}\cite{ren2017slimdb}. 

Ren et al. note that traditional LSM-trees lack optimisations to maximise read performance on SSD and provide techniques
to make use of semi-sorted data to achieve better performance\cite{ren2017slimdb}. They use a three-level compact index
instead of the standard block index used on each level in LSM-trees. This block index makes clever use of the nature of
semi-sorted data. The first index stores the prefixes of the data together and uses \textit{Entropy Coded Tries} (ECT)
for compression, keeping memory footprint down (Q5). This links to the second level, which keeps offsets to the
individual SSTable blocks. The last index contains the suffix of the last key in each block, these can also be
compressed. Compression works in these cases because the data is quite similar to each other, for example prefixes are
generally similar. The indexes are optimised for the common access patterns, such as querying for prefixes. Such a design
can achieve better space utilisation (helping with Q3), lower average latency and less I/O amplification (Q1 and Q2).
Such a design is also highly specialised and not generalising, but illustrates some ideas on how to optimise data stores
for specific patterns. 

Ideas proposed in this section can also be used for other data patterns, specific use cases such as filesystem metadata
or for different devices altogether. It showcases that it still possible to make optimisations on the host side, as the
indexes reside in memory, without focusing on how to make better use of the flash device.

\subsection{Skewed data}
Various users of key-value stores report that their access patterns are skewed\cite{cao2020characterizing}. A small part
of the data is accessed most of the time and a large part of the data is close to never read. Skewed data has
implications for the performance of the key-value store. For example, take LSM-tree based stores. Data is flushed to L0
after a set number of operations independent of the amount of unique keys. If all of these operations only operate on the
same key, they essentially only flush one value to L0, which is highly inefficient. Further on, this effect will be
amplified by later cascading compactions (increasing Q1). Balmau et al. therefore proposed to not flush hot entries and
keep them in memory and the WAL \cite{balmau2017triad}. Gilad et al. propose using a row cache for frequently accessed
keys\cite{gilad2020evendb}. This can keep the hottest entries in memory most of the time, preventing excess I/O
(concerning Q2 and Q1).

Most key-value stores try to reduce I/O cost with the help of AMQs, see \autoref{sec:amq}. However, traditional AMQs such
as bloom filters do not account for skewed data. Ideally, such filters should prevent unnecessary lookups, even for
skewed data. Therefore Li et al. proposed ElasticBF\cite{li2019elasticbf}. A bloomfilter that considers skewed data and
can thus drastically increase read throughput for such workloads.

\subsection{SSDs with compression support}
\label{sec:comp}
\begin{figure}[h]
\centering
\begin{minipage}{0.45\textwidth}
\centering
\includegraphics[width=1\linewidth]{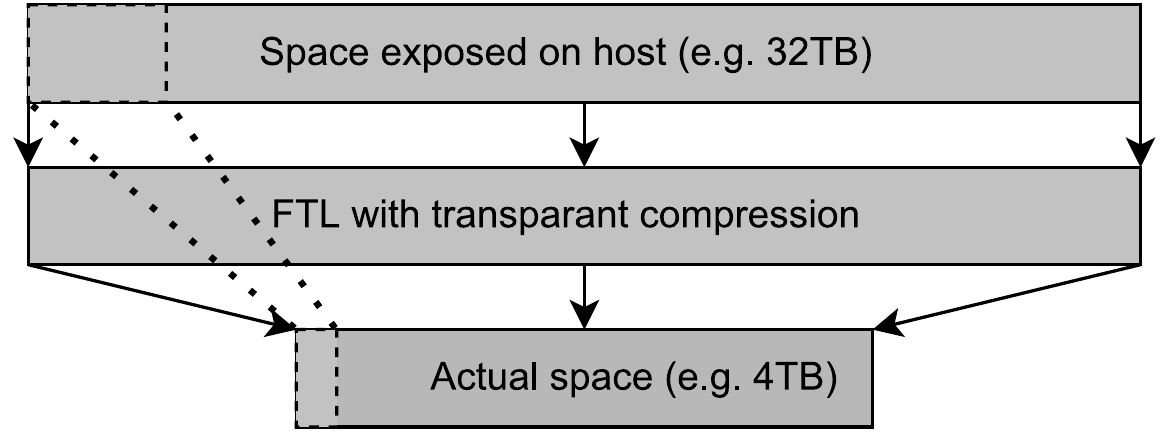}
\end{minipage}%
\caption{Example of the sparse address space when compression is used. Modified variant of a figure posted by Qiao et
al\cite{qiao2021closing}.}
\label{fig:compre}
\end{figure}

There exist flash devices with internal compression capabilities. Such devices can expose a larger address space to the
user than actually physically exists on the device. This can be done with the help of virtualisation that makes use of
said compression. This virtualisation can be directly implemented on the device with the help of an FTL. For example, a
device might expose 32 TB of logical blocks, even when the physical space might be only 4 TB. Such a design is visible in
\autoref{fig:compre}.

Chen et al. investigated the viability of using the unique properties of such devices for key-value stores in the form of
a new key-value store, \textit{KallaxDB}\cite{chen2021kallaxdb}. This design directly maps hashed keys to logical
locations on the SSD, without using a physical hash table. Instead the address space of the device functions as the
actual hash table (addressing Q7). Key-value pairs are stored at logical locations.  Logical locations are defined in
units of blocks; for example, KallaxDB reports 4 KB as a block size regardless of compression\cite{chen2021kallaxdb}.
This design would could cause under-utilisation issues on SSDs without compression capabilities because there would be
various blocks that are only partially occupied, causing fragmentation (impacting Q3). Blocks will be partially occupied
because each key-value pair has to resort to using at least an entire block and frequently key-value pairs are small,
sometimes only a few bytes\cite{cao2020characterizing}. Virtualisation that uses compression can mitigate this issue as
the address space can be made sparser. There is still under-utilisation in this case, but the actual fragmentation effect
is smaller because each logical block counts for less physical space on the device, for example reducing 4 KB to 1 KB.
The tableless design should make it possible to also run key-value stores on devices with little memory or CPU power
because there is very little logic involved on the host (mitigating Q5 and Q6 issues). Lastly, KallaxDB is shown to still
achieve adequate throughput and latency, even under compression. This compression can reduce both write and read
amplification (Q1, Q2 and Q4).

Qiao et al. showcase that such devices are also beneficial for B-trees\cite{qiao2021closing}. They supposedly reduce the
gap between LSM-trees and B-trees. To achieve this B-trees can make use of the sparse address space of the device, as the
device has a larger logical space than physical space because of the aforementioned virtualisation. This can mitigate
ordinary issues with operations such as random I/O, which were discussed in \autoref{sec:btree}. Using the compression
capabilities of the device reduces write and space amplification issues (Q1 and Q3). Therefore, Qiao et al. suggest that
it warrants at least a revisit to the role of B-tree and LSM-trees for future data management
systems\cite{qiao2021closing}. If such SSDs remain relevant, we would expect more specialised designs for such SSDs.

\section{Hybrid architectures}
\label{sec:hybrid}
\begin{table}[h!]
  \centering
  \begin{tabular}{||l l l||}
      \hline
      Problem & Solution & Examples \\
      \hline \hline
       Q1,Q4,Q7 & Flash as cache & FlashStore\cite{debnath2010flashstore}\\
       \hline
       Q1,Q2,Q7 & Multiple flash & I/O Isolation\\&devices&scheme\cite{kim2019isolation},\\&&
       SpanDB\cite{chen2021spandb},\\&&CaseDB\cite{tulkinbekov2020casedb}\\
       \hline
       Q1,Q2,Q4,& Integrating NVM & SSHKV\cite{zhan2019design},\\Q6,Q7&&
       NoveLSM\cite{kannan2018redesigning},\\&&RangeKV\cite{zhan2020rangekv},\\&&
       SplitKV\cite{han2020splitkv},\\&&prismDB\cite{raina2020prismdb}\\
       \hline
  \end{tabular}
  \caption{Overview of papers covered in \autoref{sec:hybrid}.}
  \label{tab:hybridworks}
\end{table}

We have discussed various optimisations techniques for running key-value stores solely on flash storage. Nevertheless,
flash storage is not the only component capable of storage. The idea that a key-value store will only be stored on flash,
would be a simplification of the reality.
For example, it is common for data centres, cloud providers and grid clusters alike to have a heterogeneous selection of
storage devices\cite{krish2014hats}. Alternatives to flash have their own advantages and disadvantages that can be used
for optimising key-value stores or simply keeping the total monetary cost down. This leads to hybrid designs, where
different devices are used for different parts of the key-value store (addressing Q7). This allows key-value stores to use
the advantages of different devices and mitigate their disadvantages. We will discuss a few hybrid designs that have been
proposed.

\begin{figure}[h]
\centering
\begin{minipage}{0.45\textwidth}
\centering
\includegraphics[width=1\linewidth]{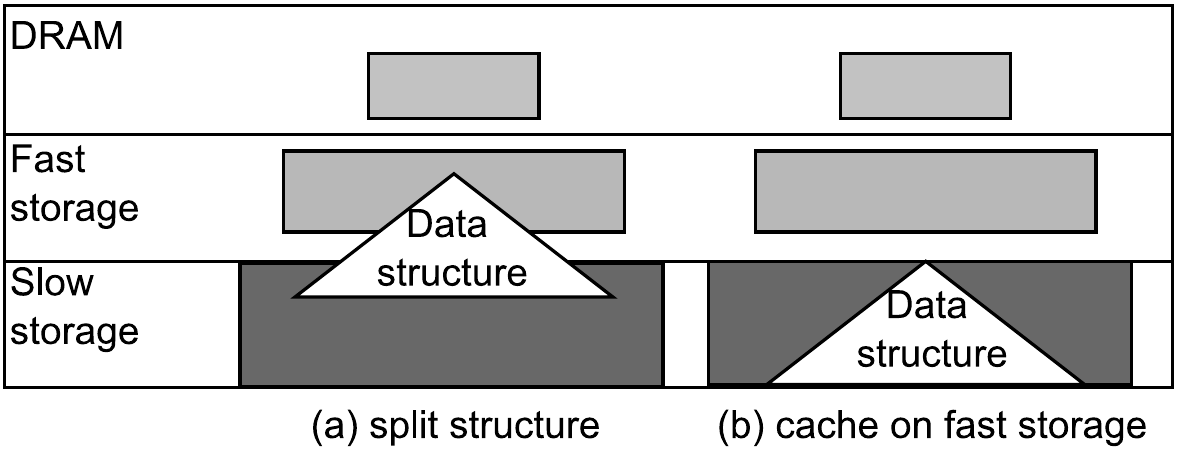}
\end{minipage}%
\caption{Two major hybrid key-value designs. The data structure can in principal be anything.}
\label{fig:tierd}
\end{figure}

A common approach we will see in such designs is that storage devices are differentiated based on their latency and
throughput. In all the approaches we will discuss the faster device is considered more expensive and the slower device
holds more data than the faster device. In such approaches, we identify two prominent designs: a design where the fast
storage is merely used as cache between DRAM and the slower storage, and designs where the main data structure is split
between the devices. The two designs are shown in \autoref{fig:tierd}(b) and \autoref{fig:tierd}(a) respectively. Ren et
al. also identified such a distinction\cite{raina2020prismdb}. We will refer to these approaches as the \textit{cache
approach} and the \textit{split approach} from now on.

\subsection{Flash as cache}
Flash has up until now been more expensive than HDDs. When the amount of data that needs to be stored becomes expansive,
it can become beneficial to store the bulk of the data on HDDs instead. In that case, flash can be used as an extra layer
of cache in between DRAM and HDD. This requires keeping hot data on SSD and eventually destaging cold data to HDD.

Such an idea is proposed by Debnath et al. in FlashStore and follows the cache approach\cite{debnath2010flashstore}. In
this design, valid and recent data is stored on flash and indexed by a hash table in DRAM. Old data is stored on HDD and
needs to be separately indexed. This also reduces the amount of data that is stored on flash, indirectly reducing Q1 and
Q4 challenges for flash.

\subsection{Integrating multiple flash devices}
Using different types of flash in a key-value design is also a hybrid design because, as discussed in \autoref{sec:flash},
there exist multiple types of flash. Some are denser and some have lower latency. We will cover a few examples to show
that differentiating flash has actual merit. 

It has been proposed to use the split design on different flash devices. For example, \textit{SpanDB} splits a LSM-tree
over multiple flash devices\cite{chen2021spandb}. It proposes to put the frequently accessed parts of the store, such as
the WAL and L0 of an LSM-tree on faster flash and the rest on slower flash. Performance is said to improve when combining
multiple devices\cite{chen2021spandb}, despite it probably being faster when only using fast flash. Nevertheless, it is
hard to truly compare such hybrids against alternatives, especially because it is dependent on the mix of storage devices
used. \textit{CaseDB} also uses a split LSM-tree design, but combines it with the idea of key-value
separation\cite{tulkinbekov2020casedb}, discussed in \autoref{sec:sep}. It stores all the metadata, bloom filters and
indexes on fast NVMe SSDs and stores the actual values on a slower SSD. CaseDB reports better write performance than
WiscKey by as much as 1.8 times and increased read performance by as much as 1.5 times\cite{tulkinbekov2020casedb}.

To overcome limits of individual flash devices, it is also possible to combine multiple physical flash devices and treat
them as one logical flash device. This allows spreading the load over multiple devices and negates limits that would occur
when only using one device (such as effects of Q1 and Q2). Such an idea is an I/O isolation scheme for SSDs proposed by
Kim et al\cite{kim2019isolation}. This is reminiscent of the scheduling logic we saw earlier in
LOCS\cite{wang2014efficient} and SILK\cite{balmau2019silk}, see \autoref{sec:ocssd} and \autoref{sec:compact}
respectively. It classifies operations and files and determines what SSD will execute a particular operation on a
particular file. To give an example: they show that is possible with such a design to only send application writes to
certain SSDs, to allow other SSDs to be reserved solely for application reads, which can significantly increase read
throughput and decrease read latency.

\subsection{Persistent memory}
Persistent memory is an emerging topic with a lot of attention from the community and various applications are already
optimised for it, such as key-value stores\cite{benson2021viper, kaiyrakhmet2019slm, zhang2021chameleondb}. These devices
can be put in memory slots just like DRAM, but are unlike DRAM able to persist data. Persistent memory can persist data
even when the power is cut off. Another interesting property is that since they are in a memory slot, they are byte
addressable (at cache line granularity). Optimising for such devices requires a radically different way of thinking. For
most these devices are at the moment still unobtainable and expensive, but they are an interesting topic nonetheless.
Further on, such devices are not necessarily flash themselves, but there have also been works that proposed to use both
flash and persistent memory in an architecture. Persistent memory goes by many names and has various implementations. We
will use the abbreviation for \textit{Non-volatile memory}, NVM, where possible, even if it is not always accurate. We
will shortly cover a few of such hybrid ideas.

We see that split and cache designs are both very common, as we uncover examples for all major data structures covered in
\autoref{sec:datastruct}. In all cases the part stored in NVM is altered to be more efficient on NVM, such as making use
of the byte addressability.
\textit{SSHKV}\cite{zhan2019design} splits a hash table, storing the metadata on NVM and storing the values on a log on
flash. \textit{SplitKV}\cite{han2020splitkv} uses a cache/split design for a B+-tree, storing hot and small items in NVM,
and cold and large values on flash (reducing Q1 and Q4 in flash). \textit{NoveLSM}\cite{kannan2018redesigning} and
\textit{RangeKV}\cite{zhan2020rangekv} both resort to splitting LSM-trees. NoveLSM keeps the mutable and immutable
memtables in NVM and RangeKV stores a modified L0, referred to as \textit{rangetabs}, in persistent memory.

Nevertheless, it can also be beneficial to use completely reworked data structures altogether. Split and cache designs
have their merits, but might not fully use the full potential of the available stack. We will cover one such design,
known as \textit{PrismDB}\cite{raina2020prismdb}, which combines \textit{Quad Level Cell} (QLC) NAND with \textit{3D
Xpoint} (Optane). This combination is remarkable, as it combines both a dense flash device and a fast NVM, combining the
benefits of both worlds. PrismDB uses a partitioned, shared-nothing architecture, where each thread has its own subset of
keys. This separation goes down from DRAM, to NVM to flash. This allows the key-value store to more efficiently use the
low latency and parallelism that NVM and the CPU can provide (addressing Q6 for NVM). In DRAM a B-tree is used, on NVM a
part of the objects and metadata are stored and the rest of the values are stored on flash. It tries to keep hot objects
on NVM and moves colder objects to flash through compaction, which is similar to various caching approaches. Data on
flash is then stored in SSTables, similar to LSM-trees.

In short, we see that NVM can be used efficiently in combination with flash devices in order to optimise key-value
stores. It can be used to split a data structure, to cache data on NVM or to create a data structure optimised for the
combination of the two.

\section{Measuring performance}
\label{sec:bench}
We have already determined what we think defines performance in \autoref{sec:background}, yet actually measuring this
performance is non-trivial. Especially with the layered design and the heterogeneity of the devices that most key-value
stores make use of. Additionally, various designs combine a multitude of ideas, add one new part, and yet test against an
alternative with significantly fewer ideas. It raises serious questions about the validity of the comparison of various
designs. Researchers have therefore opted for new ways to measure performance. Some have focused on specific data
structures, others on general problems. This section will focus on the status of this field of studies and how key-value
stores can be benchmarked in general. It will also take a look at the common workload patterns that key-value stores are
typically optimised for.

The first tool we will cover is for LSM-trees, proposed by Batsaras et al. and is known as \textit{Variable Amplification
Analysis} (VAT)\cite{batsaras2020vat}. VAT introduces an analysis to make trade-offs for LSM parameters and their effect on
among others I/O. It covers various parameters such as SSTable sizes, growth between levels, tree and forest optimisations,
key-value separation optimisations and how the effects of these can be properly measured. This immediately captures a big
part of the LSM optimisations we covered in \autoref{sec:LSM}. Such designs make it easier to reason about the
optimisations.  As we are speaking about such optimisations, it would also be a good idea to ensure that when stores are
compared, they are compared fairly. Both should be optimised. For LSM-trees there already exist some tooling to aid in
automatic optimisation, such as Monkey or Dostoevsky\cite{dayan2017monkey, dayan2018dostoevsky}. 

There also exist a myriad of other pitfalls. Fortunately, some of these have been identified. Didona et al. covered various
pitfalls that are made in measuring the performance of key-value stores with regards to flash\cite{didona2020toward}. They
also come with advice on how to solve these pitfalls. Some major findings include that it is essential to measure write
amplification from both the application and the device (AWA), tests should be run for a longer period of time, tests should
mention the \textit{state} of the SSD, test should also be done with different data set sizes, tests should consider
metadata, tests should consider \textit{overprovisioning} and tests should consider heterogeneous flash devices. The state
of the SSD refers to the various states that an SSD can be in. For example, a device can be empty or it can already be
filled partially. This has performance implications and must be noted down to make research reproducible. Overprovisioning
implies that a part of the SSD can not be used by the user, but is used by the device itself for internal operations. This
has implications for various operations such as garbage collection or out of place writes. Didona et al. also show a few
examples where depending on the device, a B-tree or an LSM-tree can perform better, hence the rule on heterogeneous
devices\cite{didona2020toward}. 

We have thus already seen an analysis tool, some automatic optimisers and a set of pitfalls. As far as we know there exists
no tool that can automatically verify all constraints and ensure validity. Similarly, there are no guarantees that all
recent studies or future studies will uphold the pitfalls. This is probably in part because of the vastly heterogeneous use
cases of key-value stores and the difficulty in setting up such tooling. For now, it is therefore simply a set of
guidelines. For future devices or stores, it would therefore be ideal if they would adhere to said guidelines and maybe add
some analysis tools such as VAT.

At the moment there do exist various benchmarks that can be used to partially test key-value stores on a few common
workloads. This can be used to show that on someone's hardware a key-value store performs better than another key-value
store for a particular use case. In general, a good benchmark makes use of a real workload, to showcase that it is also
usable in real scenarios. Such benchmarks are therefore in general provided by companies that actually use key-value stores
and have a lot of data. Therefore, the workloads made by these companies reflect their workload and might not reflect edge
cases or niche applications, making such benchmarks difficult. Instead, they approximate, which in the general case is
enough. However, for high performing solutions, this probably requires a different workload.
Some example benchmarks include Yahoo's \textit{Yahoo Cloud Serving Benchmark} (YCSB)\cite{cooper2010benchmarking} and
\textit{db\_bench}\cite{cao2020characterizing} from Meta, which is integrated in RocksDB. 

YCSB seems to be especially prevalent. Cooper et al. recognise a set of data distributions and workloads attached to said
distributions\cite{cooper2010benchmarking}. Some example distributions are \textit{uniform} and \textit{zipfian}
distributions. Uniform evenly distributes the accesses and Zipfian distributions have skewed accesses, with more accesses in
a small part of the set and a long tail tagging behind. The default workloads, named A to E, represent a set of common and
distinct use cases: update heavy(A), read heavy(B), read only(C), read latest(D) and short ranges(E). This can be used to
show case that when a key-value store is deployed on flash, it is suited for various use cases.

Cao et al. characterise the workloads of RocksDB and showcase that workloads differ wildy\cite{cao2020characterizing}. They
come with various methodologies to analyse the workloads of three common use cases. Data from Facebook, ZippyDB (distributed
key-value store) and UP2X (used for AI/ML) is traced to get the distribution. In all three workloads, the distributions
differ significantly. Again confirming that each workload is different. However, they also conclude that spatial locality is
important, making optimisations in \autoref{sec:spat} sensible.

In short, for now, measuring performance of key-value stores for flash includes using prevalent benchmarks with common
workloads. At the same time measuring flash specific metrics and adhering to a set of pitfalls. Lastly, adding some extra
analysis where necessary such as VAT. 

\section{Future Prediction}
Now that we have discussed most major contributions to key-value stores with regards to flash, we would like to discuss
what will happen to key-value stores in the future. We have seen trends with each new type of device, both in academia and
the industry. Think about key-value stores for HDDs, SMR HDDs, ordinary SSDs, NVMe SSDs and various other variants of
SSDs. Similarly, we have seen that it takes time for each technology to be adopted by the community, where some fail and
some make it, but if an idea is promising it seems to come back. Think for example about near data processing technologies
or open-channel designs; ideas that both already exist for a while.

If techniques discussed in this survey become more widely adopted, they can have significant effects in the near future.
They allow reducing write, read and space amplification, which aids in better throughput, latency and  lower wear
levelling. Further on, they allow lower CPU overheads and lower memory overheads. This would require a smaller investment
in flash storage, as the storage can be more effectively utilised and breaks done at a slower pace (wear levelling),
requiring less replacements. This in turn can reduce costs of the data center. At the same time a few of the solutions
require less of the host, allowing a reduction in investments for these devices. This also has an effect on the amount of
energy necessary, as less devices need to be used. We have already seen, that many users of key-value stores have moved to
using flash storage. We therefore expect (1) that the techniques discussed in this survey will become more widespread,
which would allow these improvements to occur. At the same time this does imply that key-value stores will become
primarily optimised for flash storage only and not for other media (yet).

Flash storage is moving in multiple directions. On one side we see a push in making flash storage denser to account for a
need to store more data in the future. For example multi-cell technologies such as  QLC and PLC NAND flash. Each of these
is becoming denser, yet less tolerant to wear-levelling and has a higher latency and lower throughput. On the other hand,
we see a push to increase the performance of flash storage. For example, Z-NAND or faster NVMe devices. We thus see both a
push to faster storage and denser storage. We expect (2) key-value stores to move along; some key-value stores that aim to
address the denser storage and some that aim to address the faster storage. This will on the long term cause either
various different key-value stores or key-value stores that become highly complex as they need to address different needs.
Developers than also need to consider what key-value store to use along with their storage decisions and requirement
decisions, adding further complexity. This requirement of different needs is going to increase on the long term, as the
needs keep expanding. On the long term this does allow using different stores and storage for the appropriate needs, which
more efficiently uses what is available.

We also see a push in reducing the amount of layers needed to address storage on flash storage. We have for example seen
open-channel SSDs and more recently ZNS SSDs. We have also seen key-value store solutions for open-channel SSDs, such as
LOCS (see \autoref{sec:ocssd}). In the short term we expect (3) more solutions for ZNS SSDs as the technology seems
promising and is already available. If this technology keeps expanding, it can on the long terms have effect on the cost
and durability of storage in the data center, as such devices can reduce wear levelling and write amplification when
properly used\cite{stavrinos2021don}. The ordinary commodity SSD, having already stood the test of time, will also not go
away anytime soon, but ZNS devices can significantly increase the performance of key-value stores, yet are at the same
time not significantly more expensive to produce.

Recently, quite a few storage technologies different from flash started to receive attention from the community. There is
an increased interest in using alternative storage media and materials\cite{chatzieleftheriou2020could}. The main intent
is to find a type of storage that better fits modern-day use cases, such as dealing with large amounts of archival data or
cloud computing. This is becoming especially important as the amount of data that is needed in the world is
ever-increasing. We also see more requirements to reduce the amount of energy needed and an interest in longevity of
storage (archiving). Some examples include glass storage in the cloud for archiving with project
Silica\cite{anderson2018glass} or an interest in storing large amounts of data in DNA\cite{li2020can}. By some,
holographic storage is also reconsidered as a viable storage medium\cite{chatzieleftheriou2020could}. We expect that more
of such media will be developed or reconsidered and that some of them will be also be considered for key-value stores in
the long term. Similarly, we expect some storage media that are already in use for key-value stores to become more
important. Such as persistent memory. We have already seen that there have been various works, both academic and
industrial, that opt to use persistent memory for key-value stores (see \autoref{sec:relatedwork}). Thus in short, we see
that there is a push to use different types of storage, all with different characteristics and some that are not even
ready for usage yet. Key-value stores being a technology to store data, will therefore both in the short, medium and long
term be considered and altered to make use of these storage media. As these technologies have different characteristics,
it is likely that we will also see different types of data structures, different from the LSM-tree and B+-tree discussed
in \autoref{sec:datastruct}. This means that we expect (4) that key-value stores will not be limited to HDD and flash
technologies and should probably on the long term also not be treated as such. This indeed implies, that key-value stores
can be used for all kinds of different needs in all kinds of different situations. 

In short, we expect that flash will remain a major storage medium for key-value stores for a long time. We expect that
novel flash-specific techniques, as discussed in this survey, will become adopted and see a widespread usage. We also
expect a part of the flash storage to move to denser solutions and a part to faster solutions. We expect key-value stores
to move along on the long term, which will require different approaches. Further on, we expect more flash storage devices
to adopt open-channel and ZNS designs and key-value stores to become more optimised for such solutions. At the same time,
we also expect other completely different storage media to be considered at the long term for key-value stores or the
other way around. This will require rethinking many designs decisions, which have been made for flash. Such redesigns are
already done for persistent memory. In the end, all of these different requirements and designs will lead to vastly
different implementations and probably different APIs.

\section{Design guidelines for development of flash-based key-value stores}
There exist various approaches that can be taken when designing key-value stores for flash. None of these approaches is
necessarily wrong or right. However, we believe that there is a lot that can be learned from already existing solutions,
such as the ones described in this survey. Based on these solutions, we come with a set of guidelines that can aid in
creating a key-value store on flash. Note that this survey does not cover all possible use cases (see \autoref{sec:open}),
but the general remarks provided in this section can still aid in developing a key-value store. Further on, in this survey
more in-depth optimisations are marked with numbers such as \textit{Q1} or \textit{Q2}, which should further aid with
finding solutions for requirements. Each optimisation should always be tested against the alternative, preferably with
benchmarking tooling.

When designing a key-value store for flash, we recommend to first take a look at the use case and requirements of the
key-value store in question. For example, should it have a high throughput, be space efficient or something else
altogether. As is discussed in \autoref{sec:bench}, different use cases typically have different data distributions.
Different distributions in turn allow for different optimisations, such as is discussed in \autoref{sec:datastruct} and
\autoref{sec:data}. Therefore, we recommend (1) to first analyse or estimate the workload of the use case. This should be
about what patterns exist in the data and the balance that exists between writes, reads, updates and deletes. When the
workload is known, look if there already exists a key-value store optimised for this workload that satisfies the
requirements. For example, key-value stores such as RocksDB\cite{dong2017optimizing} or LevelDB\cite{ghemawat2011leveldb}.
If not, this still helps with deciding on a data structure. For example, if the workload is write-heavy, data structures
such as LSM-trees or other WOIs (see \autoref{sec:datastruct}) might be an option. Depending on the workload these can be
further optimised, such as LSM-forests instead of LSM-trees (see \autoref{sec:comp}).  Similarly, if we know that the data
is skewed or certain access patterns exist it might be beneficial to look into data related optimisations~(see
\autoref{sec:data}). 

Once the use case is known, we recommend (2) looking at the storage hardware that will be used to store the data.
Different hardware, needs to be treated differently. If the storage that will be used is heterogeneous, look into how all
of these storage components can be used together effectively. For this we refer to \autoref{sec:hybrid}. For example, the
faster storage can be used as cache between DRAM and slower storage, an existing data structure can be split between
faster and slower storage, or an entirely novel data structure can be used that is optimised for a combination. Key-value
store designers should also consider whether only flash is used or also other storage media are used in this hybrid
approach. For example, if persistent memory is available it might also be beneficial to look into optimisations for
persistent memory. Designers should also look at the type of flash storage that is used. If the storage is fast, the
bottleneck will start to move more to the host. In that case we recommend (3) to also take a look at optimisations on the
side of the host. Such techniques are described in \autoref{sec:CPU}, such as making the data structures more concurrency
friendly. If we know that the CPU on the host is weak, such design considerations are also be beneficial. Therefore
consider (and benchmark), where the bottleneck of the key-value store is. Further on, designs should also address the
memory (Q5) that is available on the host. As we have seen in \autoref{sec:datastruct}, certain index data structures need
to be altered to make use of the memory available. Depending on the memory available, different data structures and
caching policies make sense.

Another aspect we recommend (4) looking at, are the interfaces between the key-value store and the storage. If the store
should be able to work on all kinds of devices and operating systems, it should adhere to a common interface, such as a
file system. However, if not and storage devices are used that allow improving the communication between storage and host,
this should be considered. This can help among others with better performance and reduced wear levelling (device
longevity). This includes optimisations that are described in \autoref{sec:interface}. For example, using open-channel
SSDs and dropping the file system and FTL to allow the key-value store to completely control the device.

Lastly, we recommend (5) to benchmark as described in \autoref{sec:bench}. For example, by adhering to a set of storage
pitfalls, using common benchmarks and using workloads provided by companies that use key-value stores in production. To be
able to use a common benchmark, it is advisable to adhere to a common interface of a benchmark such as
YCSB\cite{cooper2010benchmarking} or db\_bench\cite{cao2020characterizing}. 

In short, there a few steps to consider when deciding on a key-value store. Generally, the key-value store should be
designed for its particular use case. That means that it should consider the use case's workload, the storage devices
available, the CPU and memory on the host devices, the interfaces between the store and storage and should be benchmarked
by a common benchmark to ensure validity.

\section{Open problems and future work}
\label{sec:open}

Key-value stores are a big research field and this survey does not address all aspects and challenges that arise when
designing key-value stores. Neither did we find solutions to all challenges that we already know of from literature. The
biggest problems, which we identified as Q1-Q7, mostly reflect what is most prominent in the literature. In this section we
aim to address challenges, that we either think should be looked into in a later survey or that are still open problems for
key-value stores and storage in general.

We have not looked at how much energy is required when using key-value stores. Another survey could look into how key-value
stores on flash have an impact on energy usage. Reducing energy usage is considered important, also for storing
data\cite{tsirogiannis2010analyzing,saito2013energy, koronen2020data,dayarathna2015data}, which makes addressing this
challenge interesting. Flash storage is said to have a lower energy consumption than alternatives such as
HDDs\cite{park2011comprehensive}, which should aid developers and administrators in making decisions on what storage medium
to use. Some potential questions are: ``Do key-value stores cost a large amount of energy to maintain?", ``How does the
energy usage of key-value stores compare to other data stores?", ``How does the usage of flash impact the energy usage of
key-value stores?" and ``How can key-value stores for flash be altered to reduce their energy usage?".

Another issue we have not addressed in-depth is multi-tenancy. We have described one solution, NVMKV\cite{marmol2015nvmkv},
in \autoref{sec:native}. However, multi-tenancy is common, such as in \textit{Software as a Service} (SaaS)
solutions\cite{tsai2010towards}, and most solutions we see in the literature do not address this issue. Future work can look
at how key-value stores (on flash) address this challenge. 

We have also not addressed security and privacy concerns that arise when using key-value stores in this survey. An important
requirement for many use cases is to store data securely, so that it is hard for the non-intended to among others read,
alter or deny the data. Future surveys/work could look into how techniques proposed in this paper, are secure or can be
altered to be more secure or how to secure key-value stores on flash in general. 

Similarly, we have addressed some forms of persistence such as WAL and CoW (see \autoref{sec:commit}), but we have not
focused on these parts of key-value stores. Recovery can take a different amount of time depending on the data structure.
There is also a difference in reliability of various methods that can be used. Future work could look into reliability and
recoverability of key-value stores with regards to flash. This is important, because it ensures that solutions discussed in
this survey can be used in critical situations.

An important open problem, and a general problem in storage is how to compare solutions. For example, as discussed in
\autoref{sec:bench}, it is currently not possible to automatically verify if research upholds pitfalls that occur when
benchmarking flash. Similarly, many stores have been optimised for a particular case and device, making it hard to compare
key-value solutions or estimate what solution will work best. Further on, various solutions have made use of different
benchmarks, increasing the problem. This is a problem we expect will remain for now, and is something that needs to be
addressed. 

Future work could look into key-value stores for other storage~media than block-based flash, such as persistent memory or
\textit{key-value SSDs} (KVSSDs). KVSSD devices are made specifically for key-value stores, so are especially interesting.
Further research could also look into other types of key-value stores when used on flash, other than traditional persistent
solutions, such as ephemeral key-value stores used for caching.

Future work could also look into how flash-based key-value stores fit into the entire ecosystem that they are used in. For
example, if a database is used around key-value stores, or what tools interact with key-value stores. This is important to
get a good picture of what to optimise key-value stores for.

Lastly, future work could look into other data stores and how they can be optimised for flash. This makes it less difficult
to make an informed decision on what data store to pick apart from key-values. A related open problem is that it is at the
moment difficult to test  different solutions. Future work could look into how to do this efficiently to help with making a
decision on the data store and the storage media.

\section{Conclusion}
We conclude by answering our research question: ``What is the impact of flash storage on the design choices for key-value
stores?".
Considering the amount of flash-specific ideas that have been proposed, some already used widely in production, we can
safely say that it has had a big impact. It has become common to store key-value stores on flash storage and to optimise
these stores to get more use out of the device. To give a more precise answer to the research question, we will also answer
each sub-question individually:
\begin{itemize}
    \item  RQ1 - ``What is the current role of flash for key-value stores?":
    Flash is used in a variety of roles, ranging from the main storage medium of key-value stores, to a caching layer, to
    slow storage used for cold data. Generally, it is used to store a large amount of data that needs to remain fast to
    access. Flash is widely used for key-value stores by among others data centres, game services and webshops. 
    \item  RQ2 - ``How has flash influenced the design of key-value stores over the decade?":
    We have seen various flash specific optimisations in this survey. Key-value stores have adapted to flash specific
    challenges such as wear levelling and device parallelism. We thus see that key-value stores have started to adapt to the
    idiosyncrasies of the storage medium. The design of various key-value stores is thus directly influenced by flash
    storage. Data structures and other design decisions, all the same, have evolved as a result of flash specifics. 
    \item  RQ3 - ``What are the main challenges involved in using key-value stores on flash and how can they be mitigated?":
    The main challenges that we identified are: reducing write amplification, reducing read amplification, reducing space
    amplification, reducing garbage collection overhead, reducing memory overhead, reducing CPU overhead and how to
    integrate flash in an architecture. There exist various optimisations that tackle a few of these challenges, but none
    that tackles them all. Generally, a trade-off needs to be made, where one challenge is solved at the expense of another.
    The main challenge is then deciding what trade-off is best for the use case of a particular key-value store.
    \item  RQ4 - ``How will flash contribute to key-value stores in the future?":
    Flash will continue to remain relevant for key-value stores. We expect key-value stores to keep storing the majority of
    their data on flash for years to come. We also expect new ideas to be proposed to further optimise key-value stores for
    flash storage.
\end{itemize}

\section*{Glossary}
\begin{itemize}
    \item AHCI: Advanced Host Controller Interface
    \item AMQ: approximate membership query
    \item AWA: auxiliary write amplification
    \item CoW: copy-on-write
    \item DB: database
    \item ECT: entropy coded trie
    \item FIFO: first in first out
    \item FLSM: fragmented LSM
    \item FS: file system
    \item FTL: Flash Translation Layer
    \item GC: garbage collector
    \item HDD: hard disk drive
    \item HFTL: host-managed FTL
    \item I/O: input/output
    \item KV: key-value
    \item LBA: logical block address
    \item LRU: least recently used
    \item LSM-tree: log-structured merge-tree
    \item MLC: multi-level cell
    \item mmap: memory-map
    \item MS: multi-stage
    \item NVM: non-volatile memory
    \item NVMe: non-volatile memory Express
    \item OCSSD: open-channel SSD
    \item PCI: Peripheral Component Interconnect
    \item PCIe: PCI Express
    \item PLC: penta-level cell
    \item RA: read amplification
    \item SaaS: Software as a Service
    \item SAS: Serial Attached SCSI
    \item SATA: Serial ATA
    \item SCM: storage class memory
    \item SDF: software defined flash
    \item SLC: single-level cell
    \item SMR: shingled magnetic recording
    \item SPDK: Storage Performance Development Kit
    \item SSD: solid state drive
    \item SSTable: sorted strings table
    \item syscall: system call
    \item TLC: triple-level cell
    \item VFS: virtual file system
    \item vLog: value log
    \item vTree: value tree
    \item QLC: quad-level cell
    \item ULL: ultra low latency
    \item WA: write amplification
    \item WAL: write-ahead log
    \item WL: wear levelling
    \item WOI: write optimised index
    \item ZNS: Zoned Namespace
\end{itemize}

\bibliographystyle{acm}
\bibliography{bibliography}

\begin{thebibliography}{100}

\bibitem{ahn2015forestdb}
{\sc Ahn, J.-S., Seo, C., Mayuram, R., Yaseen, R., Kim, J.-S., and Maeng, S.}
\newblock {ForestDB}: {A} fast key-value storage system for variable-length
  string keys.
\newblock {\em IEEE Transactions on Computers 65}, 3 (2015), 902--915.

\bibitem{alsalibi2018survey}
{\sc Alsalibi, A.~I., Mittal, S., Al-Betar, M.~A., and Sumari, P.~B.}
\newblock A survey of techniques for architecting {SLC/MLC/TLC} hybrid {Flash}
  memory--based {SSDs}.
\newblock {\em Concurrency and Computation: Practice and Experience 30}, 13
  (2018), e4420.

\bibitem{anand2010cheap}
{\sc Anand, A., Muthukrishnan, C., Kappes, S., Akella, A., and Nath, S.}
\newblock Cheap and {Large CAMs} for {High Performance Data-Intensive Networked
  Systems}.
\newblock In {\em NSDI\/} (2010), vol.~10, pp.~29--29.

\bibitem{andersen2009fawn}
{\sc Andersen, D.~G., Franklin, J., Kaminsky, M., Phanishayee, A., Tan, L., and
  Vasudevan, V.}
\newblock {FAWN}: {A} fast array of wimpy nodes.
\newblock In {\em Proceedings of the ACM SIGOPS 22nd symposium on Operating
  systems principles\/} (2009), pp.~1--14.

\bibitem{andersen2010rethinking}
{\sc Andersen, D.~G., and Swanson, S.}
\newblock Rethinking flash in the data center.
\newblock {\em IEEE micro 30}, 04 (2010), 52--54.

\bibitem{anderson2018glass}
{\sc Anderson, P., Black, R., Cerkauskaite, A., Chatzieleftheriou, A., Clegg,
  J., Dainty, C., Diaconu, R., Drevinskas, R., Donnelly, A., Gaunt, A.~L.,
  et~al.}
\newblock Glass: A new media for a new era?
\newblock In {\em 10th USENIX Workshop on Hot Topics in Storage and File
  Systems (HotStorage 18)\/} (2018).

\bibitem{aritome2015nand}
{\sc Aritome, S.}
\newblock {\em {NAND} flash memory technologies}.
\newblock John Wiley \& Sons, 2015.

\bibitem{balmau2017triad}
{\sc Balmau, O., Didona, D., Guerraoui, R., Zwaenepoel, W., Yuan, H., Arora,
  A., Gupta, K., and Konka, P.}
\newblock {TRIAD}: {Creating Synergies Between Memory, Disk and Log} in {Log
  Structured Key-Value Stores}.
\newblock In {\em 2017 {USENIX} Annual Technical Conference ({USENIX}{ATC}
  17)\/} (2017), pp.~363--375.

\bibitem{balmau2019silk}
{\sc Balmau, O., Dinu, F., Zwaenepoel, W., Gupta, K., Chandhiramoorthi, R., and
  Didona, D.}
\newblock {SILK}: {Preventing Latency Spikes} in {Log-Structured Merge
  Key-Value Stores}.
\newblock In {\em 2019 {USENIX} Annual Technical Conference ({USENIX}{ATC}
  19)\/} (2019), pp.~753--766.

\bibitem{barroso2017attack}
{\sc Barroso, L., Marty, M., Patterson, D., and Ranganathan, P.}
\newblock Attack of the killer microseconds.
\newblock {\em Communications of the ACM 60}, 4 (2017), 48--54.

\bibitem{batsaras2020vat}
{\sc Batsaras, N., Saloustros, G., Papagiannis, A., Fatourou, P., and Bilas,
  A.}
\newblock {VAT}: {Asymptotic Cost Analysis} for {Multi-Level Key-Value Stores}.
\newblock {\em arXiv preprint arXiv:2003.00103\/} (2020).

\bibitem{bender2011don}
{\sc Bender, M.~A., Farach-Colton, M., Johnson, R., Kuszmaul, B.~C.,
  Medjedovic, D., Montes, P., Shetty, P., Spillane, R.~P., and Zadok, E.}
\newblock Don't thrash: how to cache your hash on flash.
\newblock In {\em 3rd Workshop on Hot Topics in Storage and File Systems
  (HotStorage 11)\/} (2011).

\bibitem{benson2021viper}
{\sc Benson, L., Makait, H., and Rabl, T.}
\newblock Viper: an efficient hybrid {PMem-DRAM} key-value store.
\newblock {\em Proceedings of the VLDB Endowment 14}, 9 (2021), 1544--1556.

\bibitem{bez2003introduction}
{\sc Bez, R., Camerlenghi, E., Modelli, A., and Visconti, A.}
\newblock Introduction to flash memory.
\newblock {\em Proceedings of the IEEE 91}, 4 (2003), 489--502.

\bibitem{bhattacharya2003asynchronous}
{\sc Bhattacharya, S., Pratt, S., Pulavarty, B., and Morgan, J.}
\newblock Asynchronous {I/O} support in {Linux} 2.5.
\newblock In {\em Proceedings of the Linux Symposium\/} (2003), pp.~371--386.

\bibitem{bhimani2017enhancing}
{\sc Bhimani, J., Yang, J., Yang, Z., Mi, N., Giri, N.~K., Pandurangan, R.,
  Choi, C., and Balakrishnan, V.}
\newblock Enhancing ssds with multi-stream: {What?} why? how?
\newblock In {\em 2017 IEEE 36th International Performance Computing and
  Communications Conference (IPCCC)\/} (2017), IEEE, pp.~1--2.

\bibitem{bjorling2019open}
{\sc Bj{\o}rling, M.}
\newblock From open-channel {SSDs} to zoned namespaces.
\newblock In {\em Proc. Linux Storage Filesyst. Conf.(Vault)\/} (2019), p.~1.

\bibitem{bjorling2021zns}
{\sc Bj{\o}rling, M., Aghayev, A., Holmberg, H., Ramesh, A., Le~Moal, D.,
  Ganger, G.~R., and Amvrosiadis, G.}
\newblock {ZNS}: Avoiding the block interface tax for flash-based {SSDs}.
\newblock In {\em 2021 USENIX Annual Technical Conference (USENIX ATC 21)\/}
  (2021), pp.~689--703.

\bibitem{bjorling2017lightnvm}
{\sc Bj{\o}rling, M., Gonzalez, J., and Bonnet, P.}
\newblock {LightNVM}: {The Linux} {Open-Channel} {SSD} {Subsystem}.
\newblock In {\em 15th USENIX Conference on File and Storage Technologies (FAST
  17)\/} (Santa Clara, CA, Feb. 2017), USENIX Association, pp.~359--374.

\bibitem{bortnikov2018accordion}
{\sc Bortnikov, E., Braginsky, A., Hillel, E., Keidar, I., and Sheffi, G.}
\newblock Accordion: Better memory organization for {LSM} key-value stores.
\newblock {\em Proceedings of the VLDB Endowment 11}, 12 (2018), 1863--1875.

\bibitem{cao2020characterizing}
{\sc Cao, Z., Dong, S., Vemuri, S., and Du, D.~H.}
\newblock Characterizing, {Modeling}, and {Benchmarking} {RocksDB}{Key-Value}
  {Workloads} at {Facebook}.
\newblock In {\em 18th USENIX Conference on File and Storage Technologies (FAST
  20)\/} (2020), pp.~209--223.

\bibitem{Cattell2011ScalableSA}
{\sc Cattell, R. G.~G.}
\newblock {Scalable SQL} and {NoSQL} data stores.
\newblock {\em SIGMOD Rec. 39\/} (2011), 12--27.

\bibitem{chai2019ldc}
{\sc Chai, Y., Chai, Y., Wang, X., Wei, H., Bao, N., and Liang, Y.}
\newblock {LDC}: a lower-level driven compaction method to optimize
  {SSD-oriented} key-value stores.
\newblock In {\em 2019 IEEE 35th International Conference on Data Engineering
  (ICDE)\/} (2019), IEEE, pp.~722--733.

\bibitem{chan2018hashkv}
{\sc Chan, H.~H., Liang, C.-J.~M., Li, Y., He, W., Lee, P.~P., Zhu, L., Dong,
  Y., Xu, Y., Xu, Y., Jiang, J., et~al.}
\newblock {Hashkv}: {Enabling} efficient updates in {KV} storage via hashing.
\newblock In {\em 2018 {USENIX} Annual Technical Conference ({USENIX}{ATC}
  18)\/} (2018), pp.~1007--1019.

\bibitem{chang2008bigtable}
{\sc Chang, F., Dean, J., Ghemawat, S., Hsieh, W.~C., Wallach, D.~A., Burrows,
  M., Chandra, T., Fikes, A., and Gruber, R.~E.}
\newblock Bigtable: A distributed storage system for structured data.
\newblock {\em ACM Transactions on Computer Systems (TOCS) 26}, 2 (2008),
  1--26.

\bibitem{chatzieleftheriou2020could}
{\sc Chatzieleftheriou, A., Stefanovici, I., Narayanan, D., Thomsen, B., and
  Rowstron, A.}
\newblock Could cloud storage be disrupted in the next decade?
\newblock In {\em 12th USENIX Workshop on Hot Topics in Storage and File
  Systems (HotStorage 20)\/} (2020).

\bibitem{chen2021spandb}
{\sc Chen, H., Ruan, C., Li, C., Ma, X., and Xu, Y.}
\newblock {SpanDB}: {A Fast, Cost-Effective LSM-tree Based} {KV Store} on
  {Hybrid Storage}.
\newblock In {\em 19th {USENIX} Conference on File and Storage Technologies
  ({FAST} 21)\/} (2021), pp.~17--32.

\bibitem{chen2021kallaxdb}
{\sc Chen, X., Zheng, N., Xu, S., Qiao, Y., Liu, Y., Li, J., and Zhang, T.}
\newblock {KallaxDB}: {A Table-less Hash-based Key-Value Stor}e on {Storage
  Hardware} with {Built-in Transparent Compression}.
\newblock In {\em Proceedings of the 17th International Workshop on Data
  Management on New Hardware (DaMoN 2021)\/} (2021), pp.~1--10.

\bibitem{choi2020new}
{\sc Choi, G., Lee, K., Oh, M., Choi, J., Jhin, J., and Oh, Y.}
\newblock {A New LSM-style Garbage Collection Scheme} for {ZNS SSDs}.
\newblock In {\em 12th {USENIX} Workshop on Hot Topics in Storage and File
  Systems (HotStorage 20)\/} (2020).

\bibitem{chung2009survey}
{\sc Chung, T.-S., Park, D.-J., Park, S., Lee, D.-H., Lee, S.-W., and Song,
  H.-J.}
\newblock A survey of flash translation layer.
\newblock {\em Journal of Systems Architecture 55}, 5-6 (2009), 332--343.

\bibitem{comer1979ubiquitous}
{\sc Comer, D.}
\newblock Ubiquitous {B-tree}.
\newblock {\em ACM Computing Surveys (CSUR) 11}, 2 (1979), 121--137.

\bibitem{conway2020splinterdb}
{\sc Conway, A., Gupta, A., Chidambaram, V., Farach-Colton, M., Spillane, R.,
  Tai, A., and Johnson, R.}
\newblock {SplinterDB}: {Closing} the bandwidth gap for nvme key-value stores.
\newblock In {\em 2020 {USENIX} Annual Technical Conference ({USENIX}{ATC}
  20)\/} (2020), pp.~49--63.

\bibitem{cooper2010benchmarking}
{\sc Cooper, B.~F., Silberstein, A., Tam, E., Ramakrishnan, R., and Sears, R.}
\newblock Benchmarking cloud serving systems with {YCSB}.
\newblock In {\em Proceedings of the 1st ACM symposium on Cloud computing\/}
  (2010), pp.~143--154.

\bibitem{cornwell2012anatomy}
{\sc Cornwell, M.}
\newblock Anatomy of a solid-state drive.
\newblock {\em Communications of the ACM 55}, 12 (2012), 59--63.

\bibitem{daim2008forecasting}
{\sc Daim, T.~U., Ploykitikoon, P., Kennedy, E., and Choothian, W.}
\newblock Forecasting the future of data storage: case of hard disk drive and
  flash memory.
\newblock {\em Foresight\/} (2008).

\bibitem{davoudian2018survey}
{\sc Davoudian, A., Chen, L., and Liu, M.}
\newblock A survey on {NoSQL} stores.
\newblock {\em ACM Computing Surveys (CSUR) 51}, 2 (2018), 1--43.

\bibitem{davoudian2020big}
{\sc Davoudian, A., and Liu, M.}
\newblock Big data systems: A software engineering perspective.
\newblock {\em ACM Computing Surveys (CSUR) 53}, 5 (2020), 1--39.

\bibitem{dayan2017monkey}
{\sc Dayan, N., Athanassoulis, M., and Idreos, S.}
\newblock Monkey: {Optimal} navigable key-value store.
\newblock In {\em Proceedings of the 2017 ACM International Conference on
  Management of Data\/} (2017), pp.~79--94.

\bibitem{dayan2018dostoevsky}
{\sc Dayan, N., and Idreos, S.}
\newblock Dostoevsky: {Better} space-time trade-offs for {LSM-tree} based
  key-value stores via adaptive removal of superfluous merging.
\newblock In {\em Proceedings of the 2018 International Conference on
  Management of Data\/} (2018), pp.~505--520.

\bibitem{dayarathna2015data}
{\sc Dayarathna, M., Wen, Y., and Fan, R.}
\newblock Data center energy consumption modeling: A survey.
\newblock {\em IEEE Communications Surveys \& Tutorials 18}, 1 (2015),
  732--794.

\bibitem{debnath2010flashstore}
{\sc Debnath, B., Sengupta, S., and Li, J.}
\newblock {FlashStore}: {High} throughput persistent key-value store.
\newblock {\em Proceedings of the VLDB Endowment 3}, 1-2 (2010), 1414--1425.

\bibitem{debnath2011skimpystash}
{\sc Debnath, B., Sengupta, S., and Li, J.}
\newblock {SkimpyStash}: {RAM} space skimpy key-value store on flash-based
  storage.
\newblock In {\em Proceedings of the 2011 ACM SIGMOD International Conference
  on Management of data\/} (2011), pp.~25--36.

\bibitem{debnath2010chunkstash}
{\sc Debnath, B.~K., Sengupta, S., and Li, J.}
\newblock {ChunkStash}: {Speeding Up Inline Storage Deduplication Using Flash
  Memory}.
\newblock In {\em USENIX annual technical conference\/} (2010), pp.~1--16.

\bibitem{Dgraph2022}
{\sc Dgraph}.
\newblock Badger.
\newblock \url{https://github.com/dgraph-io/badger}, last accessed on
  2022-02-10.

\bibitem{didona2020toward}
{\sc Didona, D., Ioannou, N., Stoica, R., and Kourtis, K.}
\newblock Toward a better understanding and evaluation of tree structures on
  flash {SSDs}.
\newblock {\em Proceedings of the VLDB Endowment 14}, 3 (2020), 364--377.

\bibitem{dipert1993flash}
{\sc Dipert, B., and Hebert, L.}
\newblock Flash memory goes mainstream.
\newblock {\em IEEE spectrum 30}, 10 (1993), 48--52.

\bibitem{dong2021sardinedb}
{\sc Dong, M., Zhong, H., Sun, B., Bi, S., and Cai, Y.}
\newblock {SardineDB}: {A Distributed Database} on the {Edge} of the {Network}.
\newblock In {\em Asia-Pacific Web (APWeb) and Web-Age Information Management
  (WAIM) Joint International Conference on Web and Big Data\/} (2021),
  Springer, pp.~186--193.

\bibitem{dong2017optimizing}
{\sc Dong, S., Callaghan, M., Galanis, L., Borthakur, D., Savor, T., and Strum,
  M.}
\newblock {Optimizing} {Space Amplification} in {RocksDB}.
\newblock In {\em CIDR\/} (2017), vol.~3, p.~3.

\bibitem{dong2021evolution}
{\sc Dong, S., Kryczka, A., Jin, Y., and Stumm, M.}
\newblock {Evolution} of {Development Priorities} in {Key-value Stores Serving
  Large-scale Applications}: {The RocksDB Experience}.
\newblock In {\em 19th {USENIX} Conference on File and Storage Technologies
  ({FAST} 21)\/} (2021), pp.~33--49.

\bibitem{eisenman2019flashield}
{\sc Eisenman, A., Cidon, A., Pergament, E., Haimovich, O., Stutsman, R.,
  Alizadeh, M., and Katti, S.}
\newblock Flashield: a hybrid key-value cache that controls flash write
  amplification.
\newblock In {\em 16th USENIX Symposium on Networked Systems Design and
  Implementation (NSDI 19)\/} (2019), pp.~65--78.

\bibitem{escriva2012hyperdex}
{\sc Escriva, R., Wong, B., and Sirer, E.~G.}
\newblock {HyperDex}: A distributed, searchable key-value store.
\newblock In {\em Proceedings of the ACM SIGCOMM 2012 conference on
  Applications, technologies, architectures, and protocols for computer
  communication\/} (2012), pp.~25--36.

\bibitem{fan2014cuckoo}
{\sc Fan, B., Andersen, D.~G., Kaminsky, M., and Mitzenmacher, M.~D.}
\newblock Cuckoo filter: {Practically} better than bloom.
\newblock In {\em Proceedings of the 10th ACM International on Conference on
  emerging Networking Experiments and Technologies\/} (2014), pp.~75--88.

\bibitem{Fevgas2019IndexingIF}
{\sc Fevgas, A., Akritidis, L., Bozanis, P., and Manolopoulos, Y.}
\newblock Indexing in flash storage devices: a survey on challenges, current
  approaches, and future trends.
\newblock {\em The VLDB Journal 29\/} (2019), 273--311.

\bibitem{ghemawat2011leveldb}
{\sc Ghemawat, S., and Dean, J.}
\newblock {LevelDB}, 2011.

\bibitem{gilad2020evendb}
{\sc Gilad, E., Bortnikov, E., Braginsky, A., Gottesman, Y., Hillel, E.,
  Keidar, I., Moscovici, N., and Shahout, R.}
\newblock {EvenDB}: optimizing key-value storage for spatial locality.
\newblock In {\em Proceedings of the Fifteenth European Conference on Computer
  Systems\/} (2020), pp.~1--16.

\bibitem{graefe2011modern}
{\sc Graefe, G., and Kuno, H.}
\newblock Modern {B-tree} techniques.
\newblock In {\em 2011 IEEE 27th International Conference on Data
  Engineering\/} (2011), IEEE, pp.~1370--1373.

\bibitem{gupta2017nosql}
{\sc Gupta, A., Tyagi, S., Panwar, N., Sachdeva, S., and Saxena, U.}
\newblock {NoSQL} databases: {Critical} analysis and comparison.
\newblock In {\em 2017 International conference on computing and communication
  technologies for smart nation (IC3TSN)\/} (2017), IEEE, pp.~293--299.

\bibitem{han2020splitkv}
{\sc Han, S., Jiang, D., and Xiong, J.}
\newblock Splitkv: Splitting {IO} paths for different sized key-value items
  with advanced storage devices.
\newblock In {\em 12th {USENIX} Workshop on Hot Topics in Storage and File
  Systems (HotStorage 20)\/} (2020).

\bibitem{harrington2016relational}
{\sc Harrington, J.~L.}
\newblock {\em Relational database design and implementation}.
\newblock Morgan Kaufmann, 2016.

\bibitem{he2017unwritten}
{\sc He, J., Kannan, S., Arpaci-Dusseau, A.~C., and Arpaci-Dusseau, R.~H.}
\newblock The unwritten contract of solid state drives.
\newblock In {\em Proceedings of the twelfth European conference on computer
  systems\/} (2017), pp.~127--144.

\bibitem{herlihy2008hopscotch}
{\sc Herlihy, M., Shavit, N., and Tzafrir, M.}
\newblock Hopscotch hashing.
\newblock In {\em International Symposium on Distributed Computing\/} (2008),
  Springer, pp.~350--364.

\bibitem{huen2017performance}
{\sc Huen, H., Choi, C., and Balakrishnan, V.}
\newblock Performance and endurance enhancements with multi-stream ssds on
  apache cassandra.
\newblock {\em Samsung Semiconductor
  \url{http://www.samsung.com/us/labs/collateral/index.html}\/} (2017).

\bibitem{idreos2020key}
{\sc Idreos, S., and Callaghan, M.}
\newblock Key-value storage engines.
\newblock In {\em Proceedings of the 2020 ACM SIGMOD International Conference
  on Management of Data\/} (2020), pp.~2667--2672.

\bibitem{im2020pink}
{\sc Im, J., Bae, J., Chung, C., Lee, S., et~al.}
\newblock {PinK}: {High-speed} in-storage key-value store with bounded tails.
\newblock In {\em 2020 {USENIX} Annual Technical Conference ({USENIX}{ATC}
  20)\/} (2020), pp.~173--187.

\bibitem{jin2017kaml}
{\sc Jin, Y., Tseng, H.-W., Papakonstantinou, Y., and Swanson, S.}
\newblock {KAML}: A flexible, high-performance key-value {SSD}.
\newblock In {\em 2017 IEEE International Symposium on High Performance
  Computer Architecture (HPCA)\/} (2017), IEEE, pp.~373--384.

\bibitem{jung2021gpukv}
{\sc Jung, M.-G., Lee, C.-G., Park, D., Park, S., Noh, J., Chung, W., Park, K.,
  and Kim, Y.}
\newblock {GPUKV}: an integrated framework with {KVSSD} and {GPU} through {P2P}
  communication support.
\newblock In {\em Proceedings of the 36th Annual ACM Symposium on Applied
  Computing\/} (2021), pp.~1156--1164.

\bibitem{kaiyrakhmet2019slm}
{\sc Kaiyrakhmet, O., Lee, S., Nam, B., Noh, S.~H., and Choi, Y.-r.}
\newblock {SLM-DB}:{Single-Level}{Key-Value Store} with {Persistent Memory}.
\newblock In {\em 17th USENIX Conference on File and Storage Technologies (FAST
  19)\/} (2019), pp.~191--205.

\bibitem{kang2007mu}
{\sc Kang, D., Jung, D., Kang, J.-U., and Kim, J.-S.}
\newblock $\mu$-tree: {An} ordered index structure for {NAND} flash memory.
\newblock In {\em Proceedings of the 7th ACM \& IEEE international conference
  on Embedded software\/} (2007), pp.~144--153.

\bibitem{kang2014multi}
{\sc Kang, J.-U., Hyun, J., Maeng, H., and Cho, S.}
\newblock The multi-streamed solid-state drive.
\newblock In {\em 6th {USENIX} Workshop on Hot Topics in Storage and File
  Systems (HotStorage 14)\/} (2014).

\bibitem{kannan2018redesigning}
{\sc Kannan, S., Bhat, N., Gavrilovska, A., Arpaci-Dusseau, A., and
  Arpaci-Dusseau, R.}
\newblock Redesigning {LSMs} for {Nonvolatile Memory} with {NoveLSM}.
\newblock In {\em 2018 USENIX Annual Technical Conference (USENIX ATC 18)\/}
  (2018), pp.~993--1005.

\bibitem{khan2014big}
{\sc Khan, N., Yaqoob, I., Hashem, I. A.~T., Inayat, Z., Mahmoud~Ali, W.~K.,
  Alam, M., Shiraz, M., and Gani, A.}
\newblock Big data: survey, technologies, opportunities, and challenges.
\newblock {\em The scientific world journal 2014\/} (2014).

\bibitem{kim2019isolation}
{\sc Kim, H., Yeom, H.~Y., and Son, Y.}
\newblock An i/o isolation scheme for key-value store on multiple solid-state
  drives.
\newblock In {\em 2019 IEEE 4th International Workshops on Foundations and
  Applications of Self* Systems (FAS* W)\/} (2019), IEEE, pp.~170--175.

\bibitem{kim2021optimizing}
{\sc Kim, S., and Son, Y.}
\newblock Optimizing {Key-Value Stores} for {Flash-Based SSDs} via {Key
  Reshaping}.
\newblock {\em IEEE Access 9\/} (2021), 115135--115144.

\bibitem{koh2018exploring}
{\sc Koh, S., Lee, C., Kwon, M., and Jung, M.}
\newblock {Exploring System Challenges} of {Ultra-Low Latency Solid State
  Drives}.
\newblock In {\em 10th USENIX Workshop on Hot Topics in Storage and File
  Systems (HotStorage 18)\/} (2018).

\bibitem{kondylakis2020coconut}
{\sc Kondylakis, H., Dayan, N., Zoumpatianos, K., and Palpanas, T.}
\newblock Coconut: A scalable bottom-up approach for building data series
  indexes.
\newblock {\em arXiv preprint arXiv:2006.13713\/} (2020).

\bibitem{koronen2020data}
{\sc Koronen, C., {\AA}hman, M., and Nilsson, L.~J.}
\newblock Data centres in future european energy systems—energy efficiency,
  integration and policy.
\newblock {\em Energy Efficiency 13}, 1 (2020), 129--144.

\bibitem{kourtis2019reaping}
{\sc Kourtis, K., Ioannou, N., and Koltsidas, I.}
\newblock Reaping the performance of fast {NVM} storage with {uDepot}.
\newblock In {\em 17th {USENIX} Conference on File and Storage Technologies
  ({FAST} 19)\/} (2019), pp.~1--15.

\bibitem{krish2014hats}
{\sc Krish, K., Anwar, A., and Butt, A.~R.}
\newblock hats: {A} heterogeneity-aware tiered storage for hadoop.
\newblock In {\em 2014 14th IEEE/ACM International Symposium on Cluster, Cloud
  and Grid Computing\/} (2014), IEEE, pp.~502--511.

\bibitem{kudale11575258b+}
{\sc Kudale, A.}
\newblock B+ tree {Preference} over {B Tree}.
\newblock {\em Chicago, USA http://www. academia.
  edu/11575258/B\_tree\_preference\_over\_B\_trees\/}.

\bibitem{kuszmaul2014comparison}
{\sc Kuszmaul, B.~C.}
\newblock A comparison of fractal trees to log-structured merge ({LSM}) trees.
\newblock {\em Tokutek White Paper\/} (2014).

\bibitem{lakshman2010cassandra}
{\sc Lakshman, A., and Malik, P.}
\newblock Cassandra: a decentralized structured storage system.
\newblock {\em ACM SIGOPS Operating Systems Review 44}, 2 (2010), 35--40.

\bibitem{landsman2013ahci}
{\sc Landsman, D., and Walker, D.}
\newblock {AHCI} and {NVMe} as interfaces for {SATA Express™ Devices}, 2013.

\bibitem{lee2019ilsm}
{\sc Lee, C.-G., Kang, H., Park, D., Park, S., Kim, Y., Noh, J., Chung, W., and
  Park, K.}
\newblock {iLSM-SSD}: An intelligent {LSM-tree} based key-value {SSD} for data
  analytics.
\newblock In {\em 2019 IEEE 27th International Symposium on Modeling, Analysis,
  and Simulation of Computer and Telecommunication Systems (MASCOTS)\/} (2019),
  IEEE, pp.~384--395.

\bibitem{lee2021boosting}
{\sc Lee, J., Oh, G., and Lee, S.-W.}
\newblock Boosting {Compaction} in {B-Tree Based} {Key-Value Store} by
  {Exploiting} {Parallel Reads} in {Flash} {SSDs}.
\newblock {\em IEEE Access 9\/} (2021), 56344--56353.

\bibitem{lee2008case}
{\sc Lee, S.-W., Moon, B., Park, C., Kim, J.-M., and Kim, S.-W.}
\newblock A case for flash memory {SSD} in enterprise database applications.
\newblock In {\em Proceedings of the 2008 ACM SIGMOD international conference
  on Management of data\/} (2008), pp.~1075--1086.

\bibitem{lepers2019kvell}
{\sc Lepers, B., Balmau, O., Gupta, K., and Zwaenepoel, W.}
\newblock Kvell: the design and implementation of a fast persistent key-value
  store.
\newblock In {\em Proceedings of the 27th ACM Symposium on Operating Systems
  Principles\/} (2019), pp.~447--461.

\bibitem{li2020can}
{\sc Li, B., Song, N.~Y., Ou, L., and Du, D.~H.}
\newblock Can we store the whole world's data in {DNA} storage?
\newblock In {\em Proceedings of the 12th USENIX Conference on Hot Topics in
  Storage and File Systems\/} (2020), pp.~15--15.

\bibitem{li2021differentiated}
{\sc Li, Y., Liu, Z., Lee, P.~P., Wu, J., Xu, Y., Wu, Y., Tang, L., Liu, Q.,
  and Cui, Q.}
\newblock Differentiated {Key-Value} {Storage Management} for {Balanced I/O
  Performance}.
\newblock In {\em 2021 USENIX Annual Technical Conference (USENIX ATC 21)\/}
  (2021), pp.~673--687.

\bibitem{li2019elasticbf}
{\sc Li, Y., Tian, C., Guo, F., Li, C., and Xu, Y.}
\newblock {ElasticBF}: {Elastic Bloom Filter} with {Hotness Awareness} for
  {Boosting Read Performance} in {Large} {Key-Value Stores}.
\newblock In {\em 2019 USENIX Annual Technical Conference (USENIX ATC 19)\/}
  (2019), pp.~739--752.

\bibitem{lim2011silt}
{\sc Lim, H., Fan, B., Andersen, D.~G., and Kaminsky, M.}
\newblock {SILT}: {A} memory-efficient, high-performance key-value store.
\newblock In {\em Proceedings of the Twenty-Third ACM Symposium on Operating
  Systems Principles\/} (2011), pp.~1--13.

\bibitem{liu2015read}
{\sc Liu, C.-Y., Chang, Y.-M., and Chang, Y.-H.}
\newblock Read leveling for flash storage systems.
\newblock In {\em Proceedings of the 8th ACM International Systems and Storage
  Conference\/} (2015), pp.~1--10.

\bibitem{liu2021ptierdb}
{\sc Liu, L., and Zhou, K.}
\newblock {PTierDB}: {Building Better Read-Write Cost Balanced Key-Value Stores
  for Small Data on SSD}.
\newblock In {\em 2021 Design, Automation \& Test in Europe Conference \&
  Exhibition (DATE)\/} (2021), IEEE, pp.~796--801.

\bibitem{lu2012bloomstore}
{\sc Lu, G., Nam, Y.~J., and Du, D.~H.}
\newblock {BloomStore}: {Bloom-filter} based memory-efficient key-value store
  for indexing of data deduplication on flash.
\newblock In {\em 2012 IEEE 28th Symposium on Mass Storage Systems and
  Technologies (MSST)\/} (2012), IEEE, pp.~1--11.

\bibitem{lu2017wisckey}
{\sc Lu, L., Pillai, T.~S., Gopalakrishnan, H., Arpaci-Dusseau, A.~C., and
  Arpaci-Dusseau, R.~H.}
\newblock Wisckey: {Separating keys} from values in {SSD}-conscious storage.
\newblock {\em ACM Transactions on Storage (TOS) 13}, 1 (2017), 1--28.

\bibitem{lukken2021past}
{\sc Lukken, C., and Trivedi, A.}
\newblock {Past, Present and Future of Computational Storage: A Survey}.
\newblock {\em arXiv preprint arXiv:2112.09691\/} (2021).

\bibitem{luo2020lsm}
{\sc Luo, C., and Carey, M.~J.}
\newblock {LSM-based} storage techniques: a survey.
\newblock {\em The VLDB Journal 29}, 1 (2020), 393--418.

\bibitem{ma2020low}
{\sc Ma, K., Liu, M., Li, T., Yin, Y., and Chen, H.}
\newblock A low-cost improved method of raw bit error rate estimation for nand
  flash memory of high storage density.
\newblock {\em Electronics 9}, 11 (2020), 1900.

\bibitem{mansouri2017data}
{\sc Mansouri, Y., Toosi, A.~N., and Buyya, R.}
\newblock Data storage management in cloud environments: {Taxonomy}, survey,
  and future directions.
\newblock {\em ACM Computing Surveys (CSUR) 50}, 6 (2017), 1--51.

\bibitem{marmol2015nvmkv}
{\sc Marmol, L., Sundararaman, S., Talagala, N., and Rangaswami, R.}
\newblock {NVMKV}: {A Scalable, Lightweight, FTL-aware Key-Value Store}.
\newblock In {\em 2015 {USENIX} Annual Technical Conference ({USENIX}{ATC}
  15)\/} (2015), pp.~207--219.

\bibitem{mei2018sifrdb}
{\sc Mei, F., Cao, Q., Jiang, H., and Li, J.}
\newblock {SifrDB}: {A} unified solution for write-optimized key-value stores
  in large datacenter.
\newblock In {\em Proceedings of the ACM Symposium on Cloud Computing\/}
  (2018), pp.~477--489.

\bibitem{menon2014cassandra}
{\sc Menon, P., Rabl, T., Sadoghi, M., and Jacobsen, H.-A.}
\newblock {CaSSanDra}: An {SSD} boosted key-value store.
\newblock In {\em 2014 IEEE 30th International Conference on Data
  Engineering\/} (2014), IEEE, pp.~1162--1167.

\bibitem{micheloni2010inside}
{\sc Micheloni, R., Crippa, L., and Marelli, A.}
\newblock {\em Inside {NAND} flash memories}.
\newblock Springer Science \& Business Media, 2010.

\bibitem{nath2007flashdb}
{\sc Nath, S., and Kansal, A.}
\newblock {FlashDB}: {Dynamic} self-tuning database for {NAND} flash.
\newblock In {\em Proceedings of the 6th international conference on
  Information processing in sensor networks\/} (2007), pp.~410--419.

\bibitem{nguyen2018optimizing}
{\sc Nguyen, T.-D., and Lee, S.-W.}
\newblock Optimizing mongodb using multi-streamed ssd.
\newblock In {\em Proceedings of the 7th International Conference on Emerging
  Databases\/} (2018), Springer, pp.~1--13.

\bibitem{ouyang2014sdf}
{\sc Ouyang, J., Lin, S., Jiang, S., Hou, Z., Wang, Y., and Wang, Y.}
\newblock {SDF}: {Software-defined} flash for web-scale internet storage
  systems.
\newblock In {\em Proceedings of the 19th international conference on
  Architectural support for programming languages and operating systems\/}
  (2014), pp.~471--484.

\bibitem{o1996log}
{\sc O’Neil, P., Cheng, E., Gawlick, D., and O’Neil, E.}
\newblock The log-structured merge-tree ({LSM-tree}).
\newblock {\em Acta Informatica 33}, 4 (1996), 351--385.

\bibitem{papagiannis2016tucana}
{\sc Papagiannis, A., Saloustros, G., Gonz{\'a}lez-F{\'e}rez, P., and Bilas,
  A.}
\newblock Tucana: Design and implementation of a fast and efficient scale-up
  key-value store.
\newblock In {\em 2016 {USENIX} Annual Technical Conference ({USENIX}{ATC}
  16)\/} (2016), pp.~537--550.

\bibitem{papagiannis2018efficient}
{\sc Papagiannis, A., Saloustros, G., Gonz{\'a}lez-F{\'e}rez, P., and Bilas,
  A.}
\newblock An efficient memory-mapped key-value store for flash storage.
\newblock In {\em Proceedings of the ACM Symposium on Cloud Computing\/}
  (2018), pp.~490--502.

\bibitem{park2011comprehensive}
{\sc Park, S., Kim, Y., Urgaonkar, B., Lee, J., and Seo, E.}
\newblock A comprehensive study of energy efficiency and performance of
  flash-based ssd.
\newblock {\em Journal of Systems Architecture 57}, 4 (2011), 354--365.

\bibitem{patrizio2018idc}
{\sc Patrizio, A.}
\newblock {IDC}: {Expect} 175 zettabytes of data worldwide by 2025.
\newblock {\em Network World\/} (2018).

\bibitem{pouyanfar2018multimedia}
{\sc Pouyanfar, S., Yang, Y., Chen, S.-C., Shyu, M.-L., and Iyengar, S.}
\newblock Multimedia big data analytics: A survey.
\newblock {\em ACM computing surveys (CSUR) 51}, 1 (2018), 1--34.

\bibitem{pritchett2008base}
{\sc Pritchett, D.}
\newblock {BASE}: An {Acid} alternative: In partitioned databases, trading some
  consistency for availability can lead to dramatic improvements in
  scalability.
\newblock {\em Queue 6}, 3 (2008), 48--55.

\bibitem{qiao2021closing}
{\sc Qiao, Y., Chen, X., Zheng, N., Li, J., Liu, Y., and Zhang, T.}
\newblock Closing the {B-tree} vs. {LSM-tree Write Amplification Gap} on
  {Modern Storage Hardware} with {Built-in Transparent Compression}.
\newblock {\em arXiv preprint arXiv:2107.13987\/} (2021).

\bibitem{qin2021kvraid}
{\sc Qin, M., Reddy, A.~N., Gratz, P.~V., Pitchumani, R., and Ki, Y.~S.}
\newblock {KVRAID}: high performance, write efficient, update friendly erasure
  coding scheme for {KV-SSDs}.
\newblock In {\em Proceedings of the 14th ACM International Conference on
  Systems and Storage\/} (2021), pp.~1--12.

\bibitem{raina2020prismdb}
{\sc Raina, A., Cidon, A., Jamieson, K., and Freedman, M.~J.}
\newblock {PrismDB}: {Read-aware} log-structured merge trees for heterogeneous
  storage.
\newblock {\em arXiv preprint arXiv:2008.02352\/} (2020).

\bibitem{raju2017pebblesdb}
{\sc Raju, P., Kadekodi, R., Chidambaram, V., and Abraham, I.}
\newblock {PebblesDB}: {Building} key-value stores using fragmented
  log-structured merge trees.
\newblock In {\em Proceedings of the 26th Symposium on Operating Systems
  Principles\/} (2017), pp.~497--514.

\bibitem{rasmussen2008round}
{\sc Rasmussen, R.~V., and Trick, M.~A.}
\newblock Round robin scheduling--a survey.
\newblock {\em European Journal of Operational Research 188}, 3 (2008),
  617--636.

\bibitem{ren2017slimdb}
{\sc Ren, K., Zheng, Q., Arulraj, J., and Gibson, G.}
\newblock {SlimDB}: {A} space-efficient key-value storage engine for
  semi-sorted data.
\newblock {\em Proceedings of the VLDB Endowment 10}, 13 (2017), 2037--2048.

\bibitem{risch2015introduction}
{\sc Risch, T.}
\newblock Introduction to {NoSQL Databases}.
\newblock {\em NoSQLDatabases.pdf\/} (2015).

\bibitem{rothermel1989aries}
{\sc Rothermel, K., and Mohan, C.}
\newblock {\em ARIES/NT: A Recovery Method Based on Write-Ahead Logging for
  Nested Transactions.}
\newblock IBM Thomas J. Watson Research Division, 1989.

\bibitem{rottenstreich2014bloom}
{\sc Rottenstreich, O., and Keslassy, I.}
\newblock The bloom paradox: {When} not to use a bloom filter.
\newblock {\em IEEE/ACM Transactions on Networking 23}, 3 (2014), 703--716.

\bibitem{rusinkiewicz1995specification}
{\sc Rusinkiewicz, M., and Sheth, A.~P.}
\newblock {Specification} and {Execution} of {Transactional Workflows}.
\newblock {\em Modern database systems 1995\/} (1995), 592--620.

\bibitem{saito2013energy}
{\sc Saito, T., Sato, K., Sato, H., and Matsuoka, S.}
\newblock Energy-aware i/o optimization for checkpoint and restart on a nand
  flash memory system.
\newblock In {\em Proceedings of the 3rd Workshop on Fault-tolerance for HPC at
  Extreme Scale\/} (2013), pp.~41--48.

\bibitem{seeger2009key}
{\sc Seeger, M., and Ultra-Large-Sites, S.}
\newblock {Key-Value} stores: a practical overview.
\newblock {\em Computer Science and Media, Stuttgart\/} (2009).

\bibitem{sharma2012sql}
{\sc Sharma, V., and Dave, M.}
\newblock Sql and nosql databases.
\newblock {\em International Journal of Advanced Research in Computer Science
  and Software Engineering 2}, 8 (2012).

\bibitem{sivasubramanian2012amazon}
{\sc Sivasubramanian, S.}
\newblock Amazon {dynamoDB}: a seamlessly scalable non-relational database
  service.
\newblock In {\em Proceedings of the 2012 ACM SIGMOD International Conference
  on Management of Data\/} (2012), pp.~729--730.

\bibitem{stavrinos2021don}
{\sc Stavrinos, T., Berger, D.~S., Katz-Bassett, E., and Lloyd, W.}
\newblock Don't be a blockhead: zoned namespaces make work on conventional ssds
  obsolete.
\newblock In {\em Proceedings of the Workshop on Hot Topics in Operating
  Systems\/} (2021), pp.~144--151.

\bibitem{sun2020cascaded}
{\sc Sun, H., Dai, S., and Huang, J.}
\newblock {Cascaded Write Amplification} of {LSM-tree-based Key-Value Stores}
  underlying {Solid-State Disks}.
\newblock {\em Microprocessors and Microsystems 78\/} (2020), 103217.

\bibitem{sun2018co}
{\sc Sun, H., Liu, W., Huang, J., and Shi, W.}
\newblock Co-kv: {A} collaborative key-value store using near-data processing
  to improve compaction for the lsm-tree.
\newblock {\em arXiv preprint arXiv:1807.04151\/} (2018).

\bibitem{syed2013future}
{\sc Syed, A., Gillela, K., and Venugopal, C.}
\newblock The future revolution on big data.
\newblock {\em Future 2}, 6 (2013), 2446--2451.

\bibitem{tokutek2013tokudb}
{\sc TOKUTEK, I.}
\newblock {TokuDB}: {MySQL} performance, {MariaDB} {Performance}, 2013.

\bibitem{tsai2010towards}
{\sc Tsai, W.-T., Shao, Q., Huang, Y., and Bai, X.}
\newblock Towards a scalable and robust multi-tenancy saas.
\newblock In {\em Proceedings of the Second Asia-Pacific Symposium on
  Internetware\/} (2010), pp.~1--15.

\bibitem{tsirogiannis2010analyzing}
{\sc Tsirogiannis, D., Harizopoulos, S., and Shah, M.~A.}
\newblock Analyzing the energy efficiency of a database server.
\newblock In {\em Proceedings of the 2010 ACM SIGMOD International Conference
  on Management of data\/} (2010), pp.~231--242.

\bibitem{tulkinbekov2020casedb}
{\sc Tulkinbekov, K., and Kim, D.-H.}
\newblock {CaseDB}: {Lightweight Key-Value Store} for {Edge Computing
  Environment}.
\newblock {\em IEEE Access 8\/} (2020), 149775--149786.

\bibitem{vinccon2018noftl}
{\sc Vin{\c{c}}on, T., Hardock, S., Riegger, C., Oppermann, J., Koch, A., and
  Petrov, I.}
\newblock Noftl-kv: Tackling write-amplification on kv-stores with native
  storage management.
\newblock In {\em Advances in database technology-EDBT 2018: 21st International
  Conference on Extending Database Technology, Vienna, Austria, March 26-29,
  2018. proceedings\/} (2018), University of Konstanz, University Library,
  pp.~457--460.

\bibitem{wang2014efficient}
{\sc Wang, P., Sun, G., Jiang, S., Ouyang, J., Lin, S., Zhang, C., and Cong,
  J.}
\newblock An efficient design and implementation of {LSM-tree} based key-value
  store on open-channel {SSD}.
\newblock In {\em Proceedings of the Ninth European Conference on Computer
  Systems\/} (2014), pp.~1--14.

\bibitem{wang2020temperature}
{\sc Wang, Y., Tan, J., Mao, R., and Li, T.}
\newblock Temperature-aware persistent data management for {LSM-tree} on {3-D
  NAND} flash memory.
\newblock {\em IEEE Transactions on Computer-Aided Design of Integrated
  Circuits and Systems 39}, 12 (2020), 4611--4622.

\bibitem{wu2003efficient}
{\sc Wu, C.-H., Chang, L.-P., and Kuo, T.-W.}
\newblock An efficient {B-tree} layer for flash-memory storage systems.
\newblock In {\em International Conference on Real-Time and Embedded Computer
  Systems and Applications\/} (2003), Springer, pp.~409--430.

\bibitem{wu2018kvssd}
{\sc Wu, S.-M., Lin, K.-H., and Chang, L.-P.}
\newblock {KVSSD}: {Close integration} of {LSM trees} and flash translation
  layer for write-efficient {KV store}.
\newblock In {\em 2018 Design, Automation \& Test in Europe Conference \&
  Exhibition (DATE)\/} (2018), IEEE, pp.~563--568.

\bibitem{xanthakis2021parallax}
{\sc Xanthakis, G., Saloustros, G., Batsaras, N., Papagiannis, A., and Bilas,
  A.}
\newblock Parallax: {Hybrid Key-Value Placement} in {LSM-based} {Key-Value
  Stores}.
\newblock In {\em Proceedings of the ACM Symposium on Cloud Computing\/}
  (2021), pp.~305--318.

\bibitem{xu2015performance}
{\sc Xu, Q., Siyamwala, H., Ghosh, M., Awasthi, M., Suri, T., Guz, Z.,
  Shayesteh, A., and Balakrishnan, V.}
\newblock Performance characterization of hyperscale applicationson on nvme
  ssds.
\newblock In {\em Proceedings of the 2015 ACM SIGMETRICS International
  Conference on Measurement and Modeling of Computer Systems\/} (2015),
  pp.~473--474.

\bibitem{yang2015optimizing}
{\sc Yang, F., Dou, K., Chen, S., Hou, M., Kang, J.-U., and Cho, S.}
\newblock Optimizing {NoSQL DB} on flash: a case study of {RocksDB}.
\newblock In {\em 2015 IEEE 12th Intl Conf on Ubiquitous Intelligence and
  Computing and 2015 IEEE 12th Intl Conf on Autonomic and Trusted Computing and
  2015 IEEE 15th Intl Conf on Scalable Computing and Communications and Its
  Associated Workshops (UIC-ATC-ScalCom)\/} (2015), IEEE, pp.~1062--1069.

\bibitem{yang2014garbage}
{\sc Yang, M.-C., Chang, Y.-M., Tsao, C.-W., Huang, P.-C., Chang, Y.-H., and
  Kuo, T.-W.}
\newblock Garbage collection and wear leveling for flash memory: {Past} and
  future.
\newblock In {\em 2014 International Conference on Smart Computing\/} (2014),
  IEEE, pp.~66--73.

\bibitem{yang2017spdk}
{\sc Yang, Z., Harris, J.~R., Walker, B., Verkamp, D., Liu, C., Chang, C., Cao,
  G., Stern, J., Verma, V., and Paul, L.~E.}
\newblock {SPDK}: {A} development kit to build high performance storage
  applications.
\newblock In {\em 2017 IEEE International Conference on Cloud Computing
  Technology and Science (CloudCom)\/} (2017), IEEE, pp.~154--161.

\bibitem{zhan2020rangekv}
{\sc Zhan, L., Lu, K., Cheng, Z., and Wan, J.}
\newblock Rangekv: An efficient key-value store based on hybrid dram-nvm-ssd
  storage structure.
\newblock {\em IEEE Access 8\/} (2020), 154518--154529.

\bibitem{zhan2019design}
{\sc Zhan, L., Zhang, Y., and Yu, K.}
\newblock Design and implementation of {SCM} and {SSD} based hybrid key-value
  store.
\newblock In {\em Proceedings of the 2019 International Conference on
  Artificial Intelligence and Computer Science\/} (2019), pp.~566--572.

\bibitem{zhang2017flashkv}
{\sc Zhang, J., Lu, Y., Shu, J., and Qin, X.}
\newblock Flashkv: Accelerating kv performance with open-channel ssds.
\newblock {\em ACM Transactions on Embedded Computing Systems (TECS) 16}, 5s
  (2017), 1--19.

\bibitem{zhang2021chameleondb}
{\sc Zhang, W., Zhao, X., Jiang, S., and Jiang, H.}
\newblock {ChameleonDB}: a key-value store for optane persistent memory.
\newblock In {\em Proceedings of the Sixteenth European Conference on Computer
  Systems\/} (2021), pp.~194--209.

\end{thebibliography}

\end{document}